\documentclass[graybox]{svmult}

\usepackage{type1cm}         
\usepackage{makeidx}         
\usepackage{graphicx}        
\usepackage{multicol}        
\usepackage[bottom]{footmisc}
\usepackage{url}
\usepackage{amsthm}
\usepackage{amssymb}
\usepackage{setspace}
\usepackage{numprint} 
\npthousandsep{,}
\usepackage[authoryear]{natbib}
\usepackage{graphicx}
\usepackage{xcolor}
\usepackage[normalem]{ulem}
\useunder{\uline}{\ul}{}
\usepackage[ruled]{algorithm2e}
\SetKwInput{KwInput}{Input}
\SetKwInput{KwOutput}{Output}
\usepackage{float}
\usepackage{mathtools}
\usepackage[hidelinks]{hyperref}

\usepackage{newtxtext}       %
\usepackage[varvw]{newtxmath}       
\usepackage[authoryear]{natbib}

\makeindex             

\newcommand{\im}{\mathrm{i}}

\begin{document}

\title*{Spectral domain likelihoods for Bayesian inference in time-varying parameter models}
\author{Oskar Gustafsson, \\ Mattias Villani and\\ Robert Kohn}
\institute{Oskar Gustafsson \at Department of Statistics, Stockholm University, SE-106 91 Stockholm, Sweden \email{oskar.gustafsson@stat.su.se}
\and Mattias Villani \at Department of Statistics, Stockholm University, SE-106 91 Stockholm, Sweden \email{mattias.villani@stat.su.se}
\and Robert Kohn \at School of Economics, University of New South Wales, NSW 2052 Sydney, Australia \email{r.kohn@unsw.edu.au}}

\maketitle

\abstract*{Inference for locally stationary processes is often based on some local Whittle-type approximation of the likelihood function defined in the frequency domain. The main reasons for using such a likelihood approximation is that i) it has substantially lower computational cost and better scalability to long time series compared to the time domain likelihood, particularly when used for Bayesian inference via Markov Chain Monte Carlo (MCMC), ii) convenience when the model itself is specified in the frequency domain, and iii) it provides access to bootstrap and subsampling MCMC which exploits the asymptotic independence of Fourier transformed data. Most of the existing literature compares the asymptotic performance of the maximum likelihood estimator (MLE) from such frequency domain likelihood approximation with the exact time domain MLE. Our article uses three simulation studies to assess the finite-sample accuracy of several frequency domain likelihood functions when used to approximate the posterior distribution in time-varying parameter models. The methods are illustrated on an application to egg price data.}

\abstract{Inference for locally stationary processes is often based on some local Whittle-type approximation of the likelihood function defined in the frequency domain. The main reasons for using such a likelihood approximation is that i) it has substantially lower computational cost and better scalability to long time series compared to the time domain likelihood, particularly when used for Bayesian inference via Markov Chain Monte Carlo (MCMC), ii) convenience when the model itself is specified in the frequency domain, and iii) it provides access to bootstrap and subsampling MCMC which exploits the asymptotic independence of Fourier transformed data. Most of the existing literature compares the asymptotic performance of the maximum likelihood estimator (MLE) from such frequency domain likelihood approximation with the exact time domain MLE. Our article uses three simulation studies to assess the finite-sample accuracy of several frequency domain likelihood functions when used to approximate the posterior distribution in time-varying parameter models. The methods are illustrated on an application to egg price data.}

\section{Introduction}\label{sec:intro} 

Innovations in computing and storage technology allow low cost data collection in a variety of applications. Data from sensors, cameras, and other recording devices are naturally time-stamped observations, which has motivated methodological developments in modeling, estimation, and prediction for time series. Long time series with many observations are increasingly common, either collected over long study periods or over shorter periods at high frequency. 

Inference for a long time series can be computationally challenging, even under stationarity. Evaluating the time domain likelihood for a stationary Gaussian model without special structure has a computational complexity of $O(T^3)$, where $T$ is the length of the time series, due to the inversion of the $T \times T$ covariance matrix. A computationally more attractive alternative is the Whittle likelihood \citep{whittle1953estimation} which approximates the Gaussian likelihood in the frequency domain with a complexity of only $O(T \log T)$. The approximation is motivated by the asymptotic independence of the discrete Fourier transform of the time series, a result that has been generalized to non-Gaussian \citep{hannan1973asymptotic, peligrad2010central}, long-memory processes \citep{fox1986large} and non-linear processes \citep{Shao2007}. The Whittle likelihood requires only the spectral density of the model, which makes it particularly useful for models defined directly in the frequency domain, see e.g. \citet{rosen2012adaptspec}. The asymptotically independent data in the frequency domain has also been exploited for the bootstrap \citep{franke1992bootstrapping} and for speeding up Markov Chain Monte Carlo (MCMC) algorithms by subsampling periodogram data \citep{salomone2020spectral}.  

The MLE based on the Whittle approximation is known to be efficient in finite samples for Gaussian processes, except for short time series with high autocorrelation, see \citet{dahlhaus1988small} and \citet{contreras2006note} and for certain non-Gaussian processes \citet{contreras2006note}. There are many suggestions on how to reduce the bias of the Whittle MLE: tapering \citep{dahlhaus1988small}, prewhitening \citep{priestly1981spectral}, boundary correction \citep{subba2021reconciling}, and replacing the spectral density in the likelihood approximation with the expected periodogram \citep{sykulski2019debiased}. 

The traditional assumption of a time invariant stationary process is often unreasonable, however, particularly for long time series which typically undergo some structural changes. Much effort has therefore gone into developing more flexible models that allow the properties of the process to change over time. One popular class of models allows the process to switch between a finite set of stationary regimes or segments, see e.g. Markov switching models in \citet{hamilton1989new}, change-point models in \citet{chib1998estimation} and \citet{rosen2012adaptspec}, and dynamic mixture innovations models in \citet{gerlach2000efficient}. Another line of research develops models and theory for locally stationary process where the parameters change more smoothly over time, such that the process is locally stationary in a small interval around each time point, see \citet{dahlhaus2012locally} for a review. This framework is extended in Gaussian setting by \citet{dahlhausPolonik2009empirical} to allow for jumps in the parameters. Perhaps the most widely used models in applications are time-varying autoregressive (tvAR) models \citep{prado2010time} where the parameters follow a stochastic process over time. The parameter process is typically taken to be a random walk or an AR(1) process with Gaussian innovations, but more recent attempts use local-global shrinkage priors 
\citep{carvalho2010horseshoe} to allow for both smoothly changing parameters, periods with no change and abrupt jumps; see e.g. \citet{kalli2014time}, \citet{kowal2019dynamic}, \citet{cadonna2020triple} and \citet{knaus2023dynamic}.

The Whittle likelihood has been generalized to models with time-varying parameters, often using a variant of the original Whittle likelihood over time segments. The \emph{block Whittle likelihood} \citep{dahlhaus1997fitting} uses time-varying periodograms over sliding and partially overlapping time windows. The block Whittle MLE is asymptotically efficient under some conditions on the parameter evolution \citep{dahlhaus2012locally}. The \emph{generalized Whittle likelihood} \citep{dahlhaus2000likelihood} does not use periodograms over segments but is instead based on so-called preperiodogram data \citep{neumann1997wavelet} computed at every time point and frequency as the Fourier transform of a locally estimated autocovariance at lag $k$ from a single pair of observations $k$ steps apart. The preperiodogram is attractive since it gives maximal time- and frequency resolution and is used in several studies (e.g. \cite{van2019data}), has good asymptotic properties \citep{dahlhaus2009local,dahlhausPolonik2009empirical}, but has interfering cross-terms which can make inference more difficult \citep{sandsten2016time}. \cite{everitt2013online} used a modified preperiodogram which is smoothed in both time- and frequency direction for efficient online learning of time-varying spectral densities. The \textit{dynamic Whittle likelihood} in \citet{tang2023bayesian} is based on the \textit{moving local periodogram} introduced in \cite{hafner2017moving}. The moving local periodogram evaluates a single frequency at each point in time, instead of all Fourier frequencies, and still maintains approximate independence among most observations since neighboring time points use different frequencies. \citet{tang2023bayesian} use their approximate likelihood for Bayesian inference, prove posterior consistency and obtain contraction rates. 

The local Whittle likelihoods for time-varying parameter models compute local periodograms in relatively short moving time windows, and are therefore prone to the same small-sample problems as the Whittle likelihood in the stationary case. This chapter explores the finite sample performance of the posterior distributions from local Whittle likelihoods for time-varying parameters models through several simulation experiments. In contrast to most previous studies exploring the frequentist properties of the Whittle MLE, we focus on how well the posterior distribution for the time-varying parameters is approximated when a local Whittle likelihood is used. We compare the block Whittle likelihood in \citet{dahlhaus1997fitting} and the dynamic Whittle likelihood in \citet{tang2023bayesian}. The data generating process in the experiments is taken to be a simple time-varying AR (tvAR) process with parameter evolutions following a random walk with Gaussian innovations. The tvAR can be  estimated without spectral methods at a reasonable computational cost, and is therefore used here as it makes it possible to easily compare the results to the gold standard posterior from the time domain likelihood. We use a tvAR process that is restricted to be stable at every time period using the parameterization in \citet{Barndorff-Nielsen1973} and \citet{monahan1984note}, and we sample from the posterior distribution using Gibbs sampling with a particle MCMC \citep{lindsten2014particle} update of the time-varying parameters. We also investigate the potential of three proposed modifications for improving the local Whittle approximations: tapering \citep{dahlhaus1988small}, prewhitening \citep{priestly1981spectral}, and boundary correction \citep{subba2021reconciling}. The effect of the segment length and overlap is also explored. The chapter concludes by illustrating the differences in the tvAR posterior between the compared likelihoods and the modifications in an application to a widely used time series on egg prices, with the additional complication of having a heteroscedastic variance.

The supplementary material to this article contains additional results, with figures  referenced with the prefix S, e.g. Figure \ref{fig:DahlhausGood_DW_M30} in Section \ref{supp:exp1}. 

\section{The Whittle likelihood for stationary processes}\label{sec:whittle}

This section reviews the frequency domain concepts needed for the Whittle likelihood. A simulated time-invariant AR(1) example introduces the basic measures used to assess the performance of the Whittle likelihood as an approximation to the time domain likelihood.

Let $\{X_t\}_{t=1}^T$ be a covariance stationary zero-mean time series with covariance function $\gamma_\tau \coloneqq \mathbb{E}(X_t X_{t-\tau})$ for integer $\tau$.  The {\em spectral density} is the Fourier transform of $\{\gamma_\tau\}_{\tau=-\infty}^\infty$ \citep{lindgren2013stationary}
\begin{equation}\label{eq:spectralDens}
    f(\omega) \coloneqq \frac{1}{2\pi}\sum_{\tau = -\infty}^\infty \gamma_\tau e^{-\im \omega\tau},
\end{equation}
where $\omega \in (-\pi,\pi]$ is the {\em angular frequency}. As an example, consider the AR($p$) process
\begin{equation}\label{eq:ARmodel}
    \phi_{p}(L)(X_{t}-\mu)=\varepsilon_{t},\quad \varepsilon_t \overset{\mathrm{iid}}{\sim} N(0,\sigma_{\varepsilon}^2),
\end{equation}
where $\phi_{p}(L)\coloneqq1-\phi_{1}L-\cdots-\phi_{p}L^{p}$ is the autoregressive polynomial in the lag operator $L$, defined by $L^k X_t = X_{t-k}$. The AR($p$) process in \eqref{eq:ARmodel} has spectral density \citep{brockwell1991time}
\begin{equation}\label{eq:ARMAspectral}
    f(\omega)=\frac{\sigma_\varepsilon^2}{2\pi}\left|\phi_{p}(e^{-\im \omega})\right|^{-2}.
\end{equation}

The {\em discrete Fourier Transform} (DFT) of $\{X_t\}_{t=1}^T$ is the complex valued series 
\begin{equation}\label{eq:DFT}
    J(\omega_k) \coloneqq \sum_{t=1}^T X_t \exp(-\im \omega_k t),
\end{equation}
for $\omega_1,\ldots,\omega_T$ in the set of Fourier frequencies
$$\Omega_T = \{2\pi k/T, \text{ for } k = -\lceil T/2\rceil + 1,\ldots,\lfloor T/2 \rfloor \}.$$
The Fast Fourier Transform (FFT) algorithm computes the DFT efficiently  \citep{lindgren2013stationary}. The {\em periodogram} 
\begin{equation}
    I_T(\omega_k) \coloneqq \frac{1}{2\pi T} \vert J(\omega_k) \vert ^2
\end{equation}
is an asymptotically unbiased estimate of $f(\omega_k)$ for a stationary process with an absolutely summable covariance function \citep{brillinger2001time}.

The Whittle log-likelihood \citep{whittle1953estimation} is defined as
\begin{equation}\label{eq:whittlelikelihood}
\ell_{T}^{W}(\boldsymbol{\theta}) := -\frac{1}{2}\sum_{\omega_k \in \Omega_T} \bigg(  
    \log f_{\boldsymbol{\theta}}(\omega_k) + \frac{I_{T}(\omega_k)}{f_{\boldsymbol{\theta}}(\omega_k)}\bigg),
\end{equation}
where $\boldsymbol{\theta}$ is the vector of model parameters and $f_{\boldsymbol{\theta}}(\omega)$ is the model's spectral density. The term for $\omega = 0$ is removed since we assume that the process has zero mean; the term for $\omega=\pi$ is also typically removed for convenience since $I_{T}(\pi)$ follows a different distribution ($\chi^2_1$) than the other periodogram data points. 

The motivation for using \eqref{eq:whittlelikelihood} as an approximate likelihood comes from the following asymptotic result for the periodogram for a stationary Gaussian process \citep{whittle1951hypothesis}
\begin{equation}\label{eq:asympPeriodogram}
        {\cal I}(\omega_k)\ \overset{\mathrm{ind}}{\sim} \mathrm{Exp} \big(f(\omega_k)\big), \quad \omega_k \in \Omega_n,
\end{equation}
as $T\rightarrow \infty$, with the exponential distribution in the scale parameterization, i.e., parameterized by its mean; the symbol $\overset{\mathrm{ind}}{\sim}$ means independently distributed. Hence, the Whittle likelihood in \eqref{eq:whittlelikelihood} is obtained when the asymptotic result in \eqref{eq:asympPeriodogram} is assumed to hold for a finite sample size $T$. Whittle's original result holds under more general conditions; see \citet{hannan1973asymptotic} for potentially non-Gaussian linear processes,  \citet{peligrad2010central} for stationary ergodic processes, \citet{Shao2007} for non-linear processes with moment conditions, and \citet{peligrad2019central} for extensions to random fields.

The Whittle likelihood has the following advantages for likelihood and Bayesian inference: i) its computational cost is only $O(T \log T)$ compared to the $O(T^3)$ in the Gaussian likelihood without special structure of the covariance matrix; ii) it can be conveniently used for models directly defined for the spectral density \citep{wahba1980automatic,rosen2012adaptspec}; iii) the asymptotic independence of the periodogram data can be exploited in the bootstrap \citep{franke1992bootstrapping} and for fast subsampling MCMC \citep{quiroz2019speeding} by subsampling periodogram observations \citep{salomone2020spectral}.
 
We first illustrate the efficiency and accuracy of the Whittle likelihood as an approximation to the time domain likelihood when fitting a simple AR(1) model with time-invariant parameters. We follow the setup in \citet{contreras2006note} with the data generating process being
\begin{equation}
    x_t = \phi x_{t-1} + \varepsilon_t, \quad \varepsilon_t \sim N(0,1),
\end{equation}
so that the only parameter to estimate is $\phi$. We use a uniform prior and compute the posterior distribution over a fine grid of values for $\phi$ in the interval $[-0.999,0.999]$. Following \citet{contreras2006note}, the time domain likelihood is the exact Gaussian likelihood with the first observation following the stationary distribution of the process.

We assess the performance of the Whittle likelihood using both a frequentist efficiency measure, and a measure more adapted to Bayesian inference. Define the root mean square error (RMSE) of a point estimator $\hat{\boldsymbol{\phi}}$ of the AR parameters $\boldsymbol{\phi} = (\phi_1,\ldots,\phi_p)^T$ over $N_{\mathrm{rep}}$ replicated time series as
\begin{equation}\label{eq:RMSE}
    \mathrm{RMSE} = \sqrt{\frac{1}{N_{\mathrm{rep}}\cdot p}\sum_{i=1}^{N_{\mathrm{rep}}} \lVert \hat{\boldsymbol{\phi}}^{(i)} - \boldsymbol{\phi}\rVert^2 },
\end{equation}
where $\boldsymbol{\phi}$ contains the true AR parameters in the data generating process and $\lVert \cdot \rVert$ is the Euclidean norm. Our first performance measure is the efficiency of the Whittle MLE vs the exact time domain MLE
\begin{equation}\label{eq:efficiency}
    \text{Efficiency} = \frac{ \mathrm{RMSE}_{\mathrm{time}} } { \mathrm{RMSE}_{\mathrm{whittle}}}.
\end{equation}
Our second performance measure quantifies how much the Whittle posterior distribution of $\boldsymbol{\phi}$ deviates from the time domain posterior distribution, for a given dataset. Any measure of divergence between two distributions can be used, but we opt for an easily interpretable divergence between a set of quantiles. Let $\boldsymbol{q}_j^{(i)} = (q_{j1}^{(i)},\ldots,q_{jr}^{(i)})^\top$ be $r$ quantiles in the Whittle posterior distribution for $\phi_j$ based on dataset $i$, and let $\boldsymbol{s}_j^{(i)}$ denote the corresponding quantiles in the time domain posterior distribution. We measure how perturbed the Whittle posterior is with respect to the gold standard time domain posterior by
\begin{equation}\label{eq:perturb}
    \text{Perturb} = \sqrt{\frac{1}{N_{\mathrm{rep}}\cdot p \cdot r} \sum_{i=1}^{N_{\mathrm{rep}}} \sum_{j=1}^p \lVert \boldsymbol{q}_j^{(i)} -\boldsymbol{s}_j^{(i)} \rVert^2 },
\end{equation}
using the $r=7$ quantiles: $2.5\%$, $10\%$, $25\%$, $50\%$, $75\%$, $90\%$ and $97.5\%$.

\begin{figure}
    \centering
    \includegraphics[width=0.95\linewidth]{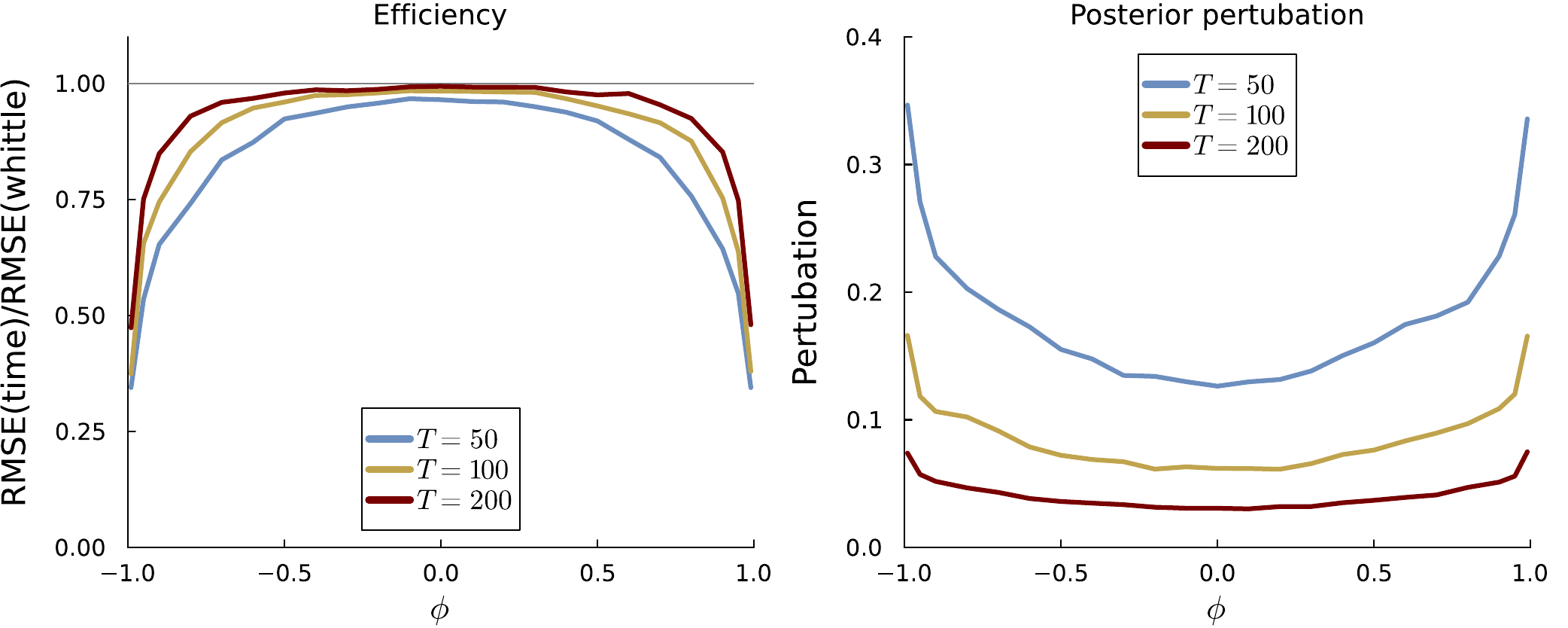}
    \caption{Efficiency and posterior perturbation when fitting a stationary AR(1). The left panel shows the efficiency in \eqref{eq:efficiency} of the Whittle MLE compared to the exact time domain MLE for point estimation of $\phi$ in the stationary AR(1) process $x_t = \phi x_{t-1} + \varepsilon_t, \varepsilon_t \sim N(0,1)$ for three different sample sizes. The right panel shows the degree of perturbation in  \eqref{eq:perturb} of the Whittle posterior from time domain posterior.}\label{fig:AR1_eff_perturb}
\end{figure}

Figure~\ref{fig:AR1_eff_perturb} plots the efficiency in \eqref{eq:efficiency} and the degree of perturbation in \eqref{eq:perturb} as a function of the true AR parameter $\phi$ for three different sample sizes. The left panel in 
Figure~\ref{fig:AR1_eff_perturb} shows that the relative efficiency of the Whittle MLE deteriorates as $\vert \phi \vert \rightarrow 1$; this extends the results in Table~6 of \citet{contreras2006note}, who report the efficiency for $T=50$ and $T=200$ at $\phi=0.2$ and $\phi=0.8$. The loss of efficiency near unit roots is also forcefully demonstrated by \citet{dahlhaus1988small} in a simulation experiment with $T=256$ observations from an AR($14$) process. The right hand panel of Figure~\ref{fig:AR1_eff_perturb} shows that the posterior perturbation from the exact time domain posterior increases with $\vert \phi \vert$. Both the efficiency and the posterior perturbation improves with $T$.

\begin{figure}
    \centering
    \includegraphics[width=0.95\linewidth]{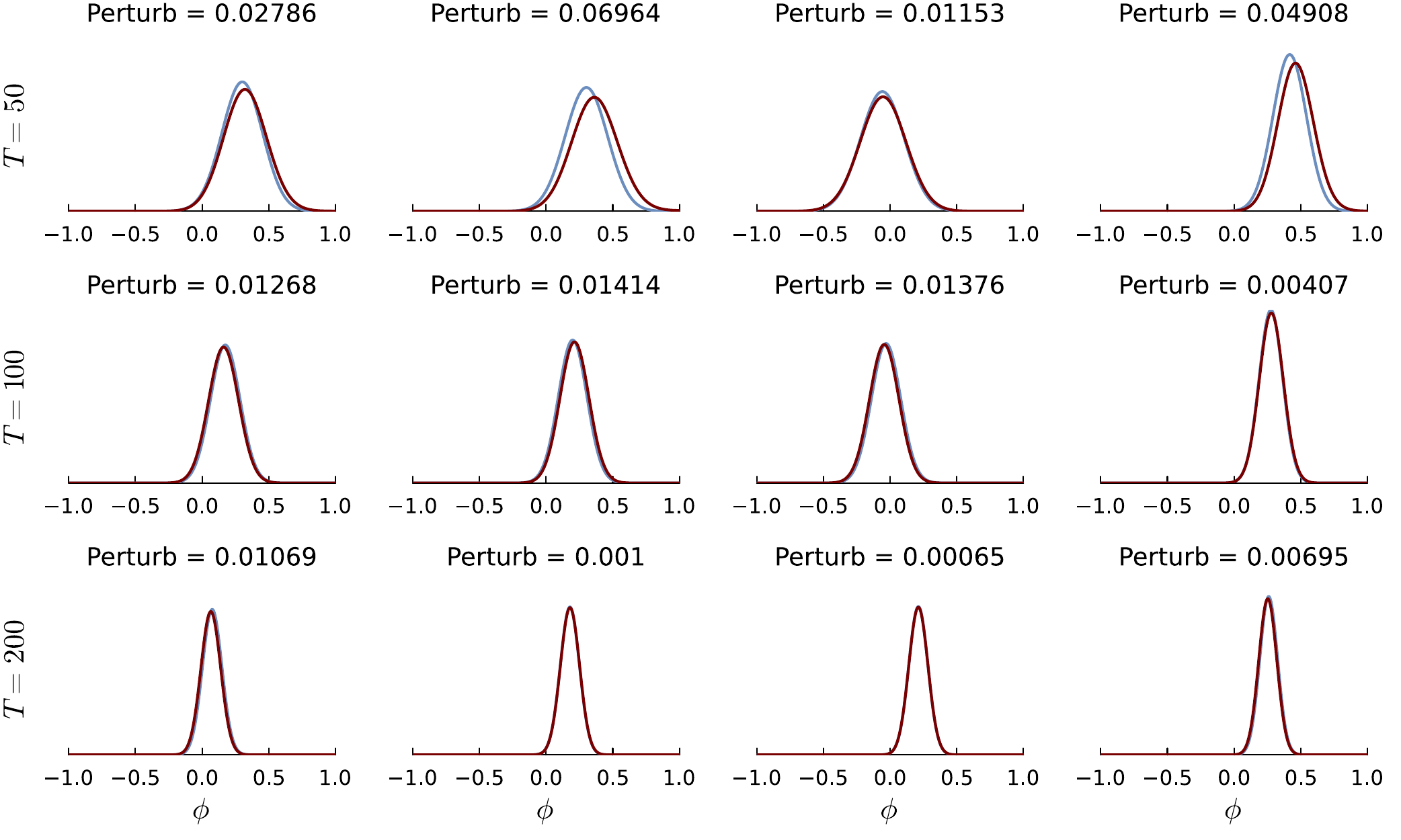}
    \caption{Illustrating the posterior perturbation of the Whittle likelihood for $\phi$ in the stationary AR(1) process, $x_t = \phi x_{t-1} + \varepsilon_t, \varepsilon_t \sim N(0,1)$, with $\phi=0.2$. Each column corresponds to a simulated dataset with $T=200$ observations and the rows use the first $T=50$, $T=100$ or all $T=200$ observations.}\label{fig:PosteriorsAR1_phi02}
\end{figure}

\begin{figure}
    \centering
    \includegraphics[width=0.95\linewidth]{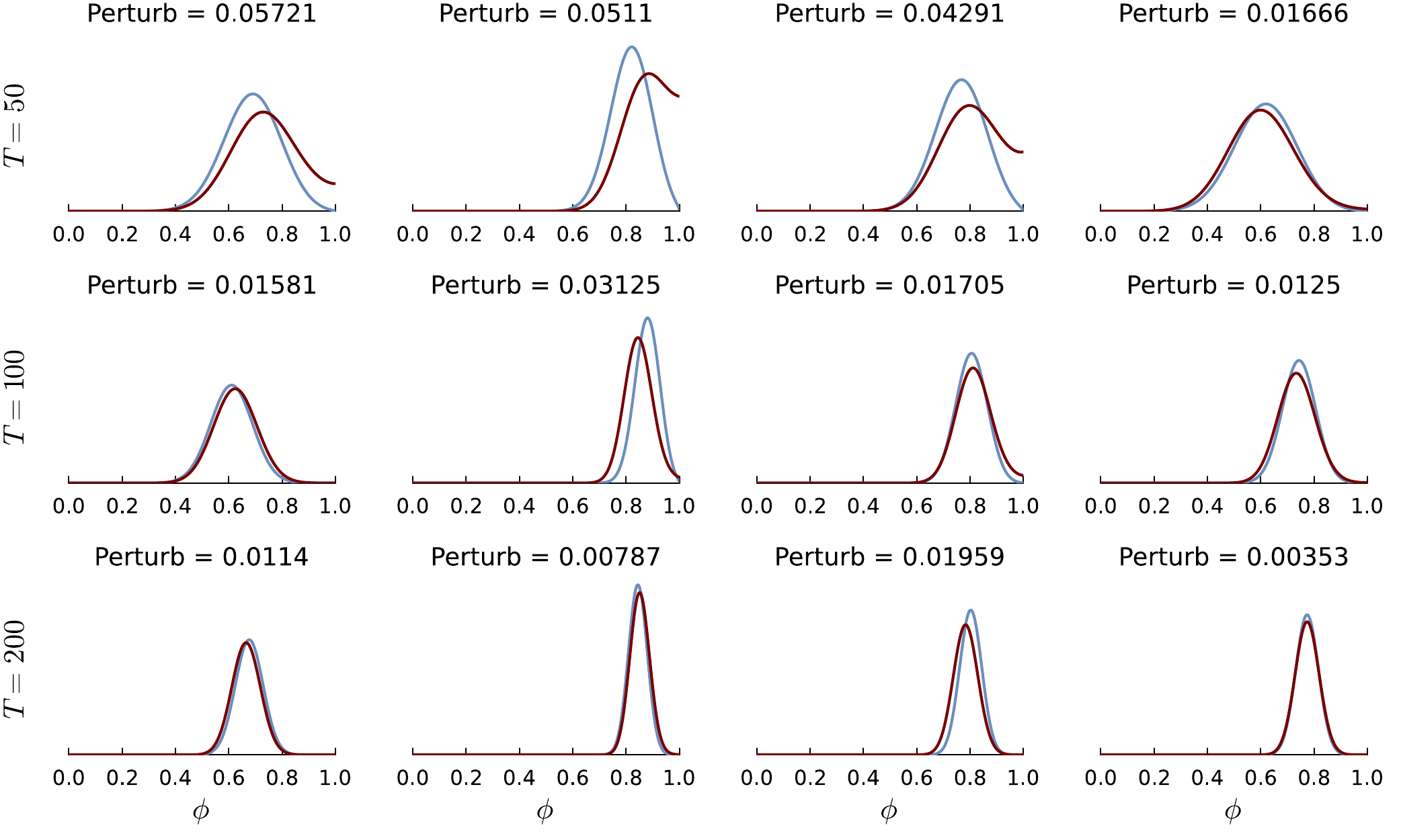}
    \caption{Illustrating the posterior perturbation of the Whittle likelihood for $\phi$ in the stationary AR(1) process, $x_t = \phi x_{t-1} + \varepsilon_t, \varepsilon_t \sim N(0,1)$, with $\phi=0.8$. Each column corresponds to a simulated dataset with $T=200$ observations and the rows use the first $T=50$, $T=100$ or all $T=200$ observations.}\label{fig:PosteriorsAR1_phi08}
\end{figure}

Figure \ref{fig:PosteriorsAR1_phi02} plots the exact and Whittle posteriors for four randomly generated datasets from the AR(1) generating process with $\phi=0.2$ for $T=50$ (top row), $T=100$ (middle row) and $T=200$ (bottom row) along with the perturbation measure in \eqref{eq:perturb} in the title. Figure \ref{fig:PosteriorsAR1_phi08} is a similar plot for $\phi=0.8$. The greater perturbation at $\phi=0.8$ is clearly visible for $T=50$, but for $T=200$, the exact and Whittle posteriors are nearly indistinguishable for all four datasets.

\section{Frequency domain likelihoods for time varying processes}\label{sec:whittlelocallystationary}

Suppose that $\{X_t\}_{t=1}^T$ is a zero-mean process with time-varying spectral density $f_{\boldsymbol{\theta}}(\omega,t)$, modeled by a vector of parameters $\boldsymbol{\theta}$. As an example, \citet{dahlhaus2012locally} uses a AR($p$) process with the $j$th AR parameter $\phi_{jt}$ evolving smoothly in time as a polynomial
\begin{equation*}
    \phi_{jt} = \theta_{j0} +  \theta_{j1}t + \ldots,  +\theta_{jr}t^r,\quad \text{ for } t=1,\ldots,T,
\end{equation*} 
for some polynomial order $r$; here $\boldsymbol{\theta} = \{\theta_{jq}\}_{j=1,\ldots,p  ; q = 0,\ldots,r}$. For the model considered in the simulation experiments in Section \ref{sec:experiments}, where the AR parameters instead evolve as random walks with Gaussian innovations we write the time-varying spectral density as $f_{\boldsymbol{\phi}_t}(\omega,t)$, hence directly indexing with $\boldsymbol{\phi}_t$, the realized AR parameters at time $t$.

\subsection{The block Whittle likelihood}\label{subsec:blockwhittle}

\begin{figure}
    \includegraphics[width=0.9\linewidth]{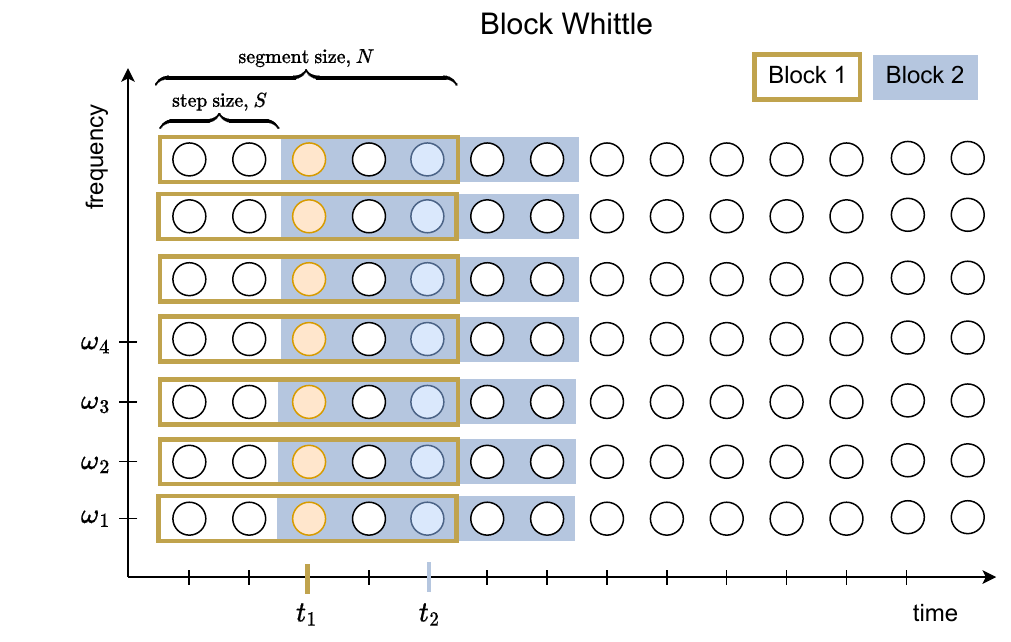}
    \includegraphics[width=0.93\linewidth]{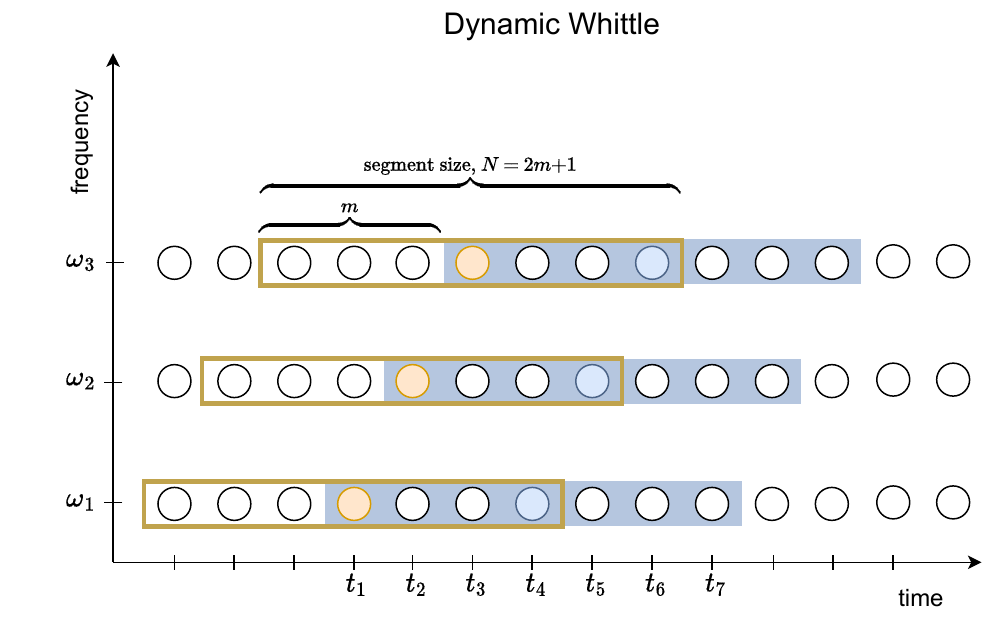}
    \caption{Illustrating the time-frequency data structure for block Whittle (top) where the time series is split up into $M$ overlapping segments, each of size $N=5$ by letting the local window jump $S=2$ observations each time. The first window is marked with an orange rectangle and the second window by a blue shaded rectangle. The dynamic Whittle (bottom) also uses segments of size $N = 2m+1$, with $m=3$, but step size $S=1$ and computes the periodogram for a single frequency in each segment; the frequencies are systematically selected starting with the first frequency $\omega_1$ in the time segment centered at time $t_1$. The filled circles mark out the time-frequency combinations where periodogram data are computed.}\label{fig:localwhittle_illustration}
\end{figure}

One of the more well known extensions of the Whittle likelihood to the time-varying case is the \emph{block Whittle} log-likelihood function introduced by \citet{dahlhaus1997fitting}, based on local periodograms computed over a sliding time window of length $N$. To reduce the dependence between the periodograms computed at neighboring time points, the window slides $S$ time steps between each computed local periodogram; see the upper part of Figure \ref{fig:localwhittle_illustration} for an illustration. We will also refer to the window length $N$ as the \emph{segment size} and $S$ as the \emph{step size}. The number of segments is $M=(T-N)/S+1$.

The block Whittle likelihood is defined as
\begin{equation}\label{eq:blockwhittle}
\ell_{T}^{BW}(\boldsymbol{\theta}) \equiv -\frac{1}{4\pi}\frac{1}{M}\sum_{j=1}^{M}\int_{-\pi}^{\pi}\left\{ \log4\pi^{2}f_{\boldsymbol{\theta}}(\omega,t_{j})+\frac{I_{T}(\omega,t_{j})}{f_{\boldsymbol{\theta}}(\omega, t_{j})}\right\} d\omega,
\end{equation}
with $t_{j}=S(j-1)+N/2$ for $j=1,\ldots,M$, and $I_{T}(\omega,t)$ is the local periodogram around $t$
\begin{equation}
I_{T}(\omega,t)\equiv \frac{1}{2\pi N}\left|\sum_{s=1}^{N}X_{t-N/2+s,T}\exp(-\im\omega s)\right|^{2}.
\end{equation}
The integral over the frequencies is computed as Riemann sum over a grid of frequencies $\omega_1,\ldots,\omega_K$. \citet{dahlhaus2012locally} defines this likelihood using a so called taper on each segment and we return to this in Section \ref{subsec:modifications}. \citet{dahlhaus2012locally} defines the block Whittle in terms of normalized time $u=t/T$ in order to use in-fill asymptotics. The block Whittle likelihood is equivalent to the model
\begin{equation}\label{eq:BW_iid_model}
    I_{T}(\omega_k,t_j) \mid f_{\boldsymbol{\theta}}(\omega_k,t_j) \overset{\mathrm{ind}}{\sim} \mathrm{Exp}(f_{\boldsymbol{\theta}}(\omega_k,t_j)), \text{ for } j=1,\ldots,M \text{ and } k=1,\ldots,K.
\end{equation}
Since the segments are partially overlapping (when $S<N$), neighboring periodogram observations $I_{T}(\omega_k,t_j)$ share some observations and are therefore not exactly independent, making the likelihood in \eqref{eq:blockwhittle} and the corresponding equivalent model in \eqref{eq:BW_iid_model} only approximate.
We note that the (assumed) independence of the periodogram data, and the corresponding sum in the Block Whittle log-likelihood approximation, makes it possible to also apply subsampling MCMC \citep{quiroz2019speeding} to locally stationary time series data, as a direct extension of \citet{salomone2020spectral} from the stationary case.

The block Whittle likelihood in \eqref{eq:blockwhittle} can be motivated as the asymptotic Kullback-Leibler divergence between the true process with spectral density $f(\omega,t)$ and the model with spectral density $f_{\boldsymbol{\theta}}(\omega,t)$ \citep{dahlhaus1997fitting}. \citet{dahlhaus1997fitting} proves consistency and asymptotic efficiency of the block Whittle MLE of $\boldsymbol{\theta}$ if $S/N \rightarrow 0$ under smoothness assumptions of the parameter evolution paths; see \citet{dahlhaus2012locally} for a large number of results on local Whittle likelihoods for a wide range of models under several different types of conditions, including time-varying ARMA models with results on the uniqueness of a time-varying spectral density. 

\citet{dahlhaus2012locally} also reviews results for the \emph{generalized Whittle log-likelihood} proposed in \citet{dahlhaus2000likelihood}, which is defined as
\begin{equation}\label{generalizedWhittle}
\ell_{T}^{GW}(\boldsymbol{\theta}) := -\frac{1}{T}\sum_{t=1}^T \frac{1}{4\pi}\int_{-\pi}^{\pi}\left\{ \log4\pi^{2}f_{\boldsymbol{\theta}}(\omega,t)+\frac{\check{J}_{T}(\omega,t)}{f_{\boldsymbol{\theta}}(\omega,t)}\right\} d\omega,
\end{equation}
where $\check{J}_T(\omega,t)$ is the preperiodogram \citep{neumann1997wavelet}
\begin{equation}
\check{J}_T(\omega,t) := \frac{1}{2\pi}\sum_{\mathclap{
       \substack{k\\
            1 \leq  \lfloor uT+0.5+k/2\rfloor, \lfloor uT+0.5-k/2\rfloor \leq T}}}
          X_{\lfloor uT+0.5+k/2\rfloor}X_{\lfloor uT+0.5-k/2\rfloor}
          \exp\left(-\im\omega k\right) .
\end{equation}

The preperiodogram can be seen as a local version of the periodogram since
\begin{equation}
I_T(\omega) = \frac{1}{T} \sum_{t=1}^T \check{J}_T(\omega,t).
\end{equation}
The MLE from \eqref{generalizedWhittle} has good asymptotic properties under weak conditions of bounded variation on the parameter evolution \citep{dahlhaus2009local,dahlhausPolonik2009empirical}. However, the perperiodogram is quite noisy with complex interfering cross-terms which can make inference more difficult \citep{sandsten2016time}. \cite{everitt2013online} therefore use a modified preperiodogram which is smoothed in both time- and frequency direction for efficient online learning of time-varying spectral densities using tvAR with similar parameter evolutions as we use in Section \ref{sec:experiments}. Our preliminary experiments show that the generalized Whittle likelihood is more challenging to deal with in a Bayesian setting, and we therefore opt for comparing the block Whittle likelihood with the dynamic Whittle likelihood presented in the next section. 

\subsection{The dynamic Whittle likelihood}\label{subsec:meyer}
\citet{tang2023bayesian} define the dynamic Whittle (DW) likelihood as
\begin{equation}
\ell_{T}^{DW}(\boldsymbol{\theta}) := -\sum_{t=1}^{T} \left\lbrace\log f_{\boldsymbol{\theta}}(\omega_{\mathrm{mod}(t)},t)+\frac{\tilde{I}(\omega_{\mathrm{mod}(t)},t)}{f_{\boldsymbol{\theta}}(\omega_{\mathrm{mod}(t)}, t)}\right\rbrace,
\end{equation}
where $\mathrm{mod}(t) = 1+((t-1) \mathrm{mod}\, m)$, $N=2m+1$ is the segment length and $\omega_1,\ldots,\omega_m$ are the positive Fourier frequencies on the segment. The moving periodogram \citep{hafner2017moving}
\begin{equation}\label{eq:movingperiodogram}
\tilde{I}(\omega_{\mathrm{mod}(t)},t) := \frac{1}{2\pi N}\left|\sum_{s=1}^{N}X_{t-m+s,T}\exp(-\im\omega_{\mathrm{mod}(t)} s)\right|^{2}.
\end{equation}
is calculated in a sliding time window of length $N$ centered at time $t$, but in contrast to the block Whittle, only one frequency $\omega_{\mathrm{mod}(t)}$ is used for each $t$; see Figure \ref{fig:localwhittle_illustration} for a visual representation. Using different frequencies in windows at two neighboring time points makes it possible to slide the window over each time point, without introducing dependence between the periodogram data. However, when the window has completed a pass over all frequencies it returns to the first frequency, which will then be partially overlapping with the previous window at the first frequency; see for example how the segments centered over $t_1$ and $t_4$ are overlapping for the same frequency $\omega_1$ in the bottom panel of Figure \ref{fig:localwhittle_illustration}.  

The dynamic Whittle likelihood is equivalent to the model
\begin{equation}
    \tilde{I}_{t} \mid f_t(\omega_{\mathrm{mod}(t)}) \overset{\mathrm{ind}}{\sim} \mathrm{Exp}\big(f_{t}(\omega_{\mathrm{mod}(t)})\big), t = 1,...,T,
\end{equation}
where $\tilde{I}_{t}:=\tilde{I}(\omega_{\mathrm{mod}(t)},t)$ and $T$ is the length of the time series, excluding the first $m$ and last $m$ observations which are needed to compute the moving periodograms in \eqref{eq:movingperiodogram} at the boundaries.
The dynamic Whittle likelihood yields a univariate observation equation with a single observation for each time point (except on the boundaries) contrasted against the block Whittle which gives roughly $T/S$ multivariate observations observed across all frequencies. 

\citet{tang2023bayesian} prove that the posterior distribution from a bivariate extension of the Bernstein-Dirichlet process prior combined with the dynamic Whittle likelihood is consistent for the time-varying spectral density; L2-norm posterior contraction rates are also given.

Importantly, both the block and dynamic Whittle log-likelihoods are sums, due to the asymptotic independence of periodogram observations. This means that one can speed up Markov Chain Monte Carlo for long time series by subsampling periodogram observations; see \citet{salomone2020spectral} and the extension to multivariate time series in \citet{villani2024spectral}. This continues to hold for the three modifications of the basic block and dynamic Whittle likelihoods presented in the next section.

\subsection{Modified frequency likelihoods}\label{subsec:modifications}
The block Whittle and dynamic Whittle likelihoods rely on periodograms from relatively short sliding time windows and are therefore likely to inherit the small-sample problems of the original Whittle likelihood for a stationary process demonstrated in Section \ref{sec:whittle}. 

We investigate three modifications of the standard Whittle likelihood, which are applied to each segment when used for time-varying processes. The first is tapering which is well-known in the literature and is recommended in \cite{dahlhaus1988small} for small samples in the stationary case, and in \cite{dahlhaus2012locally} for the block Whittle likelihood. The second approach is prewhitening \citep{priestly1981spectral}, which seems less common in the statistical literature. The last, and most recently proposed, modification is the \textit{boundary correction} in \cite{subba2021reconciling}, which performs well both as a sampling distribution for the spectral density \citep{das2021spectral}, and for parametric estimation via the Whittle likelihood \citep{subba2021reconciling}.

The debiased Whittle \citep{sykulski2019debiased} is another method for reducing the bias in the Whittle MLE. The debiased Whittle likelihood is 
\begin{equation}\label{eq:debiasedwhittle}
\ell_{T}^{W}(\boldsymbol{\theta}) := - \frac{1}{2}\sum_{\omega_k \in \Omega_T} \bigg(  
    \log \bar{f}_{\boldsymbol{\theta}}(\omega_k) + \frac{I_{T}(\omega_k)}{\bar{f}_{\boldsymbol{\theta}}(\omega_k)}\bigg),
\end{equation}
which debiases the Whittle MLE by replacing the spectral density with the expected periodogram $\bar{f}_{\boldsymbol{\theta}}(\omega) := \mathbb{E}(I_{T}(\omega))$. \citet{sykulski2019debiased} show how the expected periodogram can be computed by two FFTs, thereby retaining the $O(T \log T)$ cost of the original Whittle likelihood. However, for the original Whittle, this cost is incurred once when computing the periodogram data, which can then be used for subsequent evaluations of the likelihood for different $\boldsymbol{\theta}$ values. This is especially important for Bayesian inference with MCMC where the likelihood is typically evaluated thousands of times during the course of the algorithm. The FFTs in the debiased Whittle likelihood have to be recomputed for every new $\boldsymbol{\theta}$ value, since $\bar{f}_{\boldsymbol{\theta}}(\omega)$ depends on $\boldsymbol{\theta}$, so we will not consider this likelihood in our simulation study.

\subsection*{Tapering}\label{subsec:taper}

\citet{dahlhaus1988small} shows that the Whittle likelihood for the stationary case in \eqref{eq:whittlelikelihood} is inefficient due to bias from leakage effects in the periodogram. He suggests using a taper to correct the bias in the periodogram and defines a modified tapered Whittle likelihood. In the context of local Whittle likelihoods, this means replacing the periodogram on the segments with tapered versions
\begin{equation}
I_{T}(\omega,t)\equiv \frac{1}{2\pi H_N}\left|\sum_{s=1}^{N} h \Big(\frac{s}{N} \Big)  X_{t-N/2+s,T}\exp(-\im\omega s)\right|^{2},
\end{equation}
where $ h{\,:\,} [0,1] \rightarrow \mathbb{R} $ is a symmetric data taper with $h(x) = 1 - h(x)$ and $ H_N := \sum_{j=0}^{N-1} h^2(j/N)$ is the normalization factor. This modified Whittle likelihood is shown to be as efficient as the exact time domain MLE, and to satisfy a central limit theorem. We use a taper based on the Hanning window $h(t) = \frac{1}{2}(1-\cos(2\pi t/N))$ and apply the taper to each segment in the local Whittle likelihood.

\subsection*{Prewhitening}\label{subsec:preW}
A consequence of using the standard periodogram is that spectral power leaks to nearby frequencies, with the effect being largest at the highest peaks in the spectrum. This can hide or contaminate low-power frequencies in the spectral density, leading to distortions in the Whittle likelihood and efficiency loss for the Whittle MLE \citep{dahlhaus1988small}. Prewhitening \citep{priestly1981spectral} first removes the most dominant autocorrelation in the time domain by filtering with an estimated AR($p$) process (or ARMA($p$,$q$)) so that the residual series $e_t = \hat{\phi}_p (B) x_t$ has a periodogram from a flatter spectral density. The residual periodogram is then transformed to a periodogram for the original time series, using that the spectral density of a filtered process is
\begin{equation}
    f_x(\omega) = \vert H_{ \hat{\boldsymbol{\phi}}}(\omega) \vert ^{-2} f_e(\omega),
\end{equation}
where $H_{\hat{\boldsymbol{\phi}}}(\omega) = \hat{\phi}_p (e^{-\im \omega})$ is the transfer function of the fitted AR($p$) process. That is, the periodogram $I_x(\omega)$ used in the Whittle likelihood is computed as
\begin{equation}
    I_x(\omega_k) = \vert H_{ \hat{\boldsymbol{\phi}}}(\omega_k) \vert ^{-2} I_e(\omega_k),\quad \text{ for }k=1,\ldots,K,
\end{equation}
where $I_e(\omega)$ is the residual periodogram.
Note that the AR model used for the prewhitening does not have to be very precise as the remaining lower peaks in the spectral density will be captured by the residual periodogram $I_e(\omega_k)$.

\subsection*{Boundary-correction}\label{subsec:rao}
\citet{subba2021reconciling} express the Gaussian time domain likelihood in the frequency domain and prove that the approximation error of the Whittle likelihood is the result of boundary effects. The Gaussian likelihood can from this perspective be seen to circumvent boundary effects by predicting the time series outside its range using a best linear predictor. The approximation error increases with the persistence of the time series \citep{subba2021reconciling}, in agreement with the empirical evidence in Section \ref{sec:whittle}, \citet{dahlhaus1988small} and \citet{contreras2006note}. \citet{subba2021reconciling} suggest using the \emph{complete DFT} that augments the standard DFT with the Fourier transform of the best linear predictor to remove the discrepancy between the Whittle likelihood and the Gaussian likelihood. 

The {\em complete discrete Fourier Transform} (DFT) of $\{X_t\}_{t=1}^T$
\begin{equation}
    \label{eq:complete DFT}
    \Tilde{J}(\omega;f) = J(\omega) + \hat{J}(\omega;f)
\end{equation}
is the sum of the standard DFT $J(\omega)$ and the \emph{predictive DFT}
\begin{equation}
\label{eq:predictive DFT}
    \hat{J}(\omega;f) \coloneqq  \frac{1}{\sqrt{2\pi}} \sum_{t=1}^T X_t \sum_{\tau\leq 0} \Big(\phi_{t,T}(\tau;f)e^{\im \omega \tau}  +  e^{\im \omega T}\phi_{T+1-t,T}(\tau;f) e^{-\im (\tau-1)\omega} \Big),
\end{equation}
where $\phi_{t,T}(\tau;f)$ are coefficients of the best linear predictor of $X_\tau$ for $\tau \leq 0$ and $\tau > T$ given $\{X_t\}_{t=1}^T$ with spectral density $f$. Note that $e^{iT\omega_T}=1$, and that $\hat{J}(\omega,f)$ depends on the spectral density $f$ of the model. \citet{subba2021reconciling} suggest approximating the best linear predictor with a finite-order AR model determined by AIC. Note that the computation of $\hat{J}(\omega_k;f)$ for $k=1,\ldots,K$ is a one-time cost, performed before the MCMC. Put differently, once the completed DFT in \eqref{eq:complete DFT} is computed it can be used as fixed data for any subsequent model fitted by MCMC or by other methods.  

The example with the predictive DFT of the AR(1) model with parameter $\phi$ in \citet{subba2021reconciling} is illuminating. Since the prediction to the left of the first observation $X_1$ for an AR(1) process is $\hat X_{\tau,T} = \phi^{\vert\tau\vert +1 }X_1$ for  $\tau \leq 0$ and to the right of the last observation $X_T$ is $\hat X_{\tau,T} = \phi^{\tau-T}X_T$ for $\tau>T$, we have
\begin{equation}
    \hat{J}(\omega;f)=\frac{\phi}{\sqrt{2\pi}}\left( \frac{1}{\phi(\omega)}X_1 + \frac{e^{\im(T+1)\omega}}{\overline{\phi(\omega)}}X_T \right),
\end{equation}
where $\phi(\omega) = 1- \phi\exp(-\im \omega)$.

The {\em complete periodogram} based on the complete DFT is then 
\begin{equation}\label{eq:completeperiodogram}
    \tilde{I}_T(\omega) \coloneqq \frac{1}{2\pi T} \hat{J}(\omega;f) \overline{J(\omega)} 
\end{equation}
which is now unbiased, but not consistent, for $f(\omega)$ \citep{subba2021reconciling}.

The boundary corrected Whittle likelihood proposed by \cite{subba2021reconciling} replaces the regular periodogram with the complete periodogram in \eqref{eq:completeperiodogram}. In a simulation study, they find that the MLE from the boundary corrected Whittle likelihood outperforms other spectral likelihoods including the standard and tapered Whittle likelihood, and the debiased Whittle likelihood \citep{sykulski2019debiased}. \cite{subba2021reconciling} recommend a hybrid using the boundary corrected likelihood together with tapering, and this is the version used in the simulation studies here.

\section{Bayesian inference}

\subsection{Time-varying locally stable AR processes}\label{subsec:tvAR}

The class of time-varying AR($p$) models is used to compare the spectral likelihood approximations in Section \ref{sec:whittlelocallystationary}. This relatively simple class makes it possible to easily compare with the time domain likelihood, which is the gold standard. The AR($p$) process is of the form
\begin{equation}\label{eq:tvAR}
    x_t = \mu_t + \sum_{j=1}^p \phi_{jt}(x_{t-j} -\mu_t) + \varepsilon_t,\quad \varepsilon_t \sim N(0,\sigma_{\varepsilon,t}^2),
\end{equation}
where $\phi_{jt}$ is the $j$th AR coefficient at time $t$. We will for simplicity assume a zero mean $\mu_t$  and that the disturbance variance is constant over time, $\sigma_{\varepsilon,t}^2 = \sigma_\varepsilon^2$ for all $t$. In the application to egg price data in Section \ref{sec:realdata} we also allow for a time-varying variance.

Following \citet{fagerberg2024time} we enforce stability of the AR process at every time period using the reparametrisation in \citet{Barndorff-Nielsen1973} and \citet{monahan1984note}. This transformation maps a set of unconstrained real-valued coefficients $\boldsymbol{\theta}=(\theta_1,\ldots, \theta_p)^\top$ to the set of stable parameters, i.e. parameters that determine a stable AR process. Let $\mathbb{S}^p \subset \mathbb{R}^p$ denote the region in parameter space where the $\mathrm{AR}(p)$ process is stable. The reparameterization is defined as follows.
\begin{definition}[stability parameterization]\label{def:stability_parameterization}
An $\operatorname{AR}(p)$ process can be restricted to be stable by a 1:1 and onto map
\begin{equation*}
    \boldsymbol{\theta} \rightarrow \mathbf{r} \rightarrow \boldsymbol{\phi}.
\end{equation*}
of unrestricted parameters $\boldsymbol{\theta} = (\theta_1,\ldots,\theta_p)^\top \in \mathbb{R}^p$ via the partial autocorrelations $\mathbf{r} = (r_1,\ldots,r_p)^\top \in (-1,1)^p$ to the stable AR parameters $\boldsymbol{\phi}=(\phi_1,\ldots,\phi_p)^\top \in \mathbb{S}^p$.
The mapping from $\mathbf{r}$ to $\boldsymbol{\phi}$ is given by setting $\phi_{1,1} = r_1$ followed by the recursion
\begin{equation*}
\phi_{k,j} = \phi_{k-1,j}+r_{k}\phi_{k-1,k-j},\;\text{ for } k=2,\ldots,p \text{ and } j=1,\ldots,k-1,
\end{equation*}
and finally returning $\boldsymbol{\phi} = (\phi_{p,1},\ldots,\phi_{p,p})^\top$, where $\phi_{p,p}=r_p$.
\end{definition}

We follow \citet{monahan1984note} and use the map
\begin{equation}\label{eq:monahan}
    r_k = \frac{\theta_k}{\sqrt{1+\theta_k^2}},\quad \text{ for }k=1,\ldots,p,
\end{equation}
from the unrestricted parameters to the partial autocorrelations. The stability parameterization in Definition \ref{def:stability_parameterization} is imposed at every time period in the time-varying parameter setting. We use $\boldsymbol{\phi}_t = \mathbf{g}(\boldsymbol{\theta}_t)$ to denote the composite map $\boldsymbol{\theta}_t \rightarrow \mathbf{r}_t \rightarrow \boldsymbol{\phi}_t$.

\subsection{Likelihood and parameter evolution}\label{subsec:statespace}

The unrestricted parameters $\boldsymbol{\theta}_t$ in the fitted AR models are assumed to follow a multivariate random walk with Gaussian innovations, as in the large literature on time-varying AR and vector AR models, see e.g. \citet{prado2010time}. Given the frequency domain approximations of the likelihood discussed in Section \ref{sec:whittlelocallystationary} we can write the model as a state-space model with $\boldsymbol{\theta}_t$ as state vector. In the block Whittle case:
\begin{align}\label{eq:statespacemodel}
    I\left( \omega_k, t_j \right)\, &\overset{\text{ind}}{\sim} \, \mathrm{Exp}\left( f_{\boldsymbol{\phi}_{j},\sigma_\varepsilon}(\omega_k) \right), \quad \text{ for } k=1,\ldots,K\text{ and } j=1,\ldots,M \nonumber\\
    \boldsymbol{\phi}_j &= \mathbf{g}(\boldsymbol{\theta}_j) \nonumber \\
    \boldsymbol{\theta}_j &= \boldsymbol{\theta}_{j-1} + \boldsymbol{\eta}_j, \quad \boldsymbol{\eta}_j \overset{\mathrm{iid}} \sim N(\mathbf{0},\boldsymbol{Q}) \nonumber \\
    \boldsymbol{\theta}_0 &\sim N(\mathbf{a}_0,\mathbf{P}_0), 
\end{align}
where $f_{\boldsymbol{\phi}_j,\sigma_\varepsilon}(\omega)$ is the spectral density of the assumed AR model at time $t_j$. In the case with the dynamic Whittle likelihood the observation equation of \eqref{eq:statespacemodel} changes to 
\begin{equation}
    \tilde{I}_{t} \mid f_{\boldsymbol{\phi}_t}(\omega_{\mathrm{mod}(t)}) \overset{\mathrm{ind}}{\sim} \mathrm{Exp}\big(f_{\boldsymbol{\phi}_t}(\omega_{\mathrm{mod}(t)})\big),\quad \text{ for } t = 1,...,T,
\end{equation}
and the parameter evolution in \eqref{eq:statespacemodel} is indexed by $t=1,\ldots,T$.

The state evolution for $\boldsymbol{\theta}_t$ with covariance matrix $\boldsymbol{Q}$ in \eqref{eq:statespacemodel} is linear and Gaussian, but the observation equation is non-Gaussian due to the exponentially distributed periodogram data, and non-linear as a result of the stability parameterization. This prevents using the usual Forward Filtering Backward Sampling (FFBS) algorithm \citep{carter1994gibbs,fruhwirth1994data} based on the Kalman filter, and we instead rely on particle MCMC, in particular the PGAS algorithm in \citet{lindsten2014particle}. The remaining parameters are inferred using standard Gibbs sampling updates.

\subsection{Bayesian inference by particle MCMC within Gibbs sampling}

We derive the Gibbs sampling steps for the block Whittle likelihood; the updating steps for the dynamic Whittle likelihood follow from obvious changes.
The full conditional distribution for $\sigma_\varepsilon^2$ can be obtained as follows. The observation equation under the tvAR process is
\begin{equation}
      I\left( \omega_k, t_j \right) \overset{\text{ind}}{\sim} \mathrm{Exp}\left( f_{\boldsymbol{\phi}_{j},\sigma_\varepsilon}(\omega_k) \right) =  \mathrm{Exp}\left( \frac{\sigma^2}{2\pi} \left| \phi_{p,j}(e^{-\im\omega_k}) \right|^{-2} \right),
\end{equation} 
where $\phi_{p,j}(L)\coloneqq1-\phi_{1j}L-\cdots-\phi_{pj}L^{p}$ is the AR($p$) lag polynomial at time $t_j$. Since the exponential distribution is closed under scaling, we have 
\begin{equation}
    \tilde{I}\left( \omega_k, t_j \right)  \overset{\text{ind}}{\sim} \mathrm{Exp}\left( \sigma^2 \right),
\end{equation}
where $\tilde{I}\left( \omega, t \right) = 2\pi \left| \phi_{p,t}(e^{-\im\omega}) \right|^{2} I\left( \omega, t \right)$.
Using the conjugate prior for the exponential likelihood, $\sigma^2 \sim \text{inv-gamma}(\alpha,\beta)$, the conditional posterior distribution is
\begin{equation}\label{IG}
\sigma^2|\boldsymbol{\phi}_{0:T},I(\boldsymbol{\omega},1{:}M) \sim  \text{inv-gamma}\left(\alpha + KM,  \sum_{k=1}^K \sum_{j=1}^M \tilde{I}\left( \omega_k, t_j \right) + \beta\right),
\end{equation}
where $I(\boldsymbol{\omega},1{:}M)$ is the periodogram data over all frequencies $\boldsymbol{\omega}=(\omega_1,\ldots,\omega_K)^\top$ and time segments centered at $t_j,\text{ for } j=1,...,M$.

The prior for $\boldsymbol{Q}$ is $\text{inv-Wishart}(\nu_0, \boldsymbol{S}_0)$. The full conditional for $\boldsymbol{Q}$ is the same as in the time domain:
\begin{equation}\label{IW}
    \boldsymbol{Q}|\boldsymbol{\phi}_{0:M}, \sigma^2, I(\boldsymbol{\omega},1{:}M) \sim  \text{inv-Wishart}\left( M +\nu_0, \boldsymbol{S}_0 + \boldsymbol{\phi}^\top_{1:M}\boldsymbol{\phi}_{1:M} \right) 
\end{equation}
Since the prior for the initial state is Gaussian $\boldsymbol{\theta}_0 \sim N(\boldsymbol{a}_0,\mathbf{P}_0)$ and $\boldsymbol{\theta}_1 = \boldsymbol{\theta}_0 + \boldsymbol{\eta}_1$ from the state transition equation, standard Gaussian updating gives that the full conditional posterior for $\boldsymbol{\theta}_0$ is
\begin{equation}\label{eq:condPostInitialState}
 \boldsymbol{\theta}_0 \vert \boldsymbol{\phi}_{1:M}, \sigma^2, I(\boldsymbol{\omega},1{:}M) \sim N(\boldsymbol{m}, \boldsymbol{\Omega}),
\end{equation}
with $\boldsymbol{\Omega}^{-1} = \boldsymbol{Q}^{-1} + \boldsymbol{P}_0^{-1}$ and $\boldsymbol{m} = \boldsymbol{\Omega}\big( \boldsymbol{Q}^{-1}\boldsymbol{\theta}_1 + \boldsymbol{P}_0^{-1}\boldsymbol{a}_0 \big)$.

The complete Gibbs sampler is shown in Algorithm \ref{Gibbs DW}.

\begin{algorithm} \label{Gibbs DW}
\SetAlgoLined
\KwInput{segment length $N$ \\
\hspace{1.02cm}step size $S$ \\
\hspace{1.04cm}time-varying periodogram, $I(\omega, 1{:}M)$\\
\hspace{1.04cm}initial value for static parameters $\sigma^2_\varepsilon, \boldsymbol{Q},\boldsymbol{\phi}_0$.
} 
\BlankLine
\For{$j = 1$ \KwTo $n$}{ 
    \BlankLine
     \hspace{0.1cm}Update $\boldsymbol{\phi}_{1:M}|\boldsymbol{Q},\sigma^2_\varepsilon,\boldsymbol{\phi}_0$ using PGAS \citep{lindsten2014particle}.\\
    \hspace{0.1cm}Update $\boldsymbol{Q}|\boldsymbol{\phi}_{1:M},\sigma^2_\varepsilon,\boldsymbol{\phi}_0$ using the inverse Wishart in \eqref{IW}.  \\
    \hspace{0.1cm}Update $\sigma^2_\varepsilon| \boldsymbol{Q},\boldsymbol{\phi}_{1:M},\boldsymbol{\phi}_0$ using the inverse gamma in \eqref{IG}.    \\
    \hspace{0.1cm}Update $\boldsymbol{\phi}_0|\sigma^2_\varepsilon, \boldsymbol{Q},\boldsymbol{\phi}_{1:M}$ using the multivariate normal in \eqref{eq:condPostInitialState} and set $\boldsymbol{\phi}_0 = \mathbf{g}(\boldsymbol{\theta}_0)$.  \\

    \BlankLine
} 
\KwOutput{draws from the joint posterior for $p(\boldsymbol{\phi}_{0:M},\sigma^2_\varepsilon, \boldsymbol{Q} \vert I(\boldsymbol{\omega},1{:}M))$.}
\BlankLine
\caption{Gibbs sampling with block Whittle likelihood on $M$ segments. \label{alg:gibbs}}
\end{algorithm}

\section{Experiments}\label{sec:experiments}
This section evaluates the likelihood in three simulation settings with the data generating processes as an AR process with parameters following three different deterministic paths: 
\begin{itemize}
    \item \textbf{Experiment 1} - locally stationary time-varying AR(2) from \citet{dahlhaus2012locally}.
    \item \textbf{Experiment 2} - stationary AR(3) process with time-invariant parameters.
    \item \textbf{Experiment 3} - locally stationary AR(2) where the parameters come close to the non-stable/explosive region several times during the observed time period.
\end{itemize}

One hundred realizations of length $T=1500$ are simulated from the data generating processes in each experiment. For each generated time series, we run the Gibbs sampler in Algorithm \ref{alg:gibbs} to sample from the posterior based on the time domain likelihood, which is the target posterior. We then run the Gibbs sampler on the same datasets for both the block and dynamic Whittle approximate likelihoods, and also for these likelihoods combined with one of three improvements in Section \ref{subsec:modifications}: tapering, prewhitening and boundary correction. Each Gibbs run is based on $12000$ draws, from which $2000$ are discarded as burn-in. The remaining draws are thinned by a factor of $2$, leaving $5000$ draws for inference.

The dynamic Whittle likelihood gives rise to $m$ missing values in the inferred parameter evolution at the beginning and end of the series. With the block Whittle likelihood the posterior for the parameters is only observed at every $S$:th time point. Sophisticated interpolation methods are available \citep{hafner2017moving,dahlhaus2012locally}, but we use a simple piecewise constant extrapolation of the evolution between observed posterior estimates, see e.g.~Figure \ref{fig:EggBCnoTapParEvolDW25}. 

To compute the periodograms when prewhitening and boundary-correction are used we need a model on each segment to filter or make predictions outside the boundary. We use an AR model with the number of lags determined by the Hannan-Quinn information criteria \citep{hannan1979determination}, with coefficients estimated using Burg's method.  

Since we are mainly interested in the similarity to the time domain likelihood, we do not put any particular effort into finding good prior settings, and use the same prior for all data generating processes. This prior is $\boldsymbol{Q}\sim IW(10, 0.005\cdot(10-p-1)I_p)$, where $p$ is the number of lags in the AR model. For the block Whittle likelihood, we multiply the prior covariance matrix $\boldsymbol{Q}$ by the step size, $S$, to allow for similar parameter evolution over the whole time span as in the dynamic Whittle likelihood. The prior on the error variance is $\sigma_\varepsilon^2\sim IG(0.01,0.01)$. The unrestricted AR parameters at time $t=0$ follow a multivariate normal prior with mean $\mathbf{a}_0 = \mathbf{0}_p$ and covariance matrix $\mathbf{P}_0=10\cdot \boldsymbol{I}_p$. 

\subsection{Experiment 1 - Dahlhaus time-varying AR(2)}

The first experiment uses the AR(2) process 
\begin{equation}
    y_t = \phi_{1,t}y_{t-1} + \phi_{2,t}y_{t-2} + \varepsilon_t, \quad \varepsilon_t\sim N(0,1),
\end{equation}
in \citet{dahlhaus2012locally}, where the first AR parameter evolves like a cosine and the second AR parameter is time invariant:
\begin{align*}
    \phi_{1,t} =& -1.8\cos\left(1.5 - \cos(4 \pi t_u)\right), \text{for }t_u=(t-1)/(T-1), \\
    \phi_{2,t} =& -0.9,
\end{align*}
for $t=1,\dots,T$.
The process is always relatively far from having a unit root; the third experiment explores the near unit root case.

\begin{figure}
    \centering
    \includegraphics[width=0.7\linewidth]{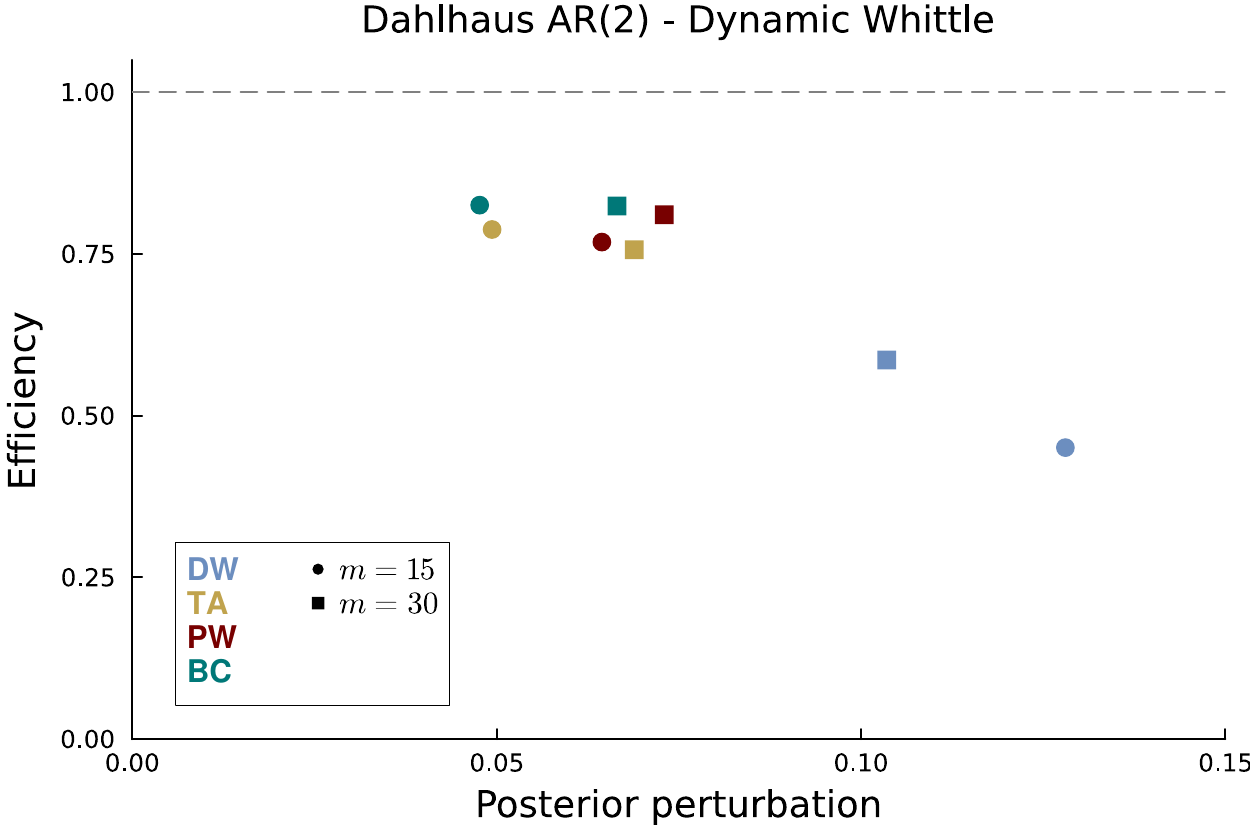}
    \includegraphics[width=0.7\linewidth]{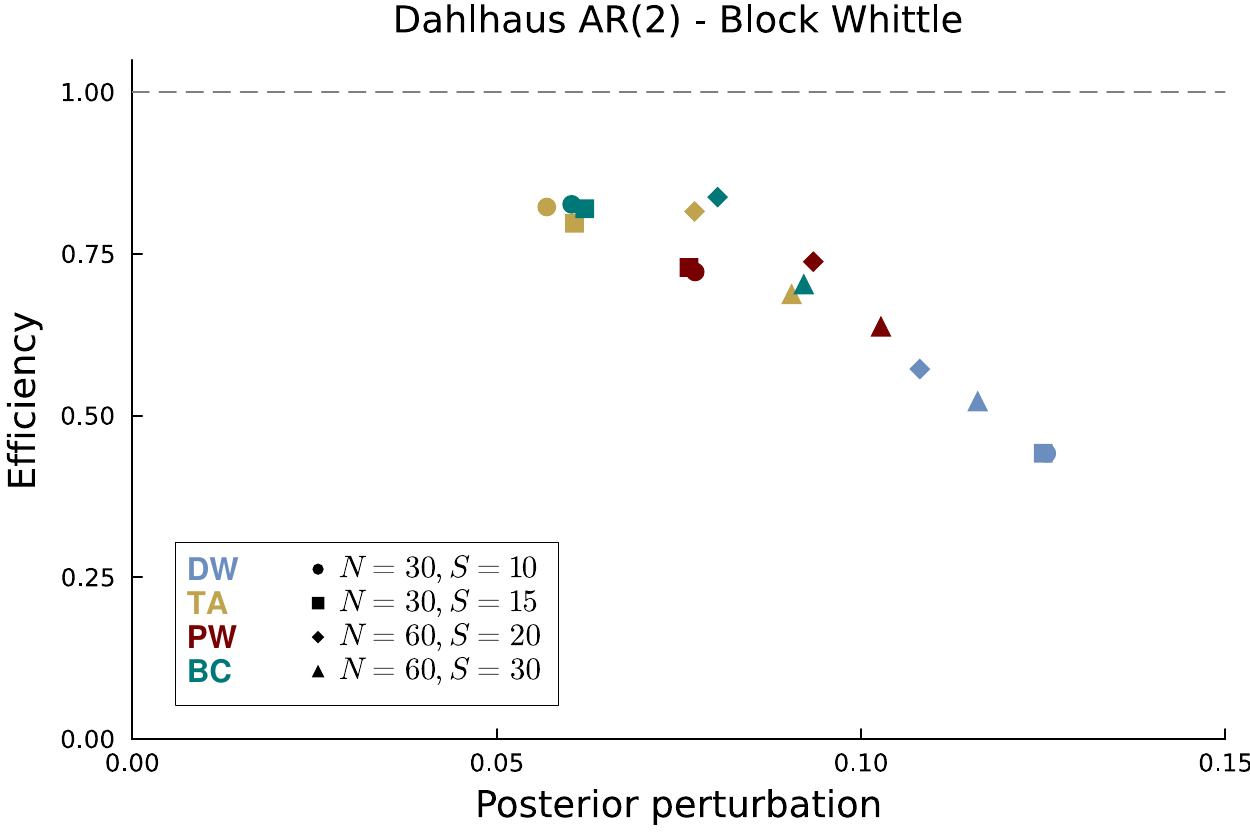}
    \caption{Experiment 1 - Dahlhaus AR(2) process. Efficiency and posterior perturbation in the posterior for the AR parameters for the dynamic Whittle (top) and block Whittle (bottom) likelihoods, and with tapering (TA), prewhitening (PW) and boundary correction (BC) applied.}\label{fig:dahlhausRMSE}
\end{figure}

Figure~\ref{fig:dahlhausRMSE} plots the efficiency in \eqref{eq:efficiency} for the posterior median estimate and the posterior perturbation measure in \eqref{eq:perturb} for different combinations of approximate likelihood (dynamic Whittle, DW and block Whittle, BW) and the three modifications: tapering (TA), prewhitening (PW) and boundary correction (BC); since there is now a time-varying step, the MSE measures in \eqref{eq:RMSE} and \eqref{eq:perturb} are averages also over time. The top panel of Figure~\ref{fig:dahlhausRMSE} shows that the dynamic Whittle likelihood without any modification (blue circle for $m=15$ and blue square for $m=30$) can be much improved on both measures by using any of the three modifications; the best modification is BC here (green markers), but tapering (yellow markers) is close second. Recall that BC is the recommended version in \citet{subba2021reconciling} which also includes tapering. Note also that the effect of the segment size is sizable when no modification is used, but less so when any of the three modifications are applied. The bottom panel of 
Figure~\ref{fig:dahlhausRMSE} shows similar results for the block Whittle likelihood.
  
\begin{figure}
    \centering
    \includegraphics[width=0.99\linewidth]{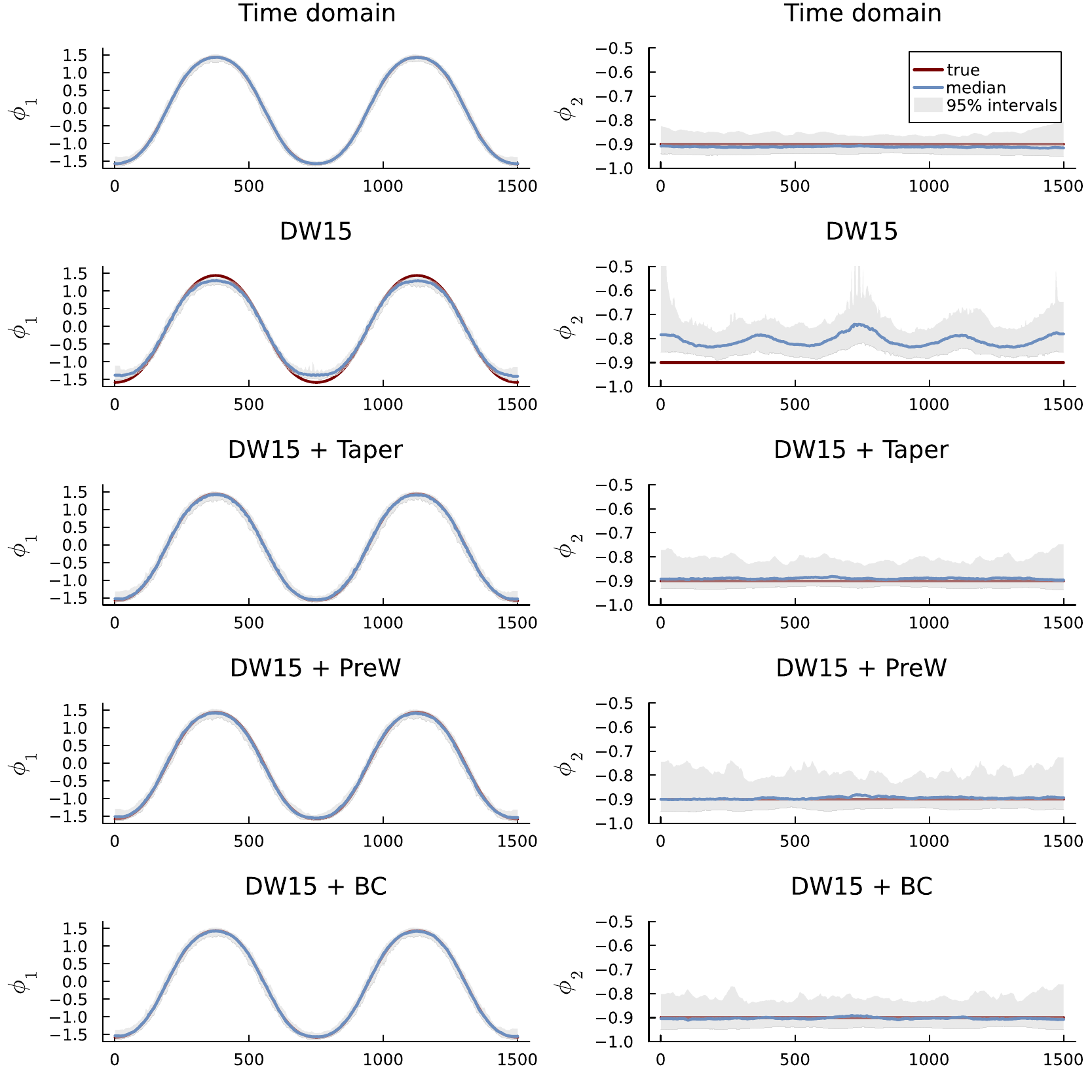}
    \caption{Experiment 1. Sampling distribution for the posterior median estimates of the parameters using time domain likelihood, and the dynamic Whittle likelihood with $m=15$, and each of the modifications.}\label{fig:DahlhausSampDist_DW_M15}
\end{figure}

\begin{figure}
    \centering
    \includegraphics[width=0.99 \linewidth]{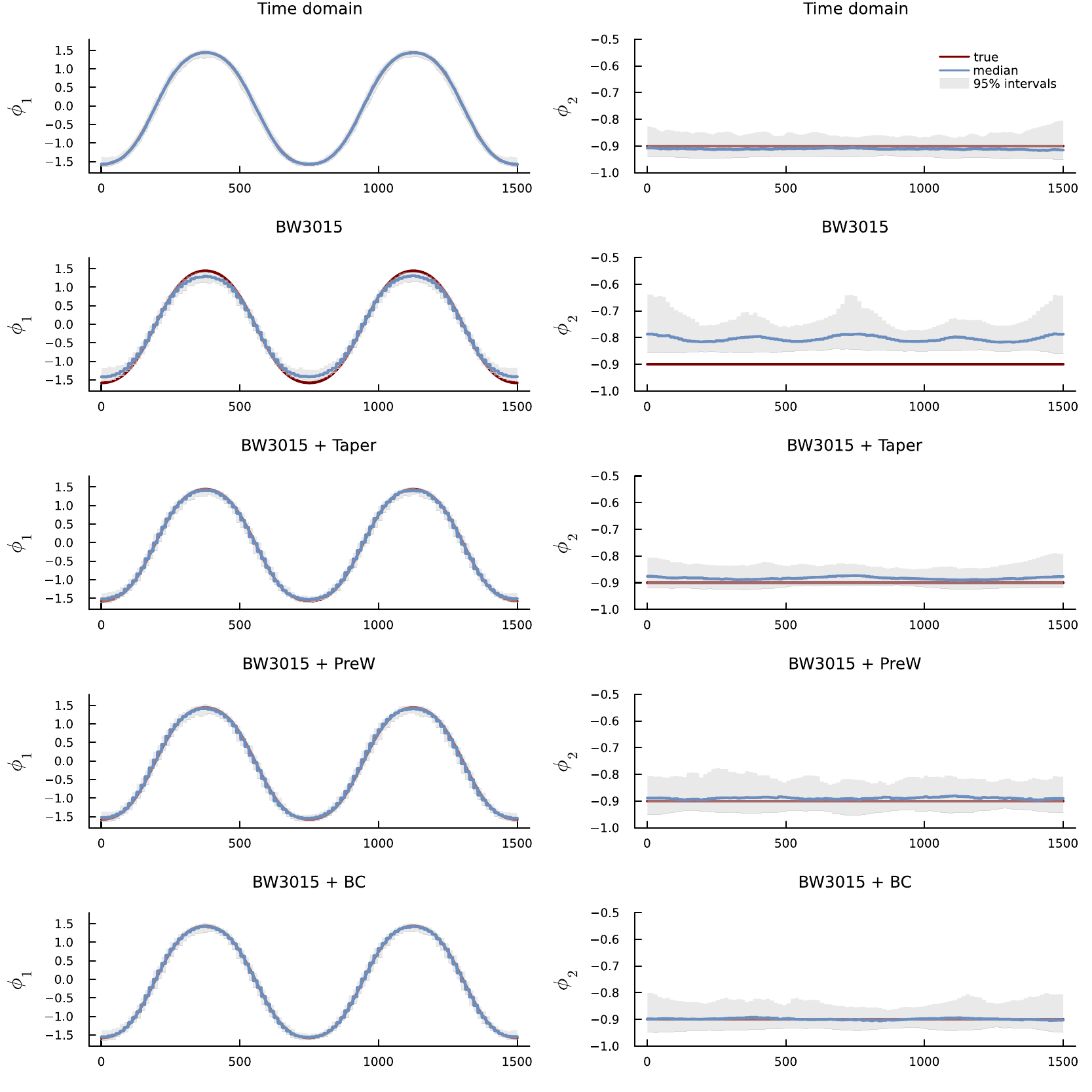}
    \caption{Experiment 1. Sampling distribution for the posterior median estimates of the parameters using time domain likelihood, and the block Whittle likelihood with $N=30$, $S=15$, and each of the modifications.}\label{fig:DahlhausSampDist_BW3015}
\end{figure}

Figure \ref{fig:DahlhausSampDist_DW_M15} plots the repeated sampling distribution of the posterior median estimate of $\phi_{1t}$ and $\phi_{2t}$ over the $100$ replicated time series for the time domain likelihood (top row) and the dynamic Whittle likelihood with $m=15$ giving a segment size of $N=2m+1=31$ (second row) with each of the three modifications (bottom three rows). The bias in the dynamic Whittle posterior median estimate is clear while the three modifications successfully remove the bias. The sampling variability in the Whittle estimates from the modifications is slightly larger than those from the time domain posterior. 
Figure~\ref{fig:DahlhausSampDist_BW3015} plots the sampling distributions for the block Whittle likelihood with $N=30$, $S=15$ and are qualitatively similar to the ones from the dynamic Whittle in 
Figure~\ref{fig:DahlhausSampDist_DW_M15}, except that tapering seems to work better for the block Whittle with no bias and low variance; this matches the result in the lower panel of Figure~\ref{fig:dahlhausRMSE}.

Figures \ref{fig:DahlhausGood_DW_M30} and \ref{fig:DahlhausBad_BW_6030} in the Supplementary material show that the posterior from the block Whittle tends to have somewhat too tight posterior intervals for the AR parameter evolution compared to the time domain posterior. The dynamic Whittle posterior seems better calibrated in this sense.

\subsection{Experiment 2 - time-invariant AR(3) process}

The second experiment explores how well the likelihood approximations perform when time-varying AR models are fitted to an AR(3) data generating process with time-invariant parameters.
\begin{equation}
    y_t = 1.4y_{t-1} - 0.7y_{t-2} + 0.2y_{t-3} + \varepsilon_t, \quad \varepsilon_t\sim N(0,1)
\end{equation}

\begin{figure}
    \centering
    \includegraphics[width=0.7\linewidth]{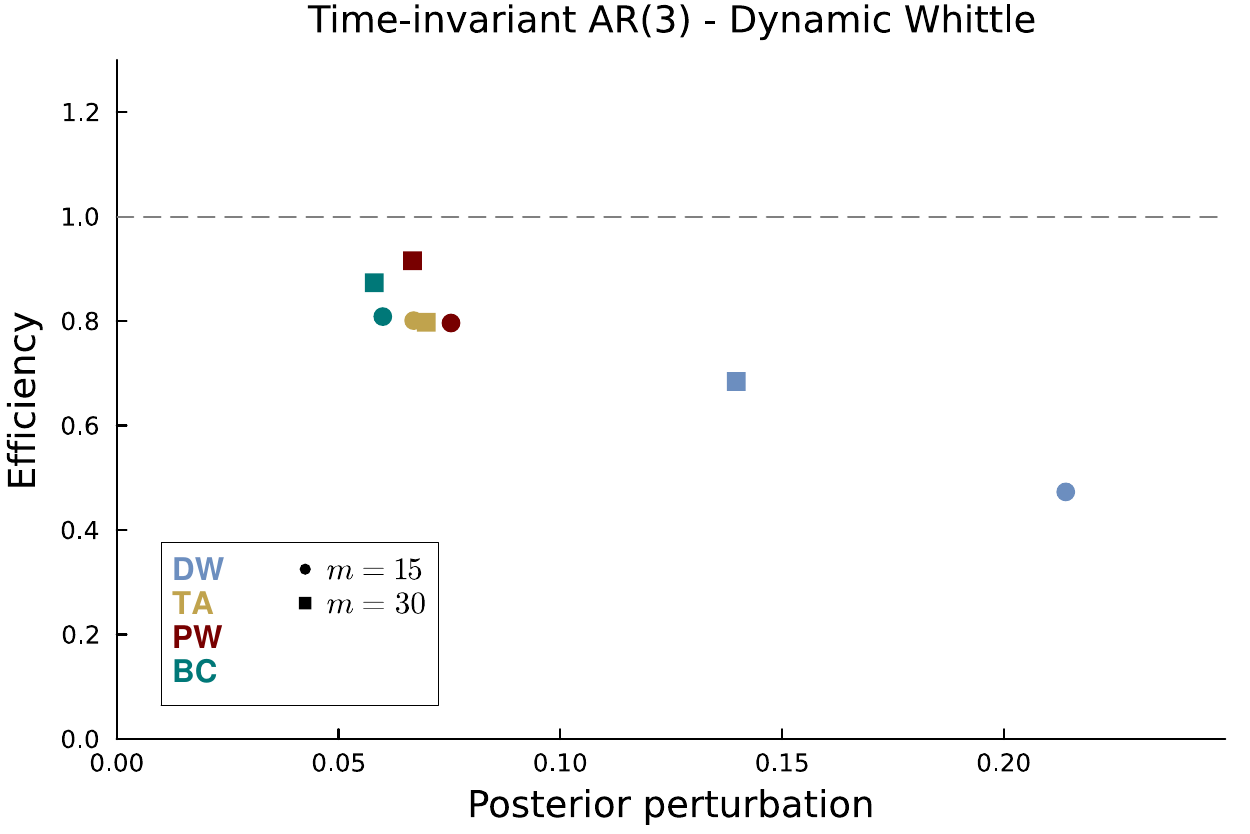}
    \includegraphics[width=0.7\linewidth]{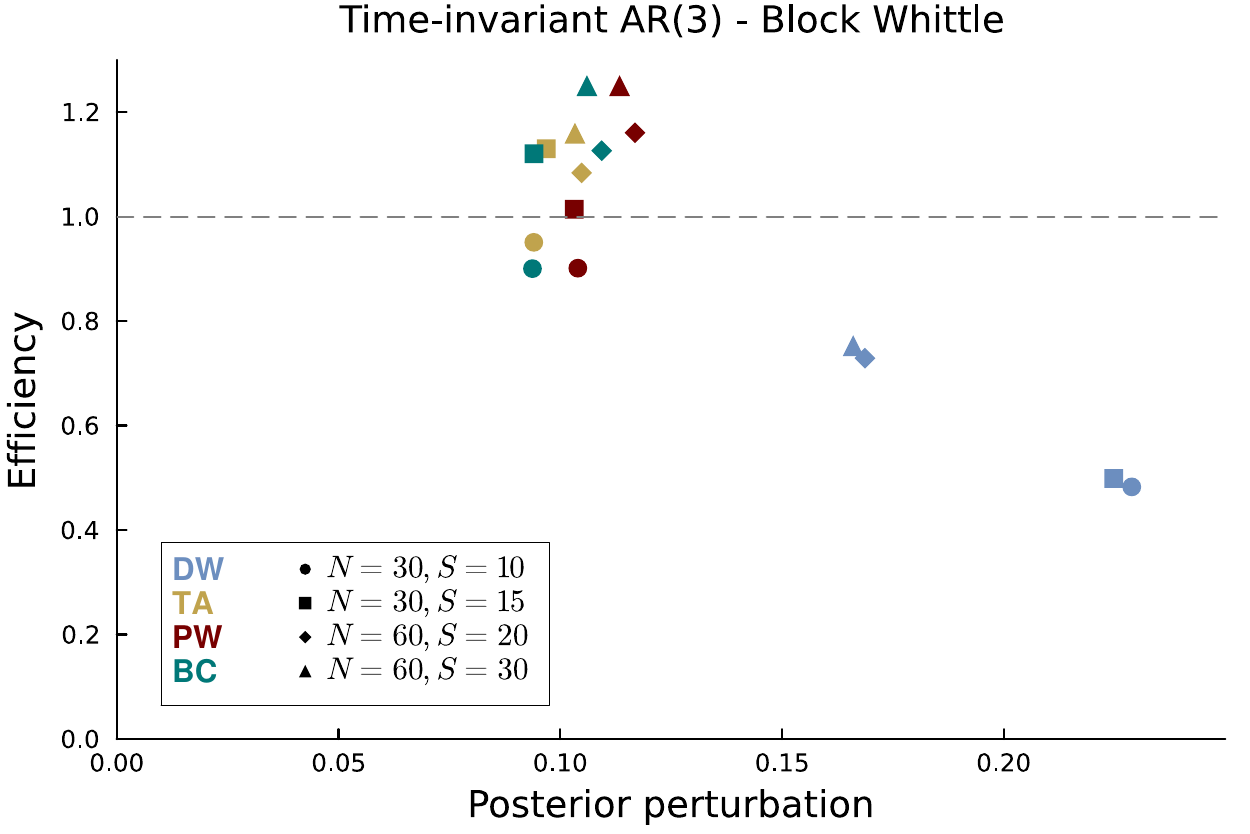}
    \caption{Experiment 2 - time-variant AR(3) process. Efficiency and posterior perturbation in the posterior for the AR parameters for the dynamic Whittle (top) and block Whittle (bottom) likelihoods, and with tapering (TA), prewhitening (PW) and boundary correction (BC) applied.}\label{fig:staticRMSE}
\end{figure}

Figure \ref{fig:staticRMSE} shows again that the raw dynamic Whittle and block Whittle can be rather inefficient and have sizable perturbations from the time domain posterior. All three modifications give substantially less perturbation and higher efficiency. The dynamic Whittle with modifications gives the smallest posterior perturbation while the modified block Whittle likelihoods have the largest efficiencies, often even a bit larger than the time domain MLE. The larger segment size performs a lot better than the smaller one for the raw Whittle-type likelihoods, which is expected given the time-invariant nature of the data. The step size in the block Whittle has virtually no effect. Both the dynamic and block Whittle likelihoods become much more robust to the segment length when any of the three modifications are applied.

Figures \ref{fig:statARSampDist_DW_M15} and \ref{fig:statARSampDist_BW_N30S15} in Section \ref{supp:exp2} show the sampling distributions of the posterior median estimate for the dynamic Whittle with $m=15$ and for block Whittle with segment size $N=30$ and step size $S=15$. The bias for the raw Whittle likelihoods is clearly visible while the sampling distributions for the three modifications are unbiased and similar. 

\subsection{Experiment 3 - time-varying AR(2) near unit root}

\begin{figure}
    \centering
    \includegraphics[width=0.99 \linewidth]{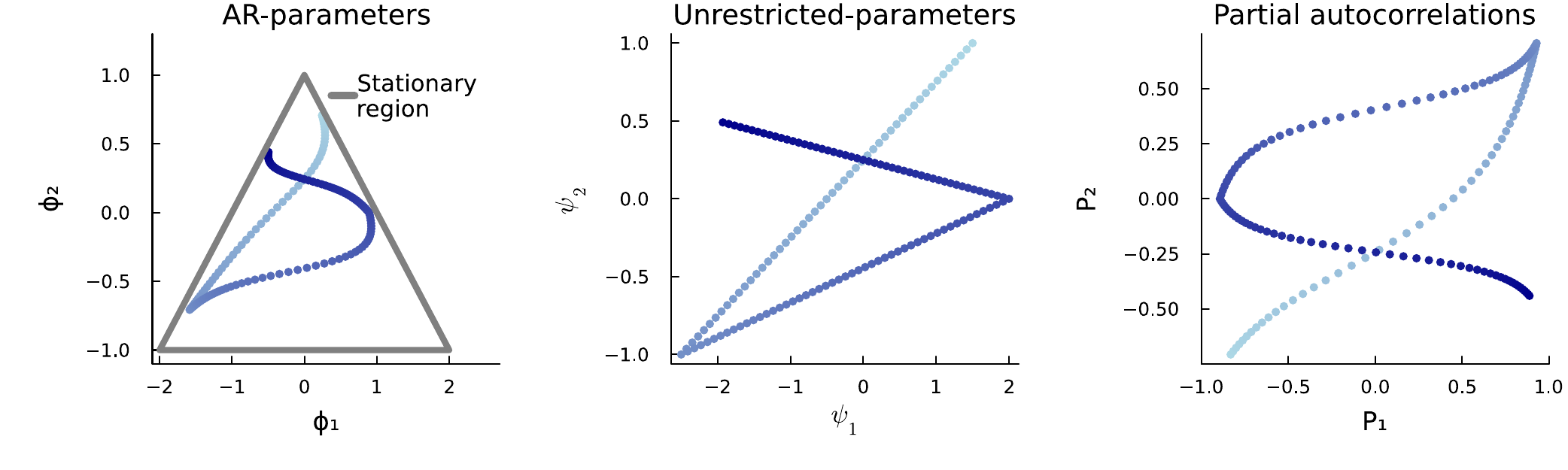}
    \caption{Parameter evolution over time where the parameters start with light blue scatters and move in steps of 10 until it ends at the dark blue color at time 1500.}\label{fig:ScatterEvolBlues}
\end{figure}

The final experiment investigates the case with parameters that give a near unit root process at certain time periods. The parameter paths are determined in the unrestricted parameter space as 
{\smaller
\begin{align*}
    \theta_{1t} = 
    \begin{cases}
        1.5 - 4(t - 1)/500,     \\             
        -2.5 + 4.5(t - 501)/500, \\         
        2 - 4(t - 1001)/500,  
    \end{cases} & \theta_{2t} = 
    \begin{cases}
        1 - 2(t - 1)/500,     &\,\, \text{ for } t=1,\ldots, 500 \\            
        -1 + (t - 501)/500,   &\,\, \text{ for } t=501, \ldots, 1000 \\        
        -0.5(t - 1001)/500,   &\,\, \text{ for } t=1001, \ldots, 1500
    \end{cases} 
\end{align*}
}%
and then transformed to the stability region via a composition of the transformations in Definition \ref{def:stability_parameterization} and \eqref{eq:monahan}. Figure \ref{fig:ScatterEvolBlues} plots the time evolution of the AR parameters. The leftmost panel shows that there are four periods where the parameters are close to the non-stable region.

\begin{figure}
    \centering
    \includegraphics[width=0.7\linewidth]{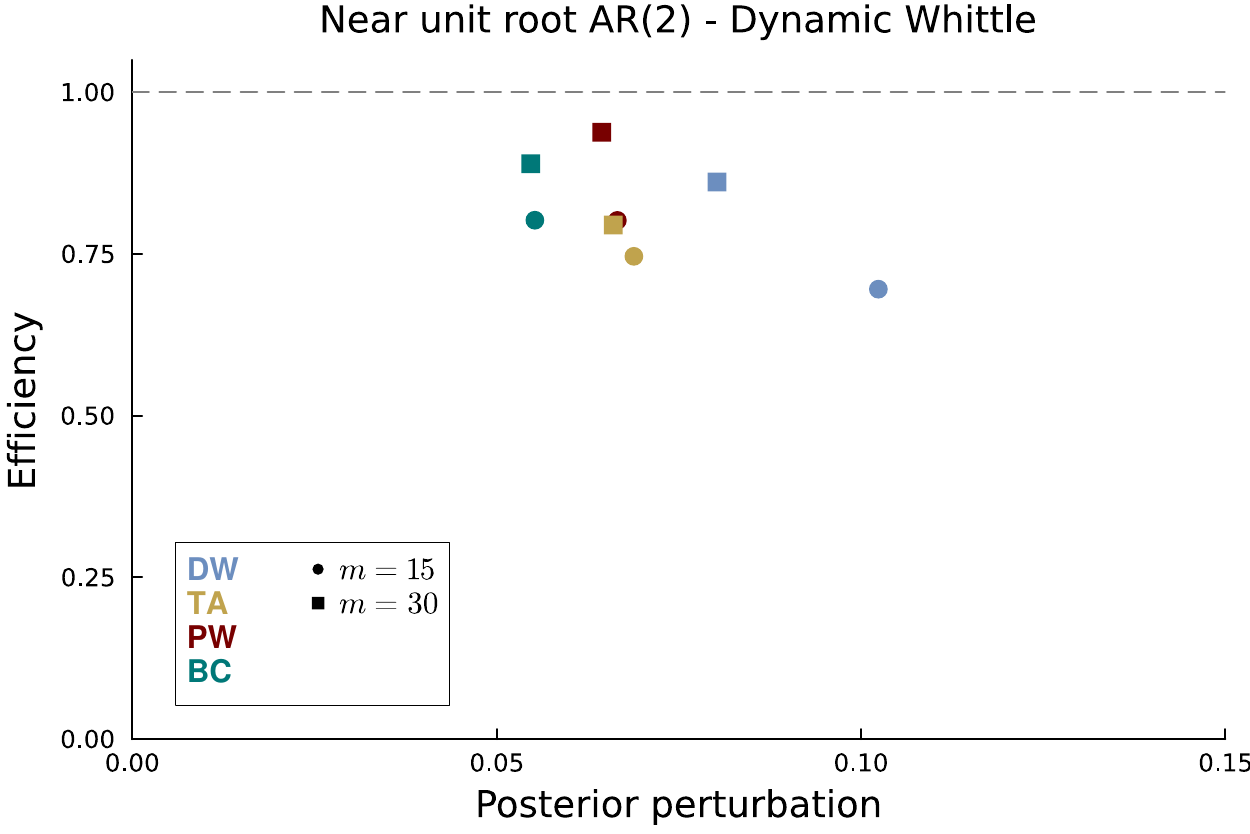}
    \includegraphics[width=0.7\linewidth]{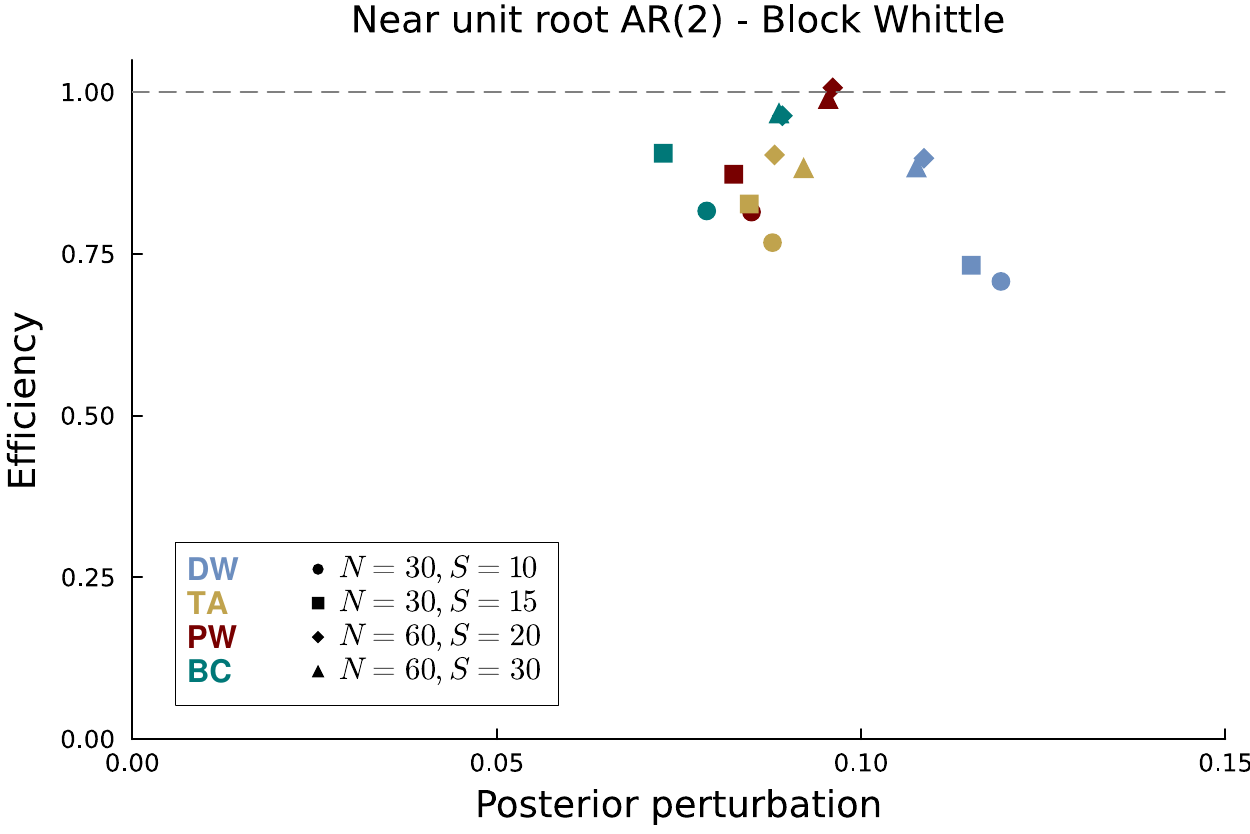}
    \caption{Experiment 3 - near unit root AR(2) process. Efficiency and posterior perturbation in the posterior for the AR parameters for the dynamic Whittle (top) and block Whittle (bottom) likelihoods, and with tapering (TA), prewhitening (PW) and boundary correction (BC) applied.}\label{fig:nearUnitRMSE}
\end{figure}

Figure \ref{fig:nearUnitRMSE} displays the efficiency and perturbation of the approximate posterior from the local Whittle likelihoods from the time domain posterior. The three modifications again lead to increased efficiency and lower perturbation when applied to the raw dynamic and block Whittle likelihoods. The dynamic Whittle performs slightly better than the block Whittle with lower posterior perturbation than the block Whittle variants. Surprisingly, the effect of the modifications is not large here, even though the process is close to unit roots on several occasions, possibly because the time spent in those regions is too short to cause problems.

\section{Application to egg price data}\label{sec:realdata}

\begin{figure}
    \includegraphics[width=0.99\linewidth]{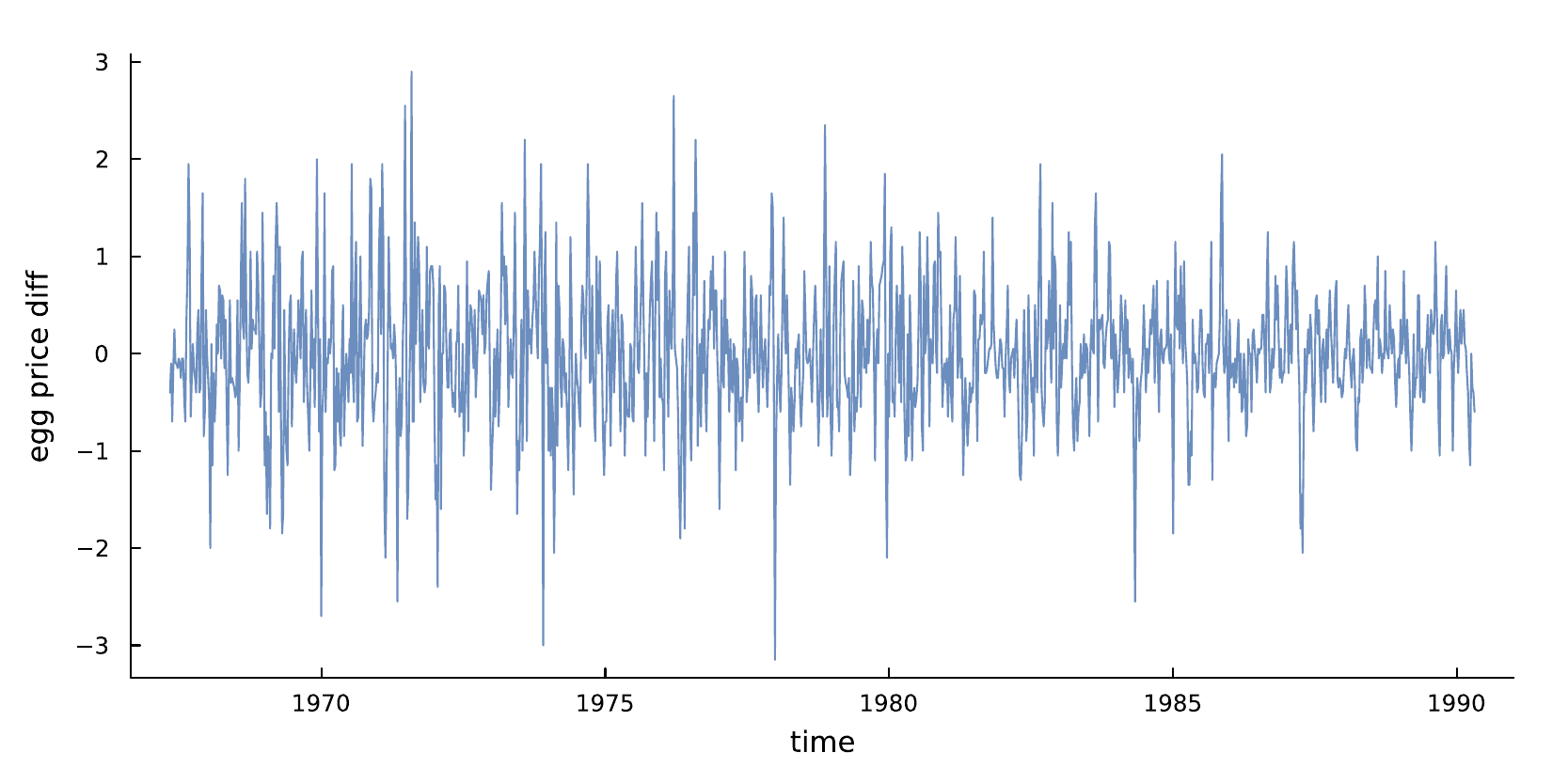}
     \caption{Weekly changes in egg prices.}\label{fig:EggDat}
\end{figure}

We analyze a data set from \citet{fan2008nonlinear}, with 1201 time series observations on weekly egg prices in the German market from April 1967 to May 1990. Following \citet{paparoditis2010validating} and \citet{tang2023bayesian}, the first differences of the series is analyzed as locally stationary. Figure \ref{fig:EggDat} shows that the change in egg prices has a decreasing variance over time. We therefore extend the tvAR model in \eqref{eq:tvAR} to have a stochastic volatility component 
\begin{equation}
    \log \sigma_{\varepsilon,t}^2 = \log \sigma_{\varepsilon,{t-1}}^2 + \zeta_t,\quad \zeta_t \overset{\mathrm{iid}}{\sim}N(0,\sigma_\zeta^2)
\end{equation}
and add $\log \sigma_{\varepsilon,t}^2$ as a new state variable at the end of the state vector.

Our aim here is to study the differences in approximating the time domain smoothing posterior of the latent AR parameters, rather than to obtain the optimal model for the data. To keep it simple, we split the time series into different non-overlapping segments and use the \texttt{auto.arima} function in R to find the optimal number of lags. The \texttt{auto.arima} function is restricted to search for models without differencing, without moving average lags, and to only to consider stationary models. Segment lengths of either 60 or 100 observations are used, and in most cases, either 1 or 2 lags are chosen. A tvAR($2$) model with stochastic volatility is therefore used.

The prior is set to $\boldsymbol{Q}\sim \text{Inv-Wishart}\left(10,\mathrm{Diag}(10^{-4}, 10^{-4},10^{-3})\right)$ and $\boldsymbol{\theta}_0\sim N\left(\mathbf{0}_{p+1},10 \cdot \boldsymbol{I}_{p+1}\right)$ for both the time domain and dynamic Whittle approaches. For the block Whittle the state variances are multiplied by the step size, $S$.

We use a segment size $N=100$ (+1 for dynamic Whittle since $N=2m+1$), following \cite{tang2023bayesian}, and step size $S=50$ for the block Whittle. Section \ref{supp:eggprices} in the supplementary material also presents results for a shorter segment size of $N=50$ combined with a step size of $S=25$.

\begin{figure}
    \includegraphics[width=0.99\linewidth]{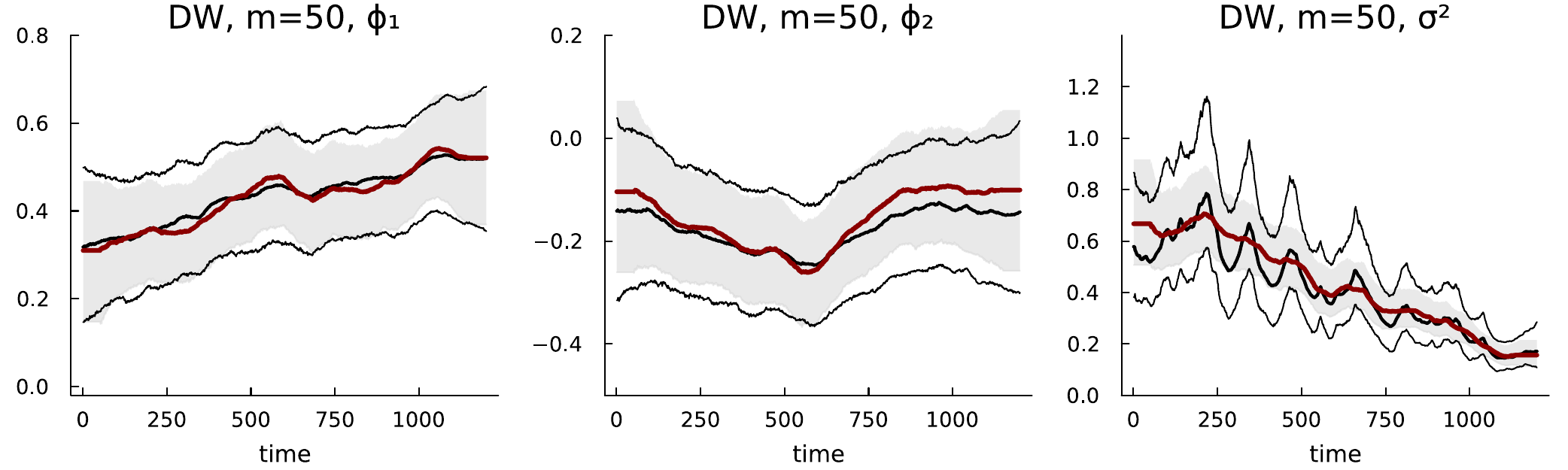}
    \includegraphics[width=0.99\linewidth]{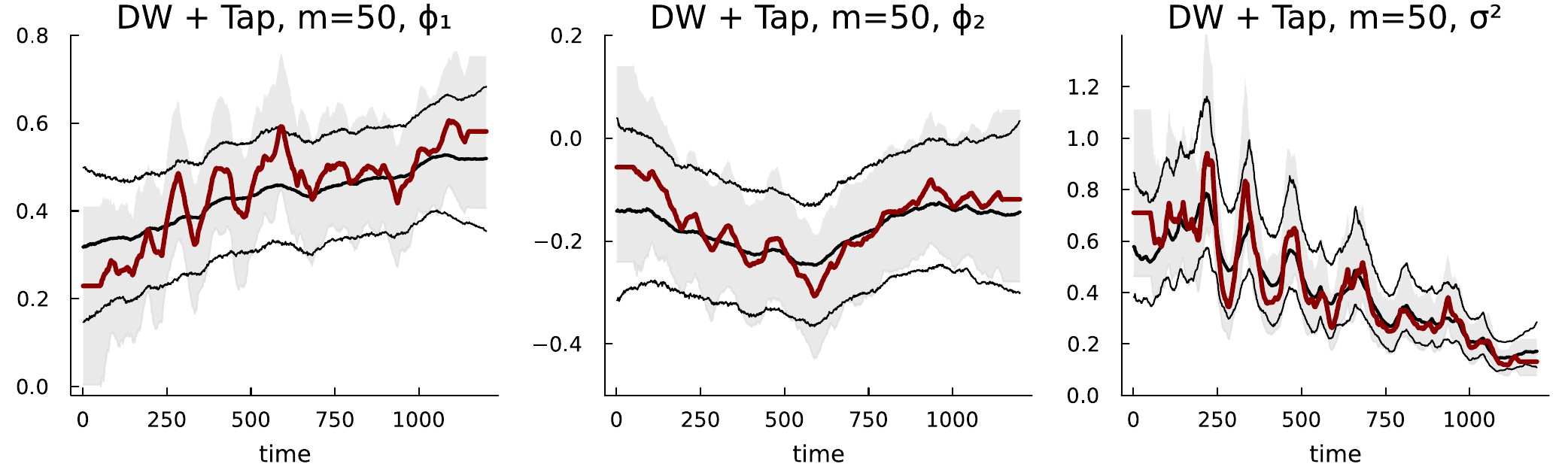}
    \includegraphics[width=0.99\linewidth]{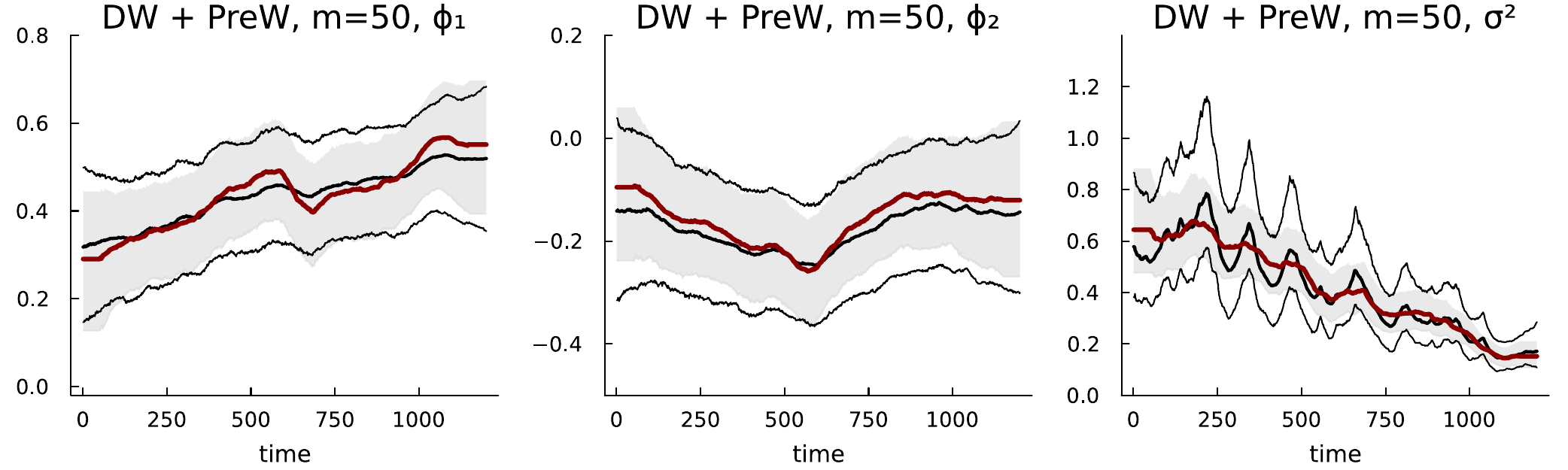}
    \includegraphics[width=0.99\linewidth]{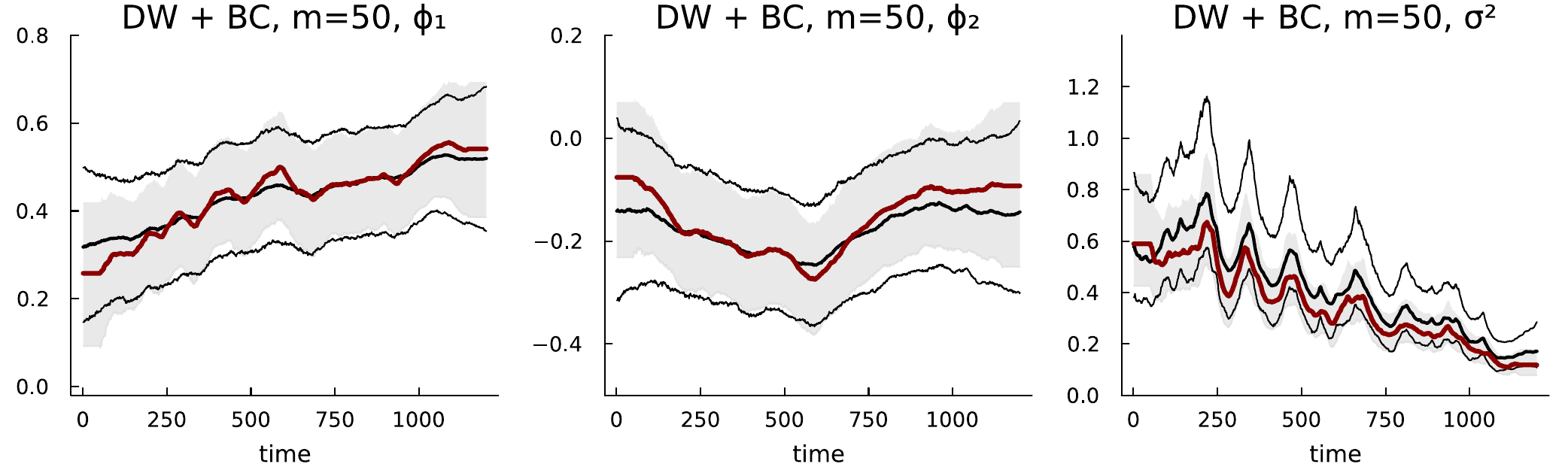}
     \caption{Egg price data. Posterior median (red) and 95\% probability bands (light-shaded grey) for the AR and variance parameter evolutions estimated from the dynamic Whittle likelihood compared to the corresponding measures for the time domain likelihood(black lines).}\label{fig:EggParEvolDW50}
\end{figure}

\begin{figure}
    \includegraphics[width=0.99\linewidth]{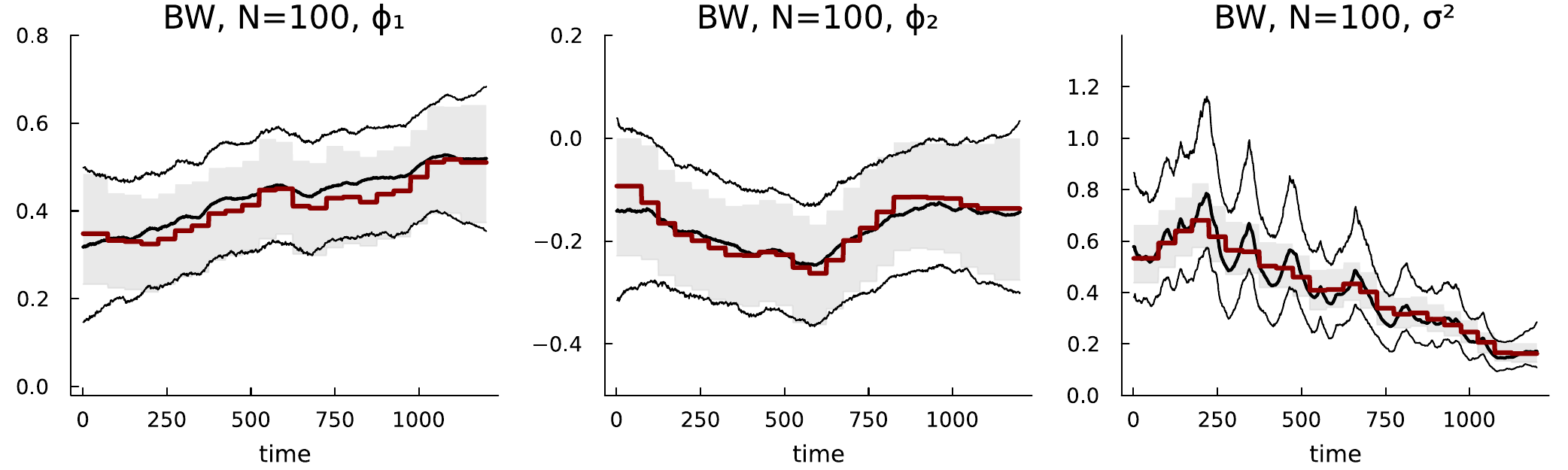}
    \includegraphics[width=0.99\linewidth]{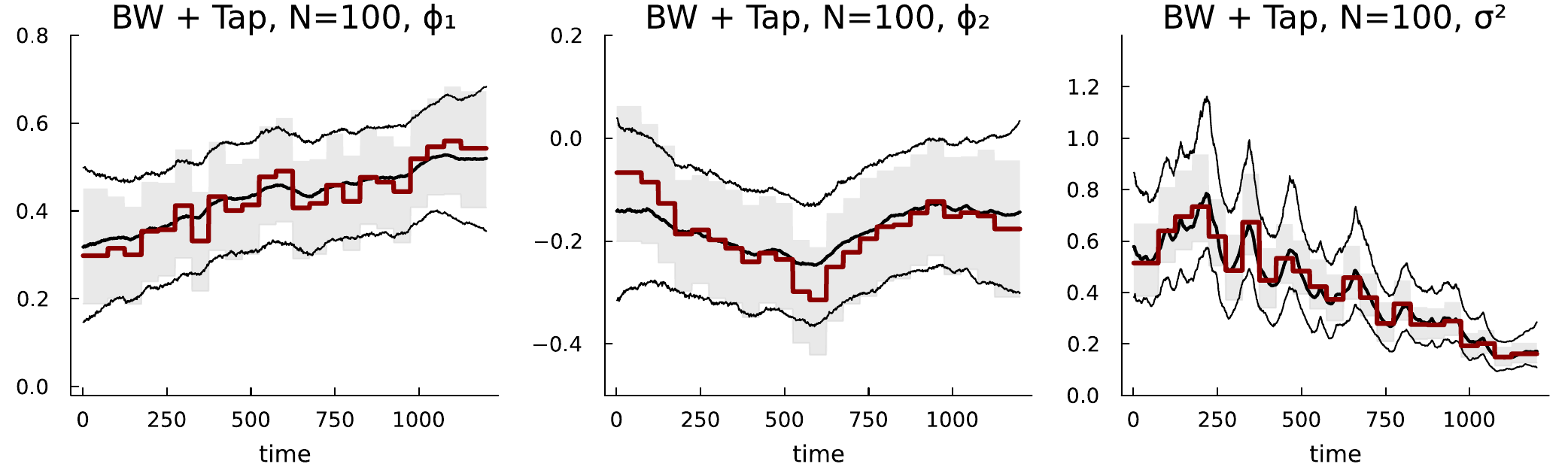}
    \includegraphics[width=0.99\linewidth]{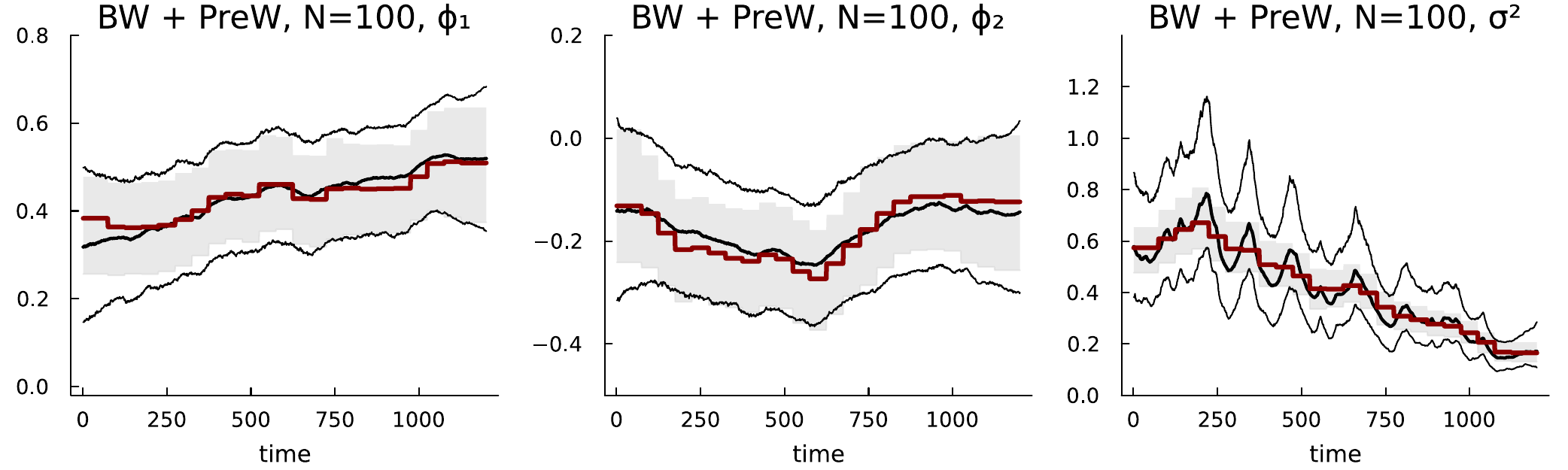}
    \includegraphics[width=0.99\linewidth]{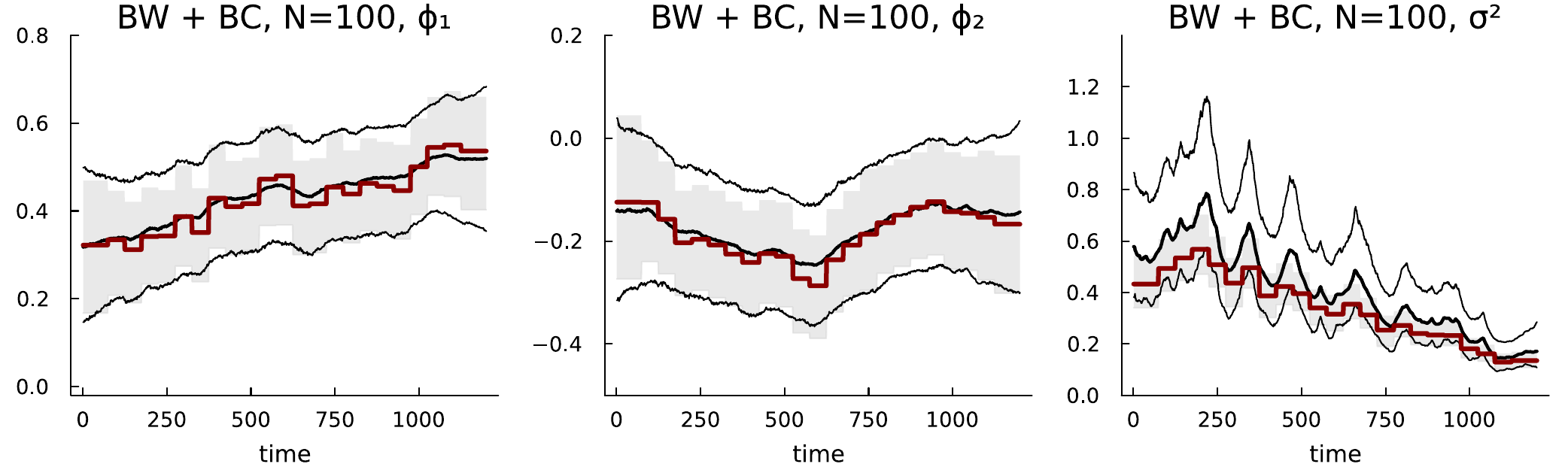}
     \caption{Egg price data. Posterior median (red) and 95\% probability bands (light-shaded grey) for the AR and variance parameter evolutions estimated from the block Whittle likelihood compared to the corresponding measures for the time domain likelihood (black lines).}\label{fig:EggParEvolBW100AR}
\end{figure}
Figures \ref{fig:EggParEvolDW50} and \ref{fig:EggParEvolBW100AR} show that the raw Whittle likelihood and the pre-whitened Whittle likelihood (red line for posterior median and shaded light gray regions for intervals) do a fairly good job when used to approximate the posterior distribution for the AR parameters from the time domain likelihood (black lines). However, the variance evolution is oversmoothed with too narrow probability bands, which is probably an effect of the fairly long data segments. Figures \ref{fig:EggParEvolDW25} and \ref{fig:EggParEvolBW50} in the supplementary material show that the smoothing posterior of the variance can be better captured when using shorter segments. 

\begin{figure}
    \includegraphics[width=0.99\linewidth]{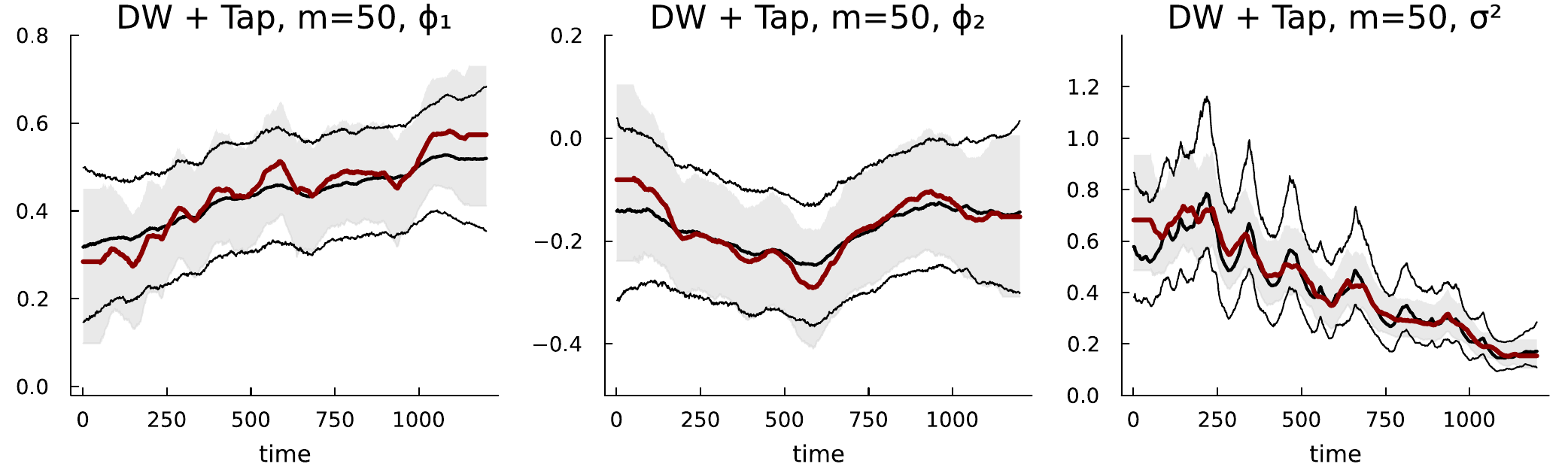}
    \includegraphics[width=0.99\linewidth]{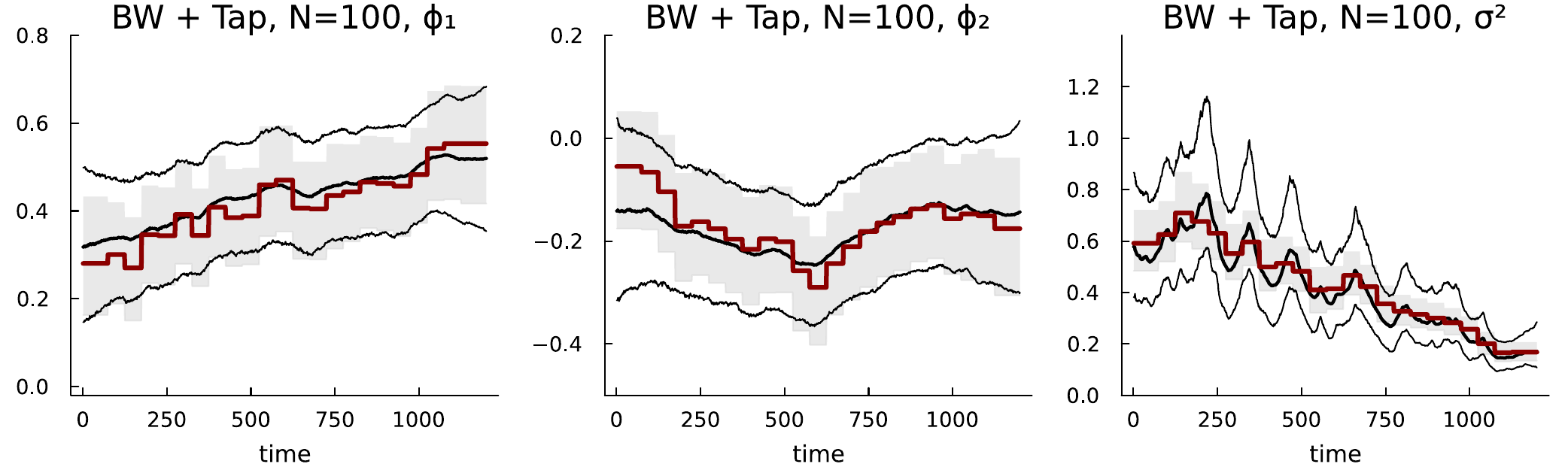}
     \caption{Egg price data. Posterior median (red) and 95\% probability bands (light-shaded grey) for the AR and variance parameter evolutions estimated from the tapered dynamic and block Whittle likelihood, \emph{with rescaling} on each segment, compared to the corresponding measures for the time domain likelihood (black lines).}\label{fig:EggRescaleParEvolDW50}
\end{figure}

The tapered likelihood produces excessively wiggly posteriors compared to their time domain counterpart. Tapering seems to exaggerate the movement of the variance. The normalized Hanning taper used here ensures that the limiting distribution of $I_T(\omega_k)$ is the same whether or not tapering is used \citep[p.128]{brillinger2001time}. Tapering can however distort the variance on a short finite segment, and the overlap between segments can then trick the model into wrongly inferring a volatility cluster. To mitigate this problem we rescale each tapered segment in the time domain to have the same empirical variance as the non-tapered segment data. This rescaling gives a much less perturbed posterior for $\sigma^2_{\varepsilon,t}$, see Figure \ref{fig:EggRescaleParEvolDW50}, although the posterior for $\phi_{1t}$ is still too wiggly. 

\begin{figure}
    \includegraphics[width=0.99\linewidth]{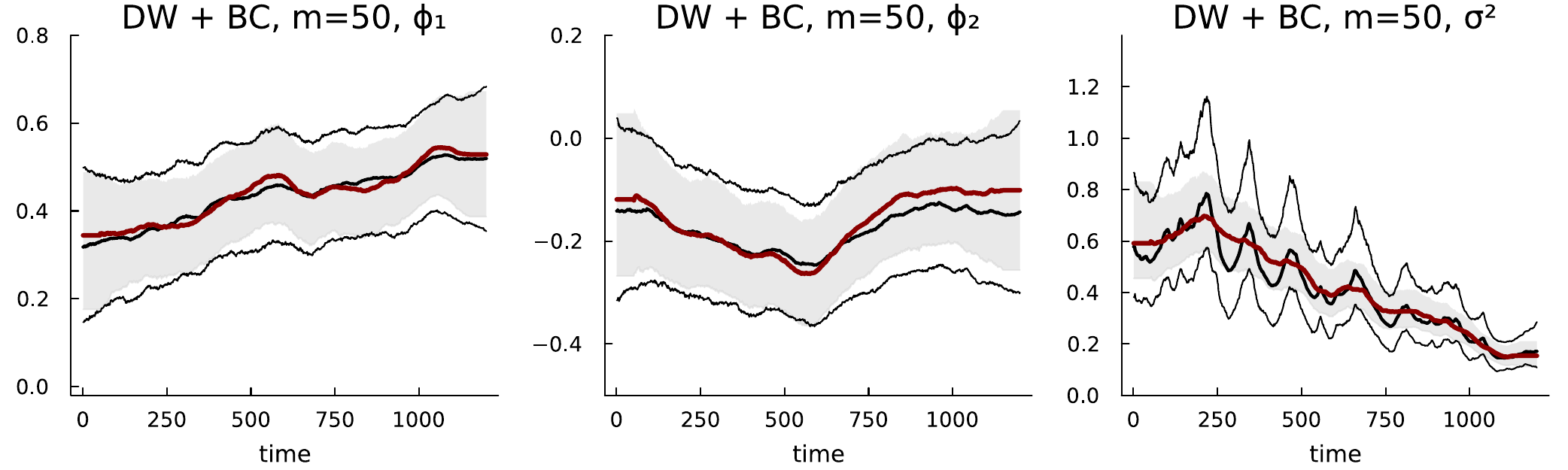}
    \includegraphics[width=0.99\linewidth]{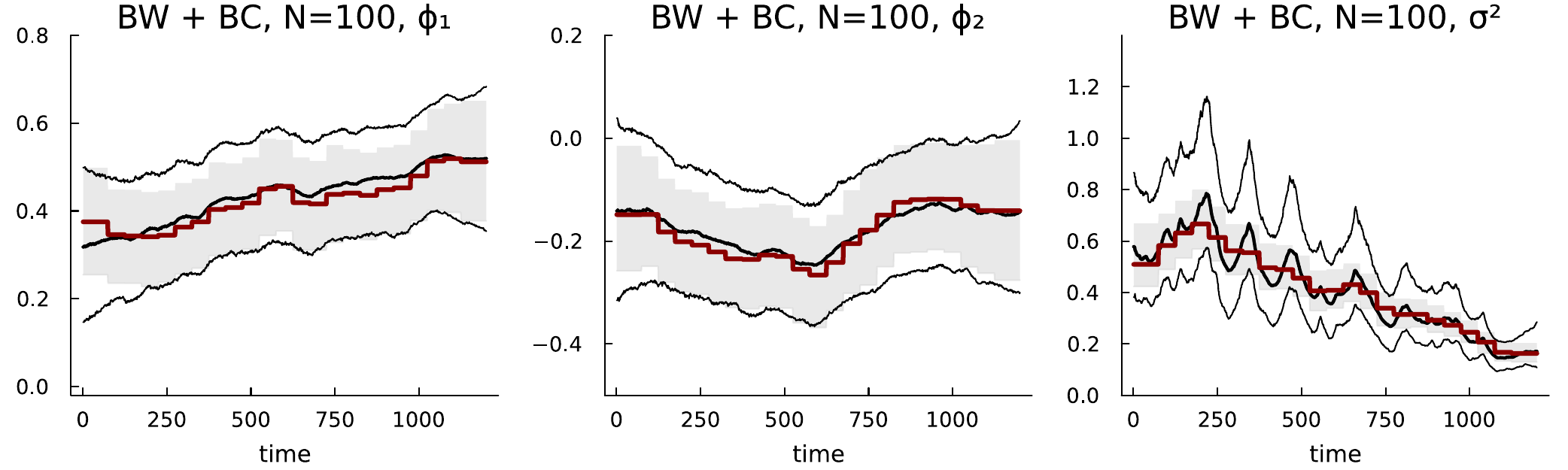}
     \caption{Egg price data. Posterior median (red) and 95\% probability bands (light-shaded grey) for the AR and variance parameter evolutions estimated from the boundary corrected dynamic and block Whittle likelihood, \emph{without tapering}, compared to the corresponding measures for the time domain likelihood (black lines).}\label{fig:EggBCnoTapParEvolDW50}
\end{figure}

The boundary corrected approach uses tapering so the same problem is to some extent also seen for this modification in Figures \ref{fig:EggParEvolDW50} and \ref{fig:EggParEvolBW50}. In addition, as noted in \cite{das2021spectral}, boundary correction induces inflated variance through the predictive DFT. One can consider a similar rescaling as for the tapered versions, but the rescaling would have to be done on the different periodogram data, and is not attempted here. We instead use boundary correction without tapering which performs very well for the AR paths and better on the log variance than BC with tapering, see Figure \ref{fig:EggBCnoTapParEvolDW50} and \ref{fig:EggBCnoTapParEvolDW25}.

\begin{figure}
    \includegraphics[width=0.49\linewidth]{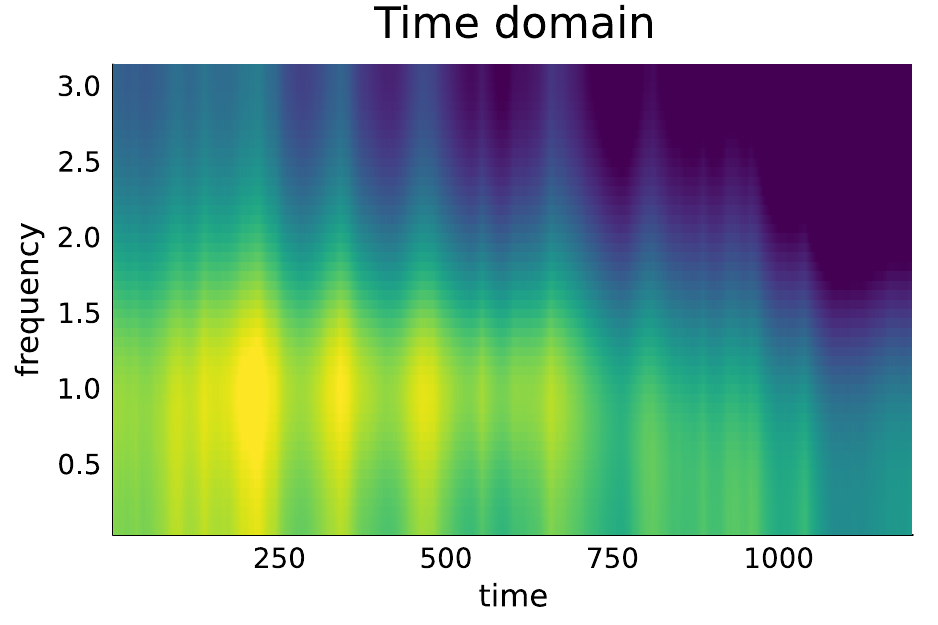}
    \includegraphics[width=0.49\linewidth]{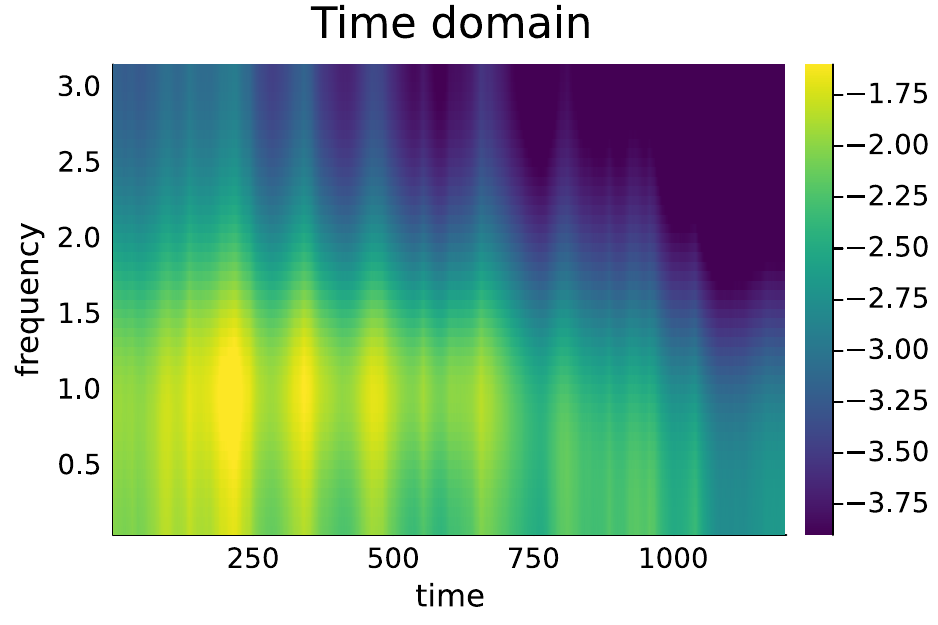}
    \includegraphics[width=0.49\linewidth]{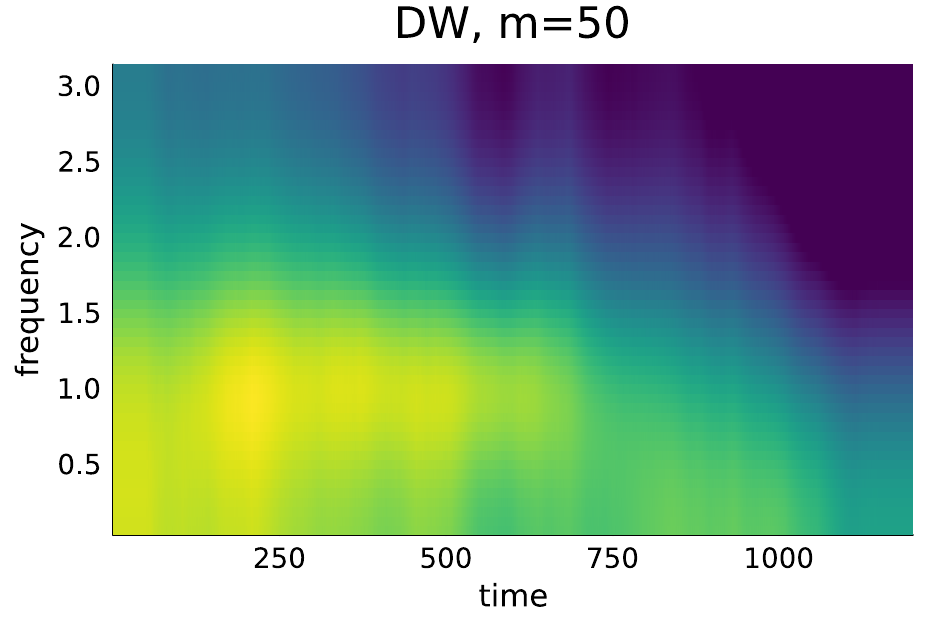}
    \includegraphics[width=0.49\linewidth]{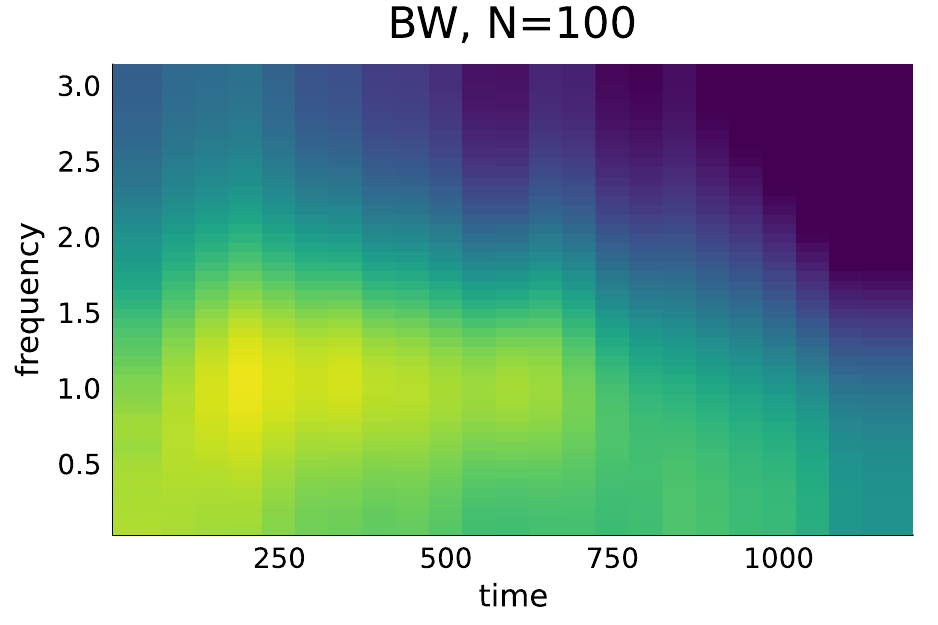}
        \includegraphics[width=0.49\linewidth]{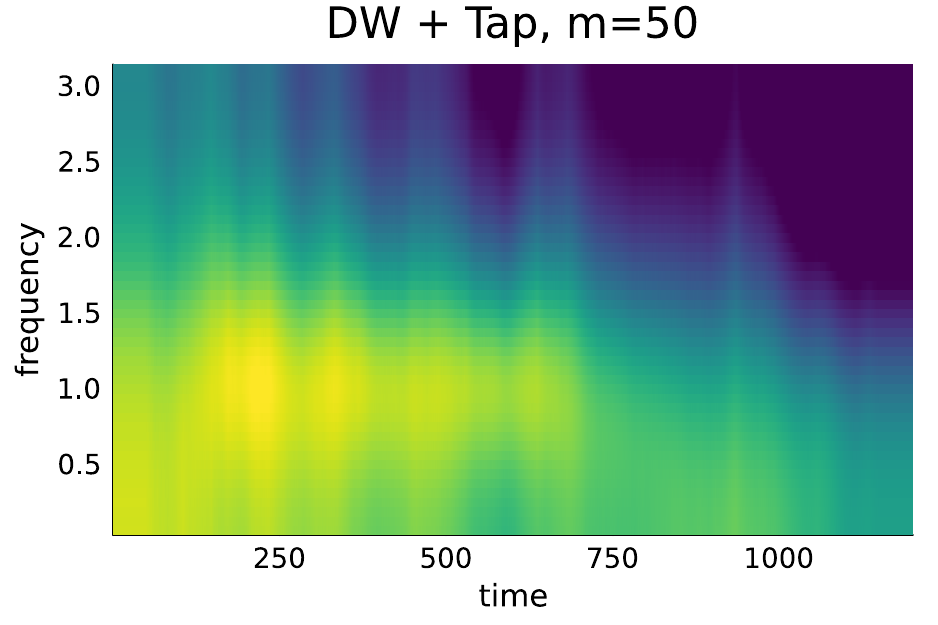}
    \includegraphics[width=0.49\linewidth]{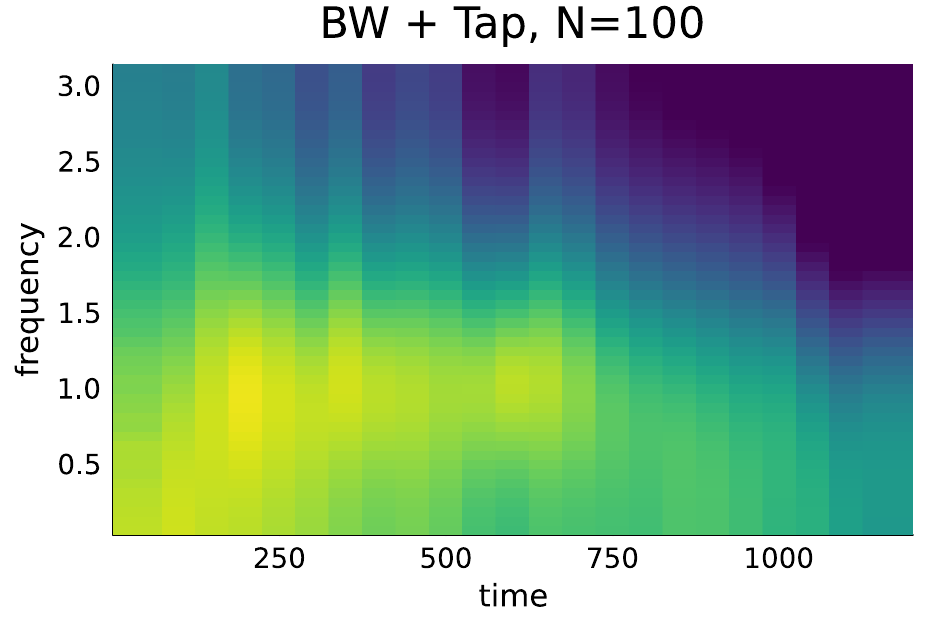}
        \includegraphics[width=0.49\linewidth]{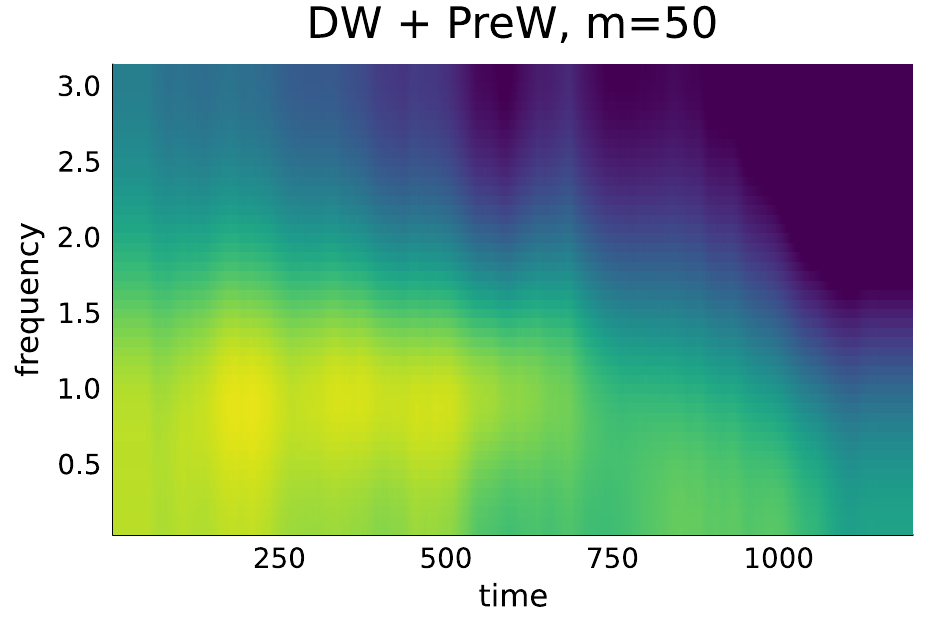}
    \includegraphics[width=0.49\linewidth]{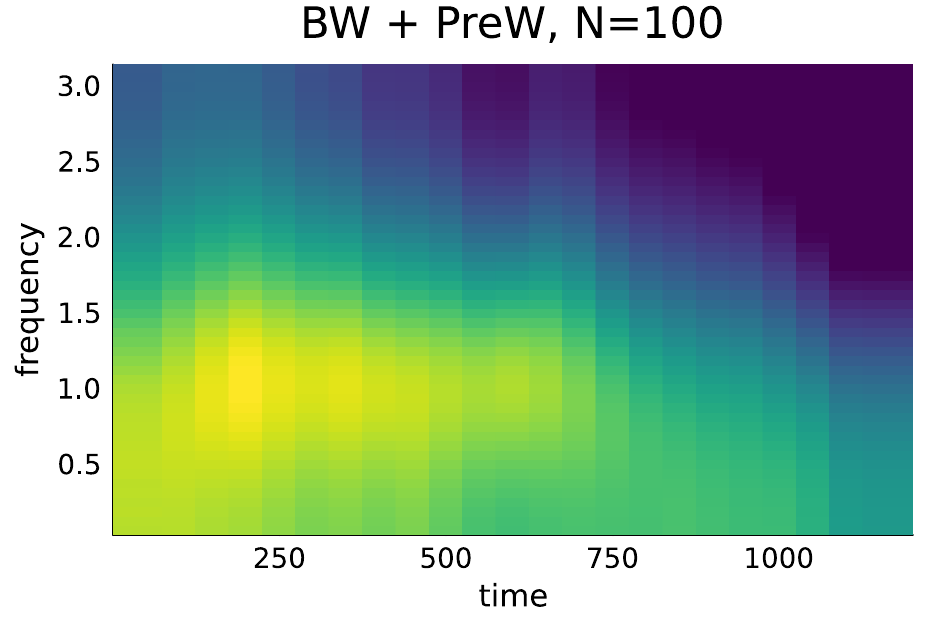}
        \includegraphics[width=0.49\linewidth]{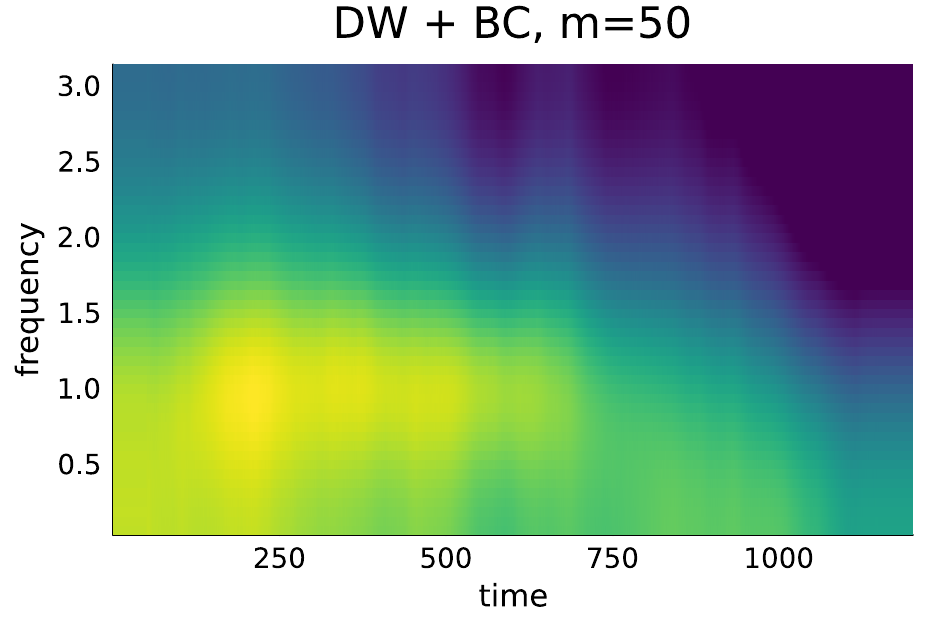}
    \includegraphics[width=0.49\linewidth]{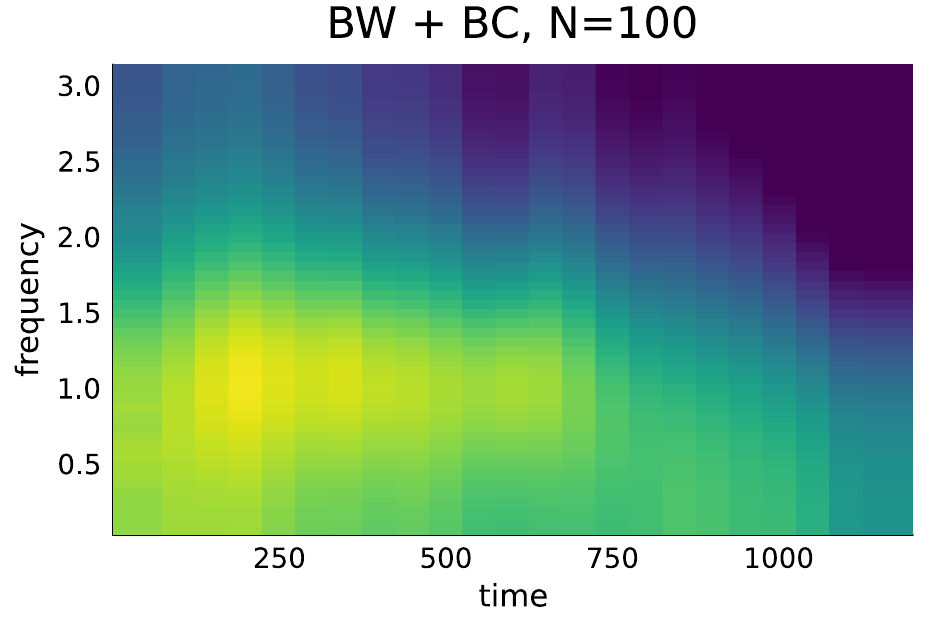}
     \caption{Egg price data. Posterior median of the log Spectrogram for the longer segment length. The tapered segment data is rescaled to match the non-tapered variance and boundary correction is used without tapering.}\label{fig:EggSpecto50100}
\end{figure}

\begin{figure}
    \includegraphics[width=0.49\linewidth]{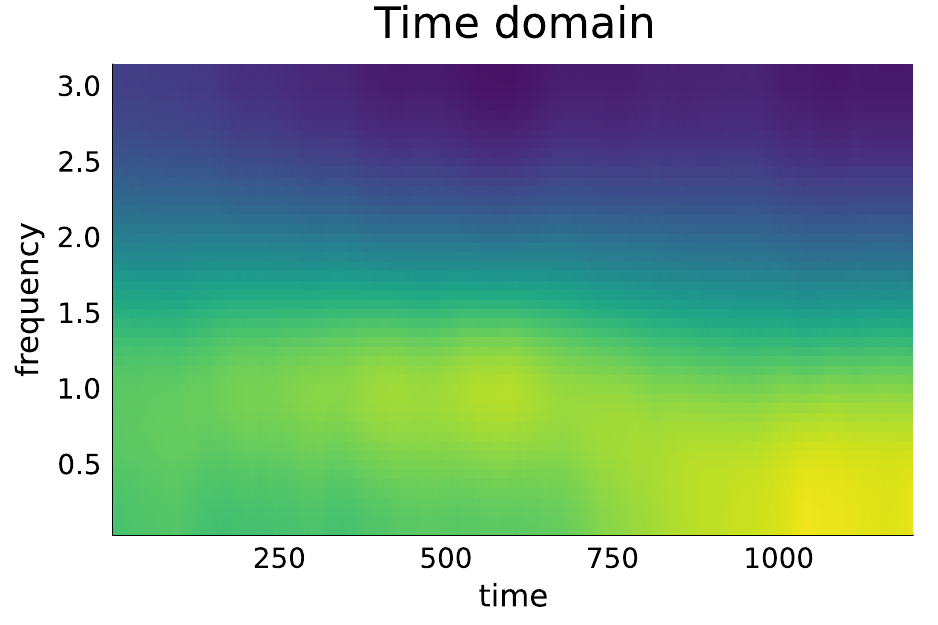}
    \includegraphics[width=0.49\linewidth]{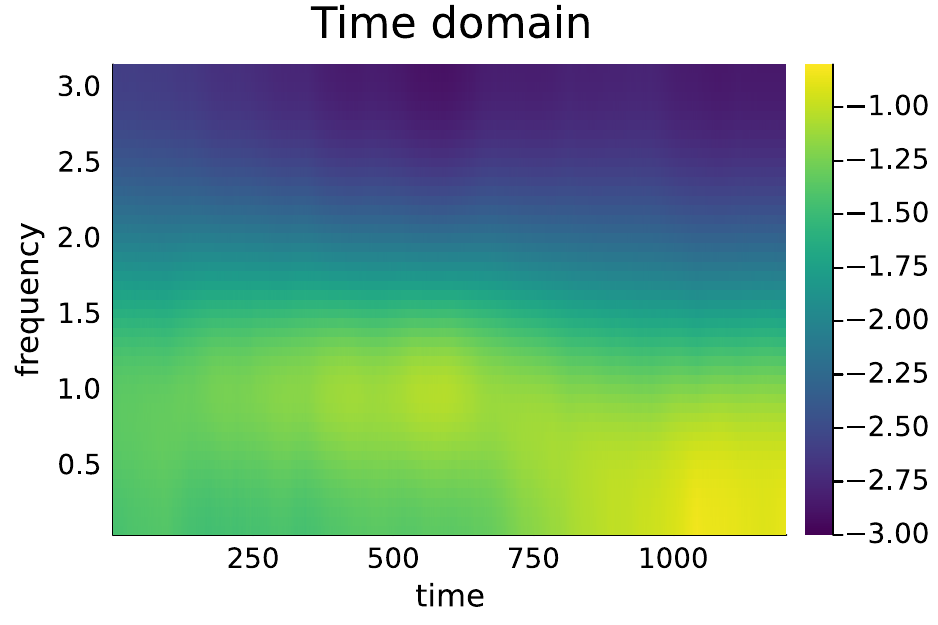}
    \includegraphics[width=0.49\linewidth]{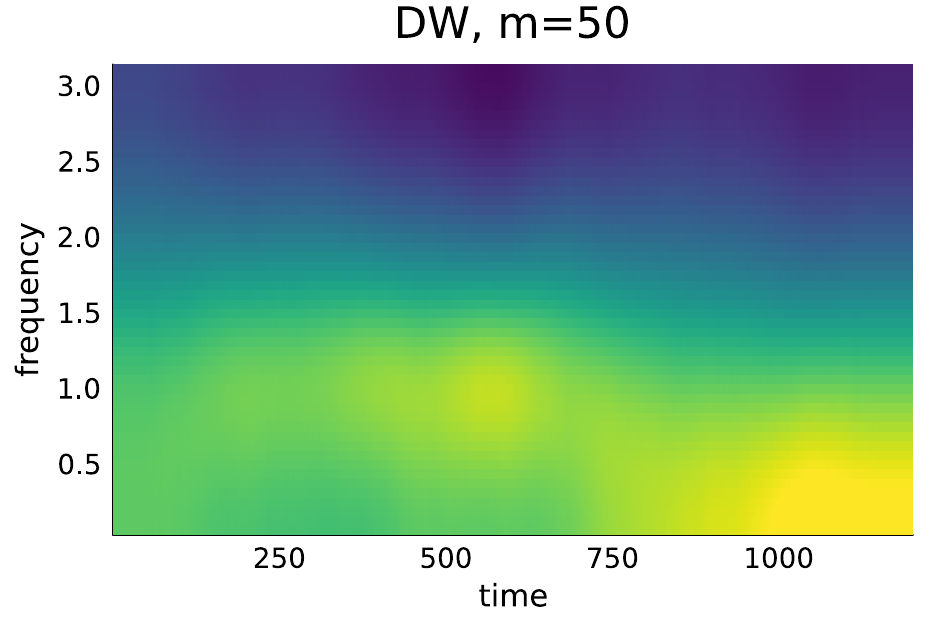}
    \includegraphics[width=0.49\linewidth]{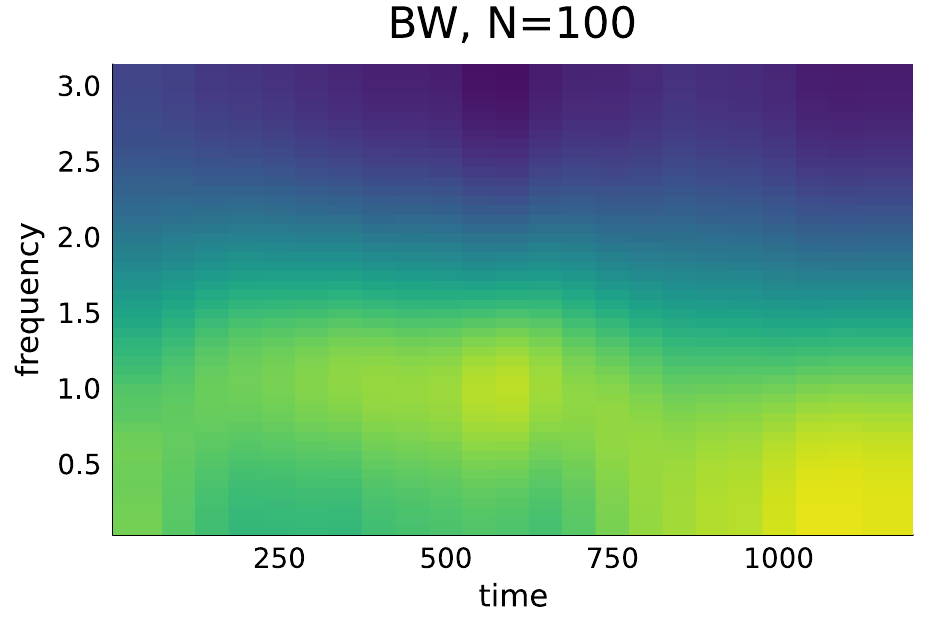}
    \includegraphics[width=0.49\linewidth]{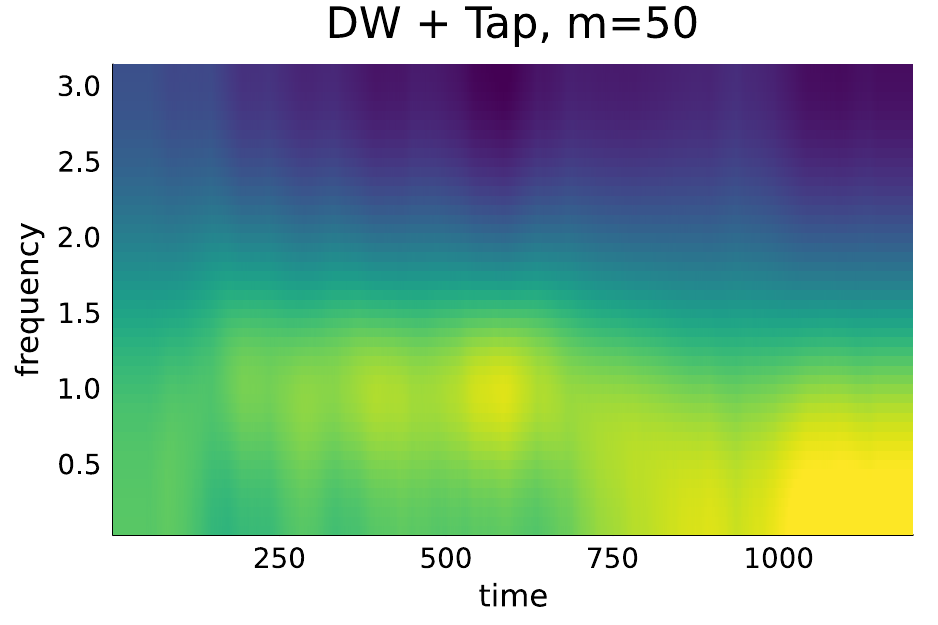}
    \includegraphics[width=0.49\linewidth]{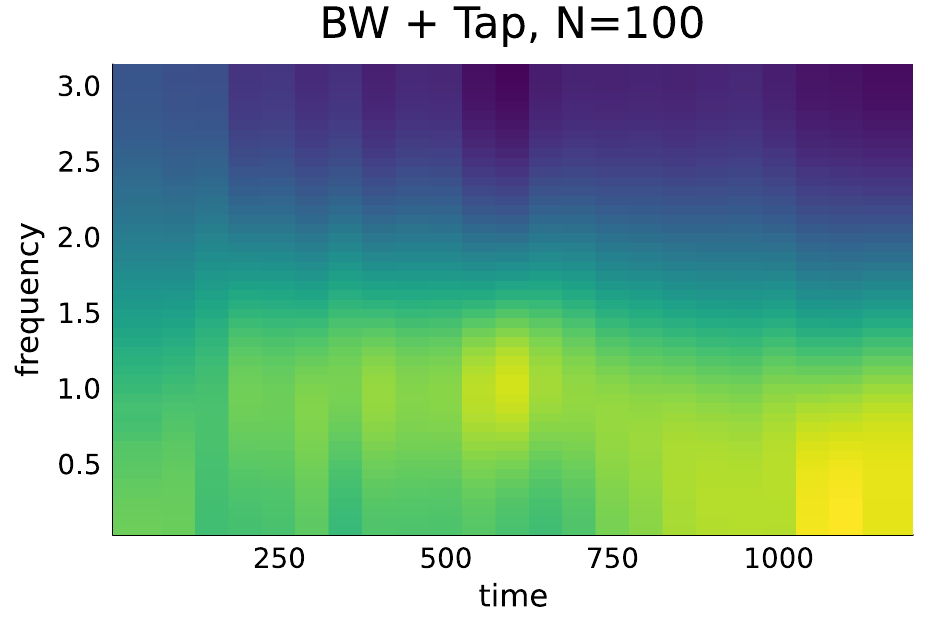}
    \includegraphics[width=0.49\linewidth]{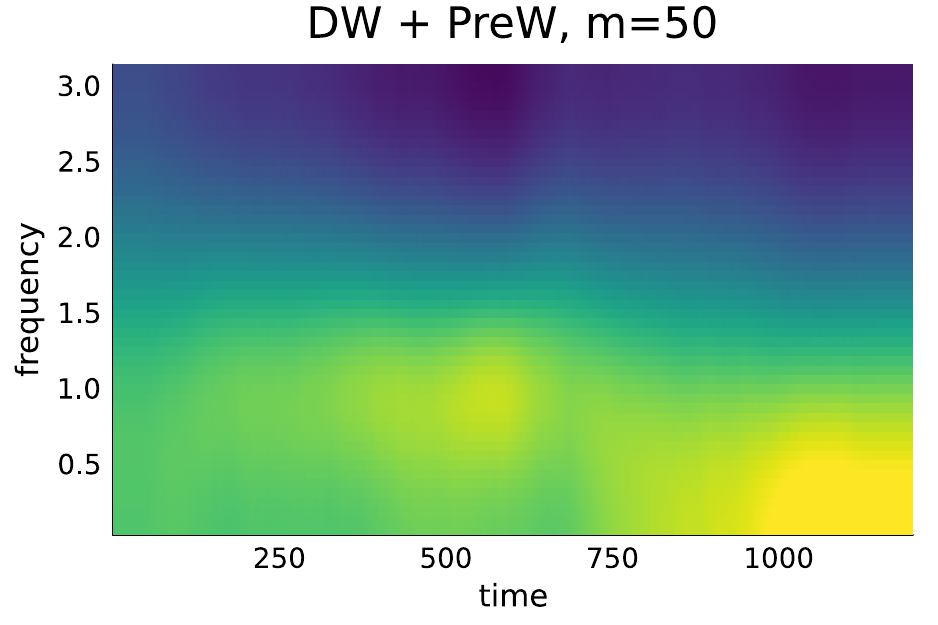}
    \includegraphics[width=0.49\linewidth]{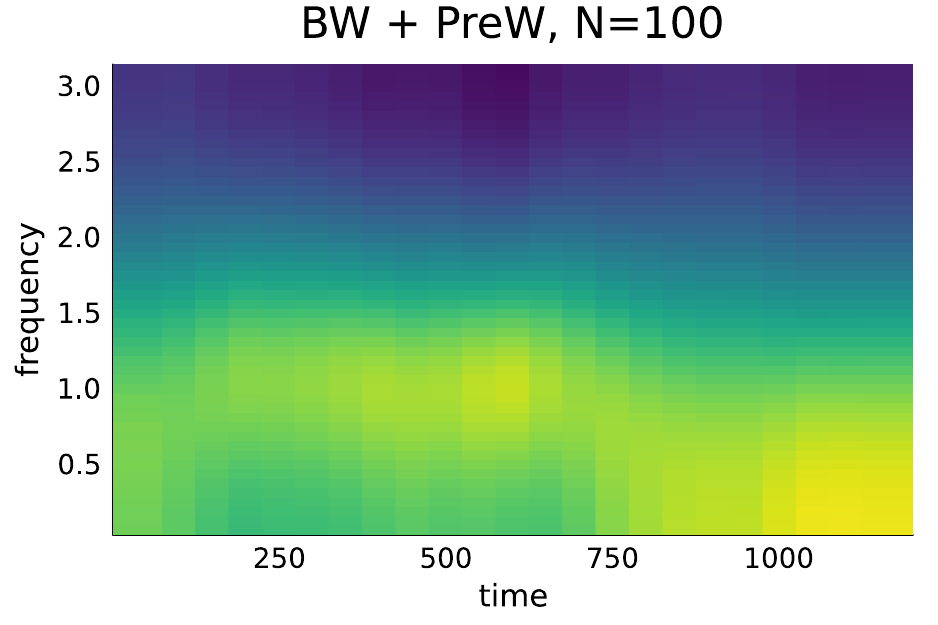}
    \includegraphics[width=0.49\linewidth]{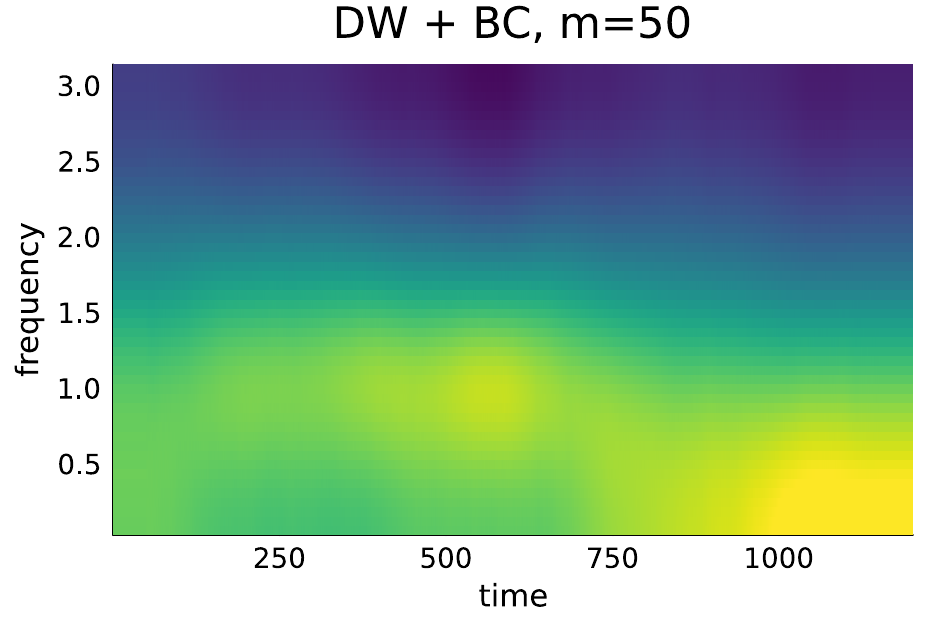}
    \includegraphics[width=0.49\linewidth]{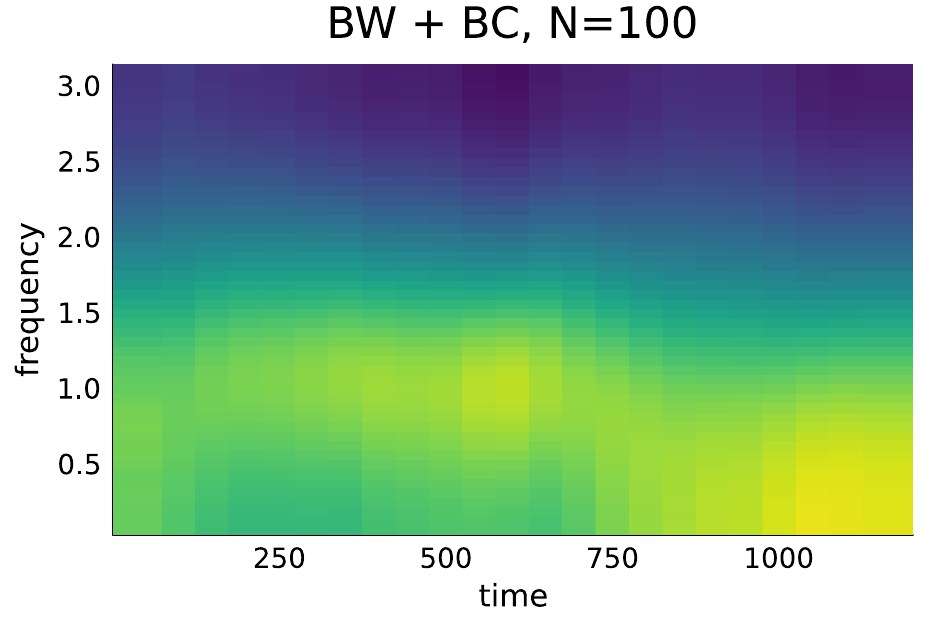}
     \caption{Egg price data. Posterior median of the log of the squared transfer function for the longer segment length. The tapered segment data is rescaled to match the non-tapered variance and boundary correction is used without tapering.}\label{fig:EggTrans50100}
\end{figure}

Figure \ref{fig:EggSpecto50100} plots the posterior median for the log spectral density. To isolate the time-variation in the AR parameter from the time-variation in the variance, Figure \ref{fig:EggTrans50100} plots the posterior median of the log of the squared transfer function, $\log \vert \phi_p (e^{-\im \omega}) \vert ^2$. The raw and prewhitened Whittle likelihoods approximate the time domain median of the log of the squared transfer function fairly well while the log spectrogram is over-smoothed.

\section{Conclusions}\label{sec:conclusions}

We evaluate the accuracy of the posterior distribution from the block and dynamic Whittle likelihood approximations for time-varying parameter processes. The comparisons are made for the popular time-varying AR model, which are parameterized to be stable at every time period. Due to the non-Gaussian distribution of the periodogram data, and the nonlinearity stemming from the stability parameterization, the posterior distribution is sampled by Gibbs sampling with a particle MCMC step for the parameter evolutions.

The quality of the approximations are compared in three simulation experiments along two dimensions: i) the (relative) efficiency of point estimates of the AR parameters from the approximate posteriors compared to the time domain estimates, and ii) the degree of perturbation of the approximate posterior from the target time domain posterior. The perturbation is measured by the distance between the two distributions' quantiles.

Both the dynamic and block Whittle likelihoods perform relatively well in all experiments, with an edge for the dynamic Whittle. Both likelihood approximations are much improved by any of the three modifications: tapering, prewhitening and boundary correction. 

We apply the likelihood approximations and modifications to a previously analyzed time series of egg prices. The raw Whittle posteriors and the one from prewhitening are fairly accurate representations of the time domain posterior, at least for the AR parameters. However, the time-varying variance of the data, modeled by a stochastic volatility model, causes some problems for likelihood approximations based on tapering, which needs to be rescaled to give a more correct representation   of the posterior for the variance evolution.

\begin{acknowledgement}
Oskar Gustafsson and Mattias Villani were partially funded by the Swedish Research Council, grant 2020-02846. The computations were enabled by resources provided by the National Academic Infrastructure for Supercomputing in Sweden (NAISS), partially funded by the Swedish Research Council through grant agreement no. 2022-06725.
\end{acknowledgement}

\bibliographystyle{apalike}
\bibliography{refSpectralLike}


\title*{Supplement to 'Spectral domain likelihoods for Bayesian inference in time-varying parameter models'}
\titlerunning{Supplementary Material}

\author{Oskar Gustafsson$^{a}$\thanks{Corresponding author: oskar.gustafsson@stat.su.se. $^a$Department of Statistics, Stockholm University. 
$^b$School of Business, University of New South Wales. $^c$Data Analytics Center for Resources and Environments (DARE).}, Mattias Villani$^{a}$ and Robert Kohn$^{b,c}$}

\maketitle

\renewcommand{\theequation}{S\arabic{equation}}
\renewcommand{\thesection}{S\arabic{section}}

\renewcommand{\thelemma}{S\arabic{lemma}}
\renewcommand{\thealgocf}{S\arabic{algocf}}
\renewcommand{\thefigure}{S\arabic{figure}}
\renewcommand{\thetable}{S\arabic{table}}
\renewcommand{\thepage}{S\arabic{page}}
\renewcommand{\thetable}{S\arabic{table}}
\renewcommand{\thepage}{S\arabic{page}}
\setcounter{page}{1}
\setcounter{section}{0}
\setcounter{equation}{0}
\setcounter{algocf}{0}
\setcounter{lemma}{0}
\setcounter{table}{0}
\setcounter{figure}{0}
\numberwithin{equation}{section}

\vspace{-2cm}
\textbf{Abstract}: This Supplement contains additional results for the paper \emph{Spectral domain likelihoods for Bayesian inference in time-varying parameter models}.

%
%
\section{Additional results from Experiment 1}\label{supp:exp1}

\begin{figure}[H]
    \centering
    \includegraphics[width=0.99\linewidth]{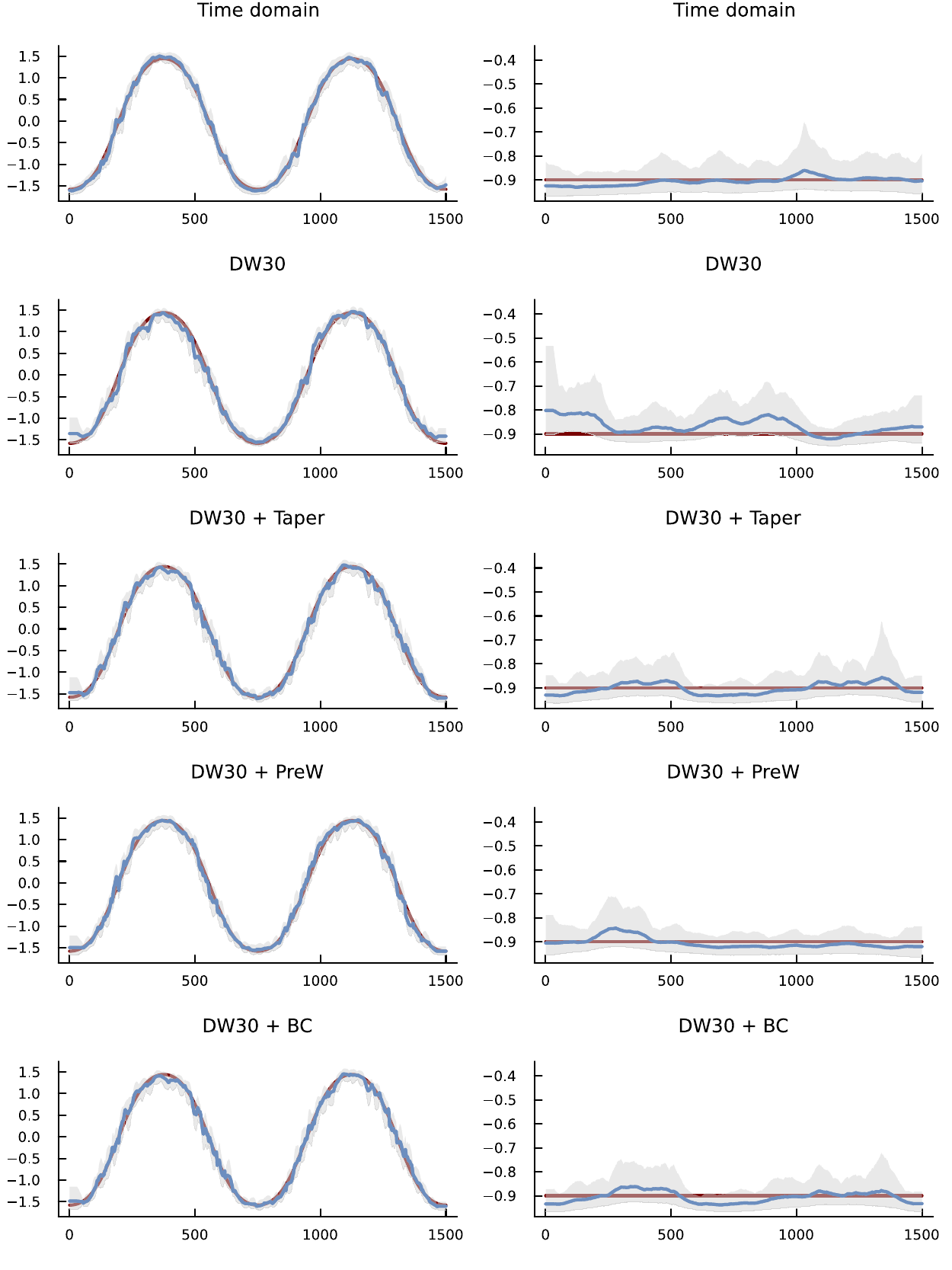}
    \caption{Experiment 1. Posterior distribution based on the 5\% best dataset for each method ranked according to the perturbation measure. Dynamic Whittle with $m=30$.}\label{fig:DahlhausGood_DW_M30}
\end{figure}

\begin{figure}[H]
    \centering
    \includegraphics[width=0.99\linewidth]{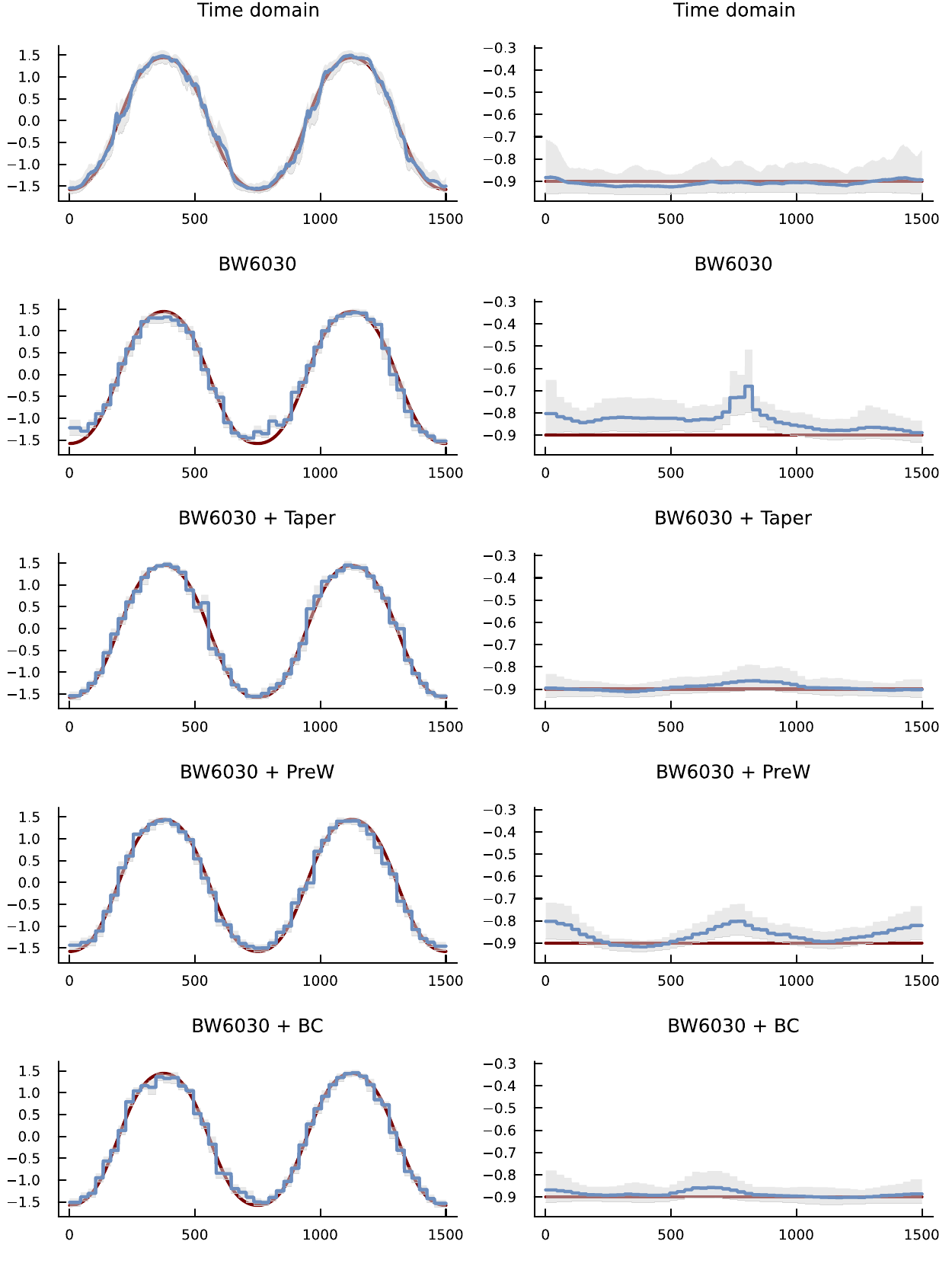}
    \caption{Experiment 1. Posterior distribution based on the 5\% worst dataset for each method ranked according to the perturbation measure. Block Whittle with $N=60$, $S=30$.}\label{fig:DahlhausBad_BW_6030}
\end{figure}

Figures \ref{fig:DahlhausGood_DW_M30} and \ref{fig:DahlhausBad_BW_6030} plots the posterior distribution for the 5\% best dataset and the 5\% worst dataset for each methods, ranked according to the perturbation measure. Note that the posteriors for different methods is therefore typically for different datasets.

\section{Additional results from Experiment 2}\label{supp:exp2}

\begin{figure}[H]
    \centering
    \includegraphics[width=0.99\linewidth]{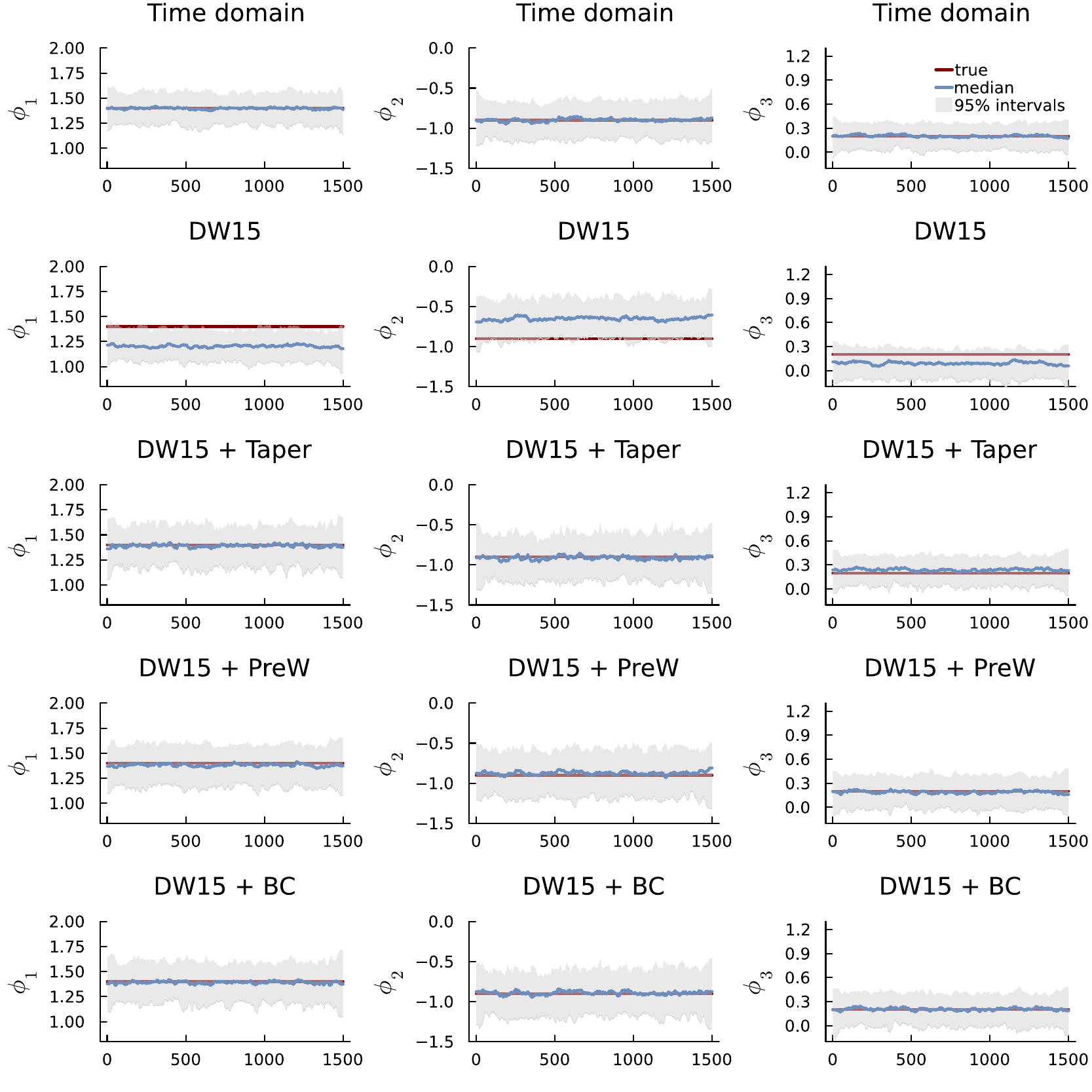}
    \caption{Experiment 2. Sampling distribution for the posterior median estimates of the parameters. Dynamic Whittle with $m=15$.}\label{fig:statARSampDist_DW_M15}
\end{figure}

\begin{figure}[H]
    \centering
    \includegraphics[width=0.99\linewidth]{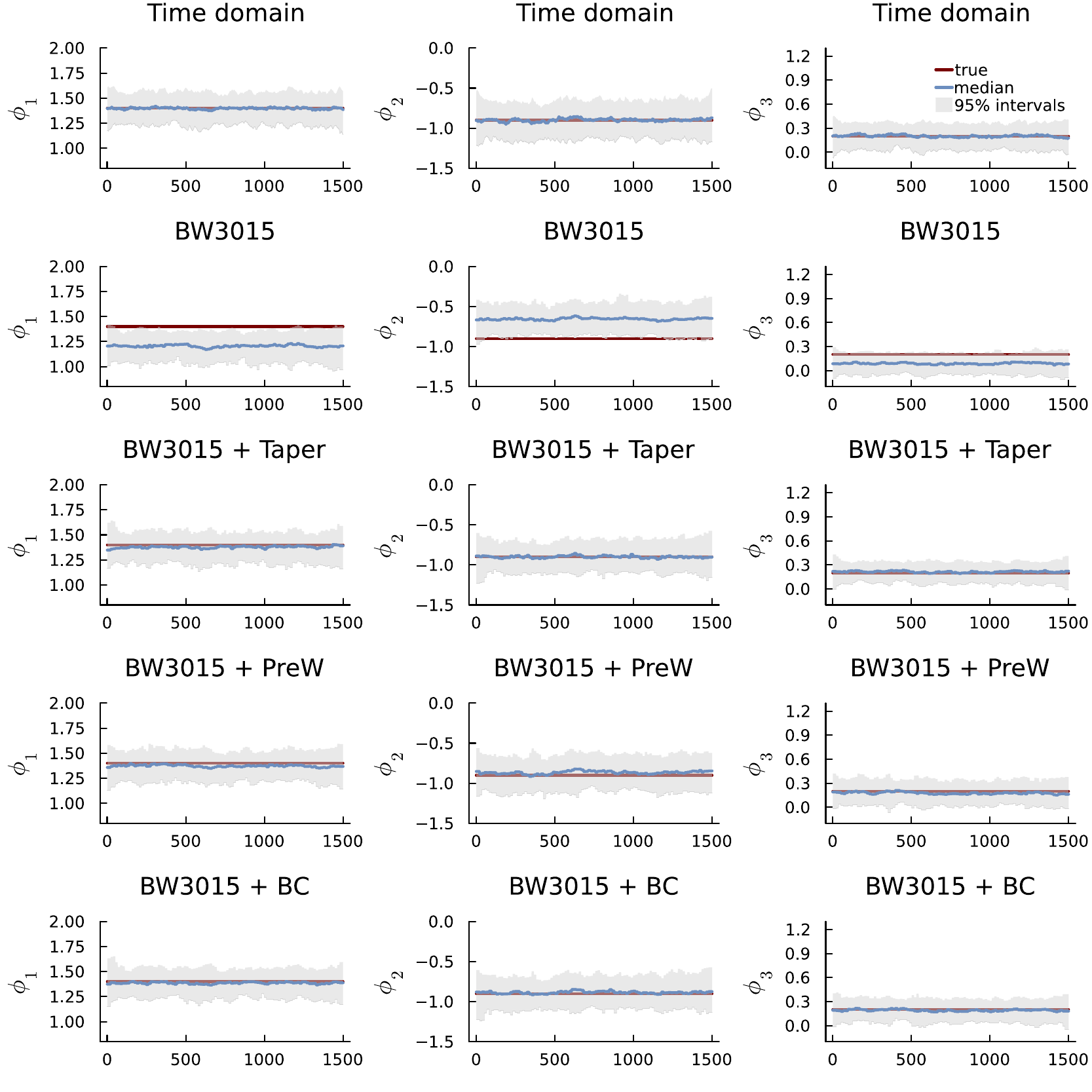}
    \caption{Experiment 2. Sampling distribution for the posterior median estimates of the parameters. Block Whittle $N=30$ and $S=15$.}\label{fig:statARSampDist_BW_N30S15}
\end{figure}

Figures \ref{fig:statARSampDist_DW_M15} and \ref{fig:statARSampDist_BW_N30S15} plots the sampling distributions for the posterior median estimates from the dynamic and block Whittle posteriors for experiment 2.

\newpage 

\section{Additional results from Experiment 3}\label{supp:exp3}

\begin{figure}[H]
    \centering
    \includegraphics[width=0.99\linewidth]{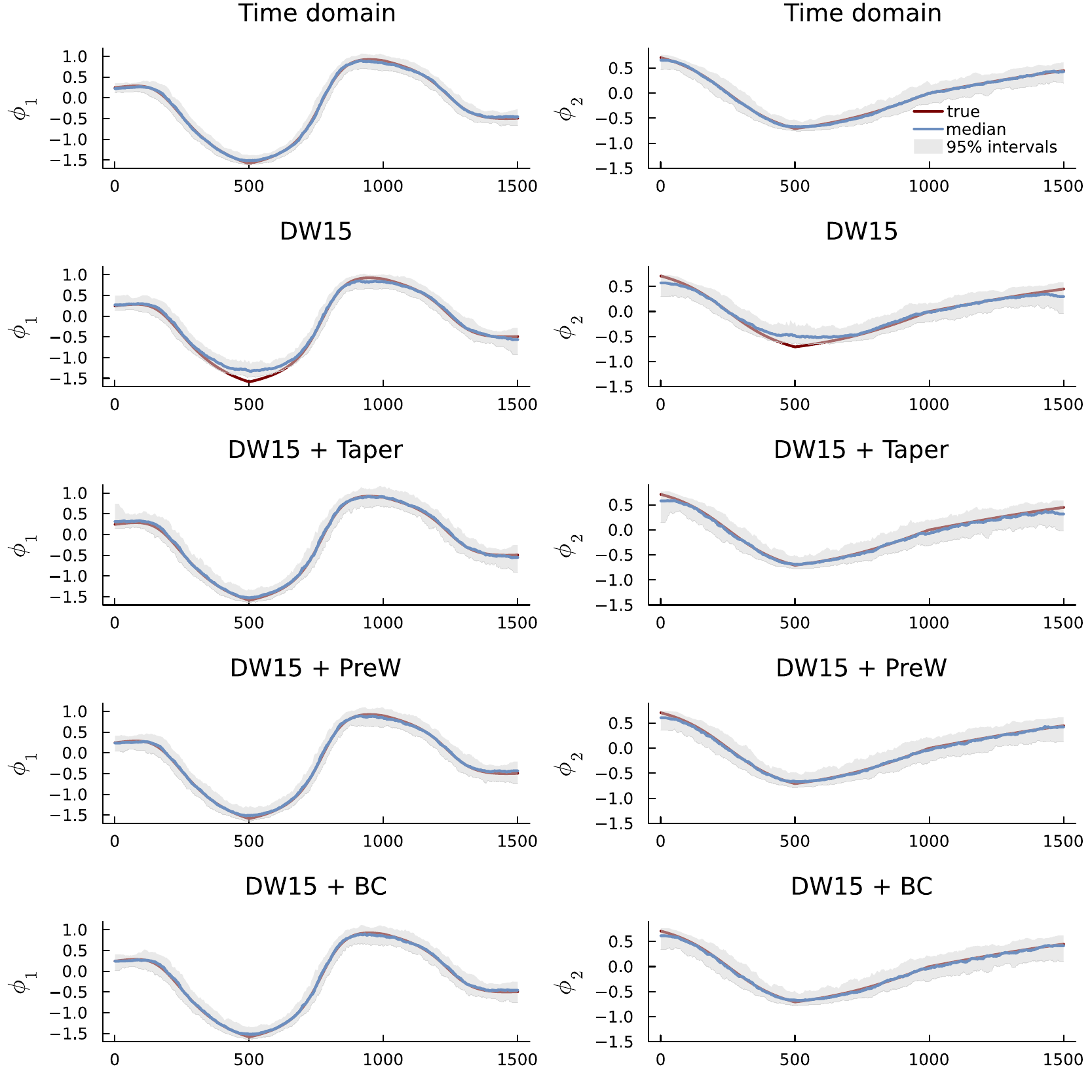}
    \caption{Experiment 3. Sampling distribution for the posterior median estimates of the parameters from the dynamic Whittle with $m=15$.}\label{fig:nearUnitSampDist_DW_M15}
\end{figure}

\begin{figure}[H]
    \centering
    \includegraphics[width=0.99\linewidth]{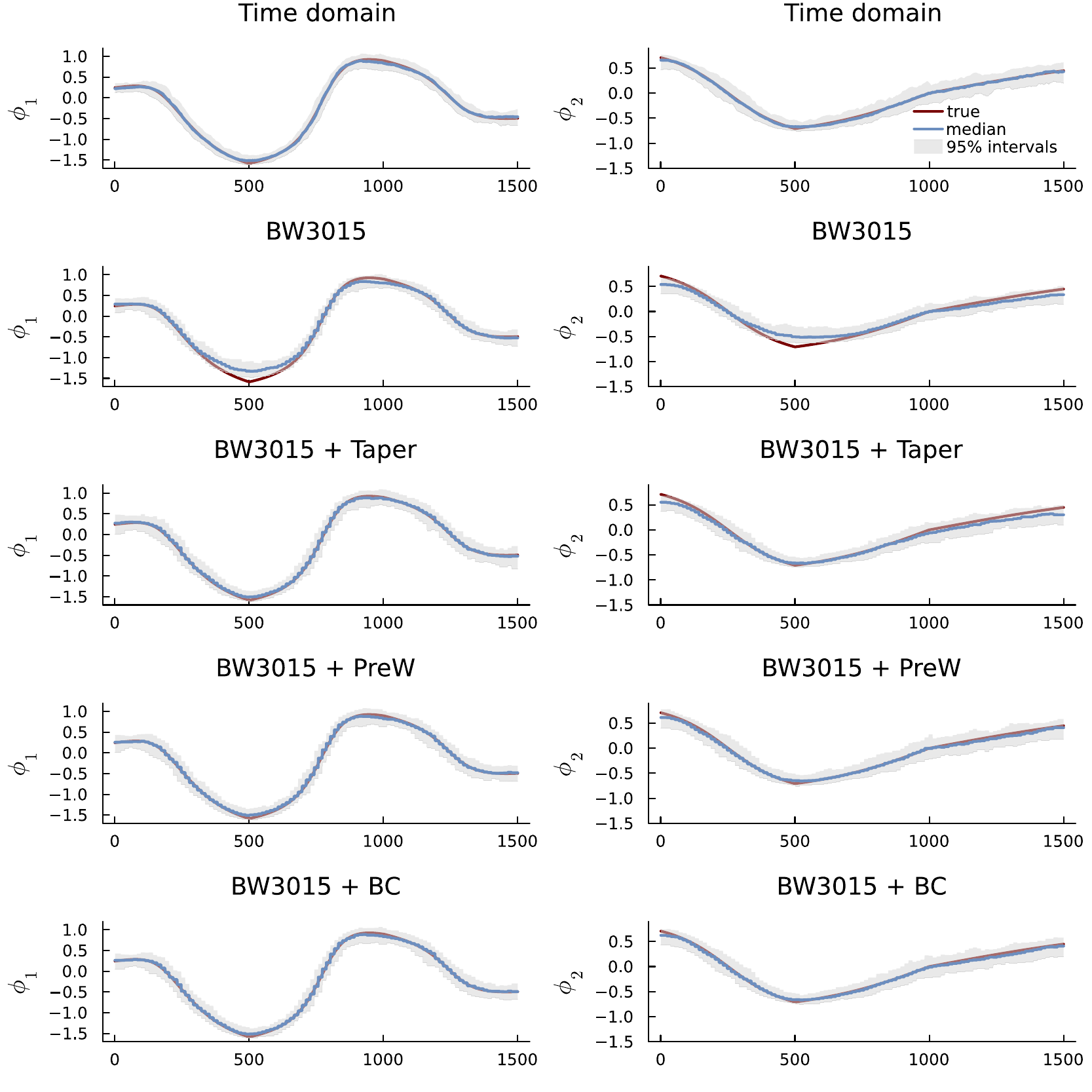}
    \caption{Experiment 3. Sampling distribution for the posterior median estimates of the parameters from the block Whittle with $N=30$ and $S=15$.}\label{fig:nearUnitSampDist_BW3015}
\end{figure}

Figures \ref{fig:nearUnitSampDist_DW_M15} and \ref{fig:nearUnitSampDist_BW3015} plots the sampling distributions for the posterior median estimates from the dynamic and block Whittle posteriors for experiment 3.

\newpage

\section{Additional results from the application to egg price data}\label{supp:eggprices}

\begin{figure}
    \includegraphics[width=0.99\linewidth]{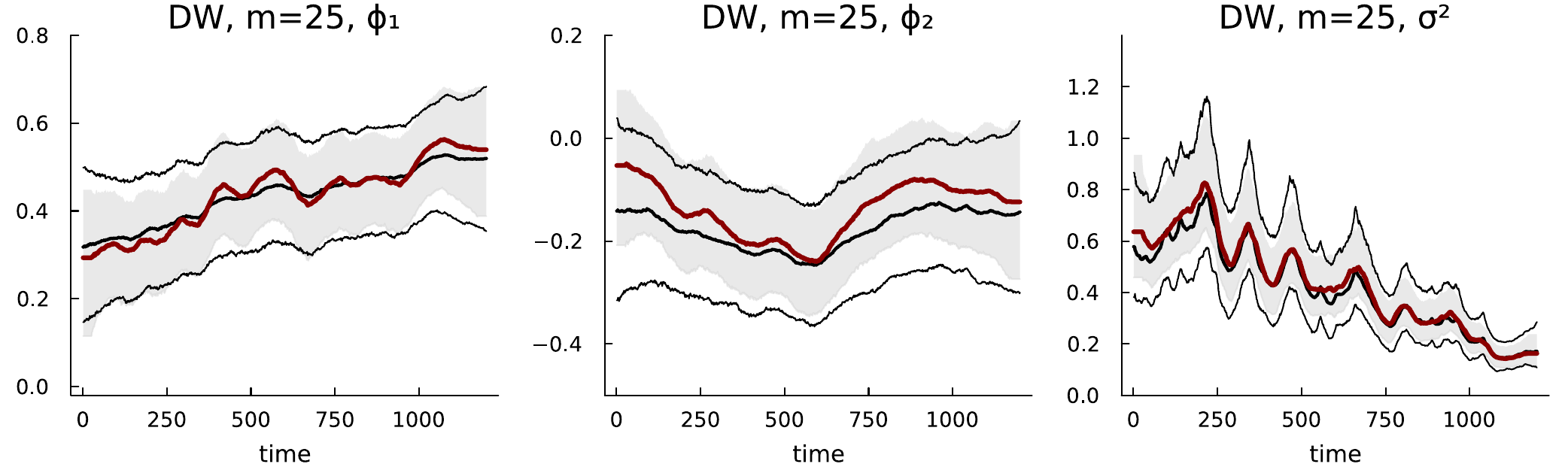}
    \includegraphics[width=0.99\linewidth]{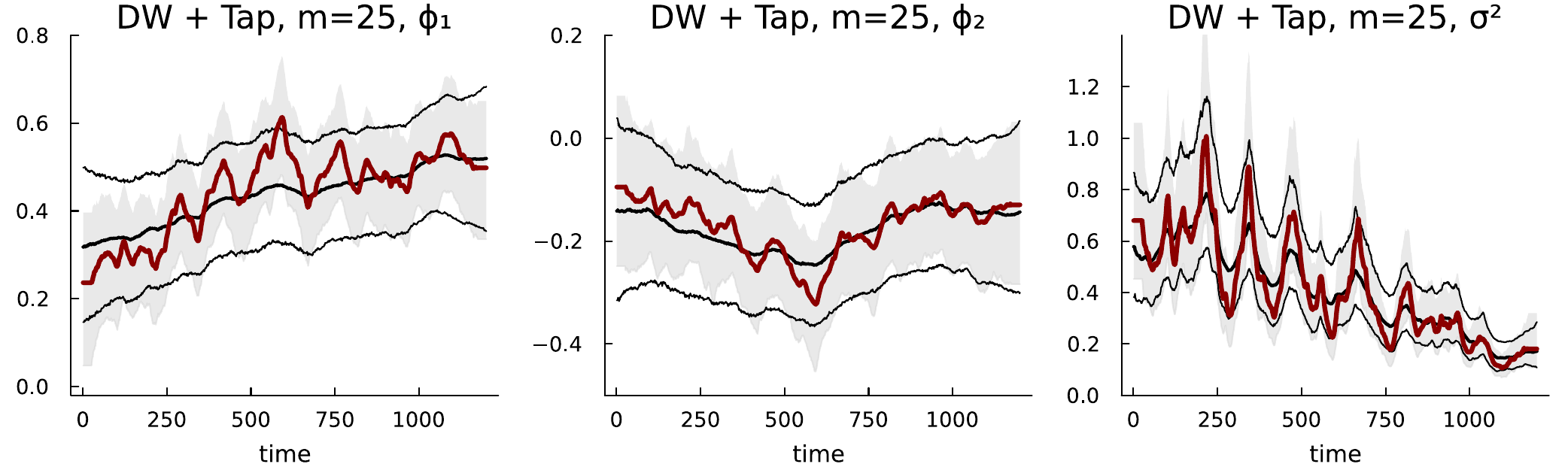}
    \includegraphics[width=0.99\linewidth]{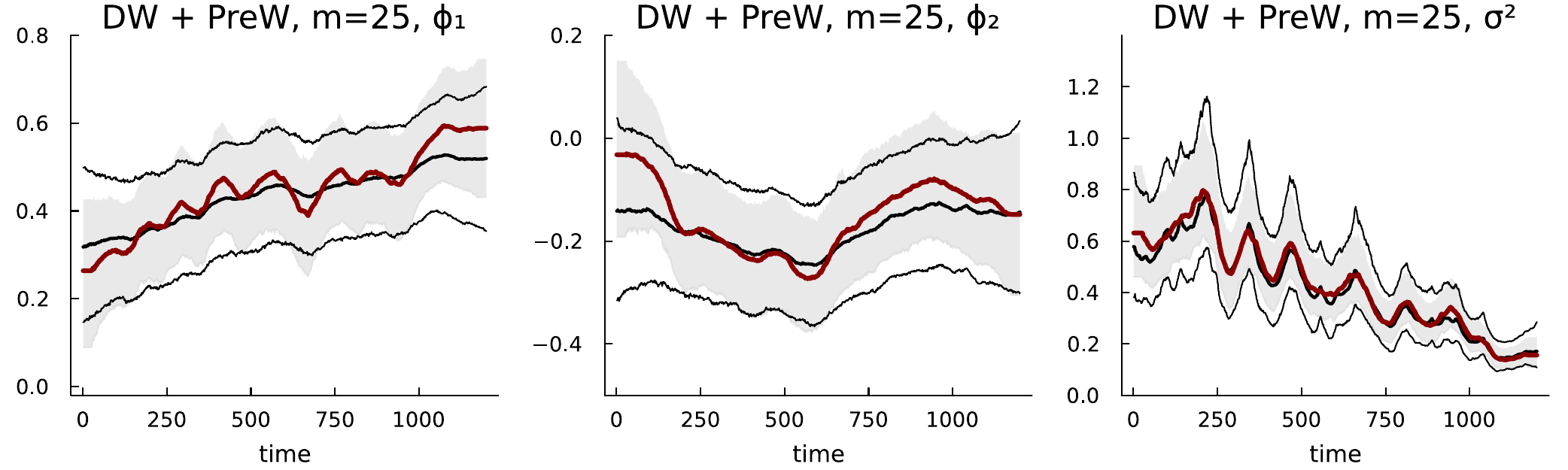}
    \includegraphics[width=0.99\linewidth]{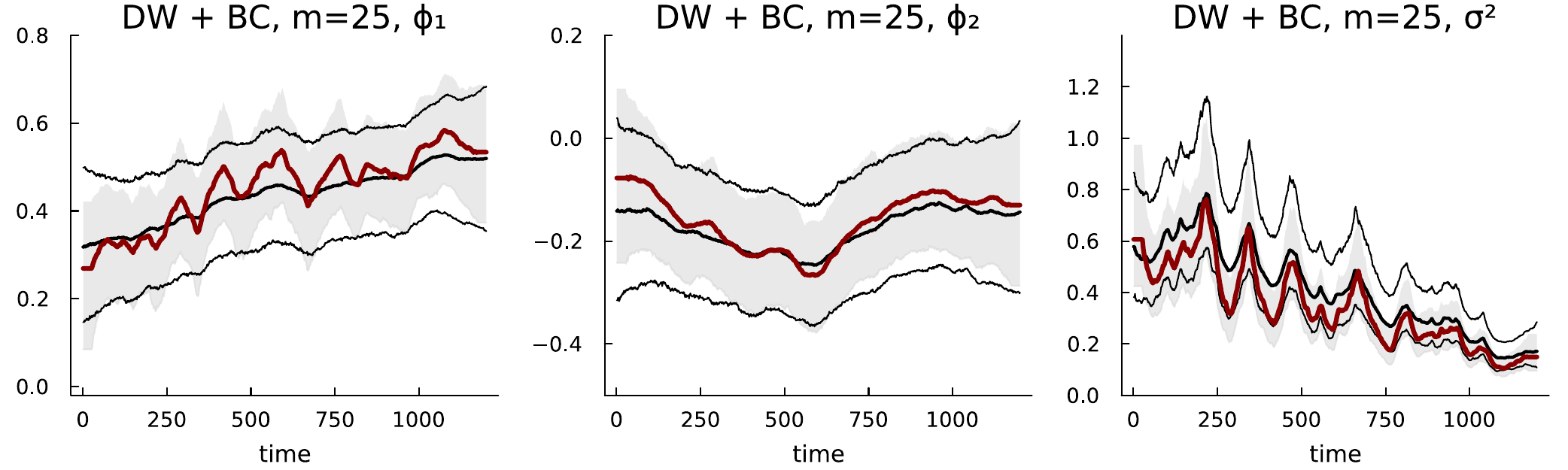}
     \caption{Egg price data. Posterior median (red) and 95\% probability bands (light-shaded grey) for the variance and AR parameter evolutions estimated from the dynamic Whittle likelihood compared to the corresponding measures for the time domain likelihood (black lines).}\label{fig:EggParEvolDW25}
\end{figure}

\begin{figure}
    \includegraphics[width=0.99\linewidth]{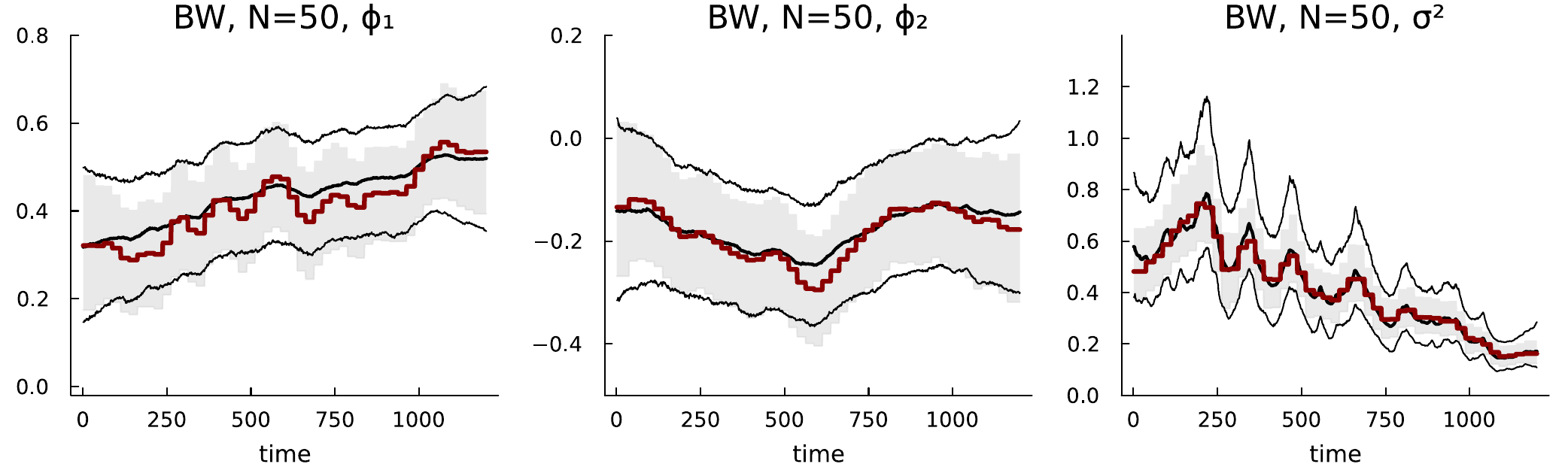}
    \includegraphics[width=0.99\linewidth]{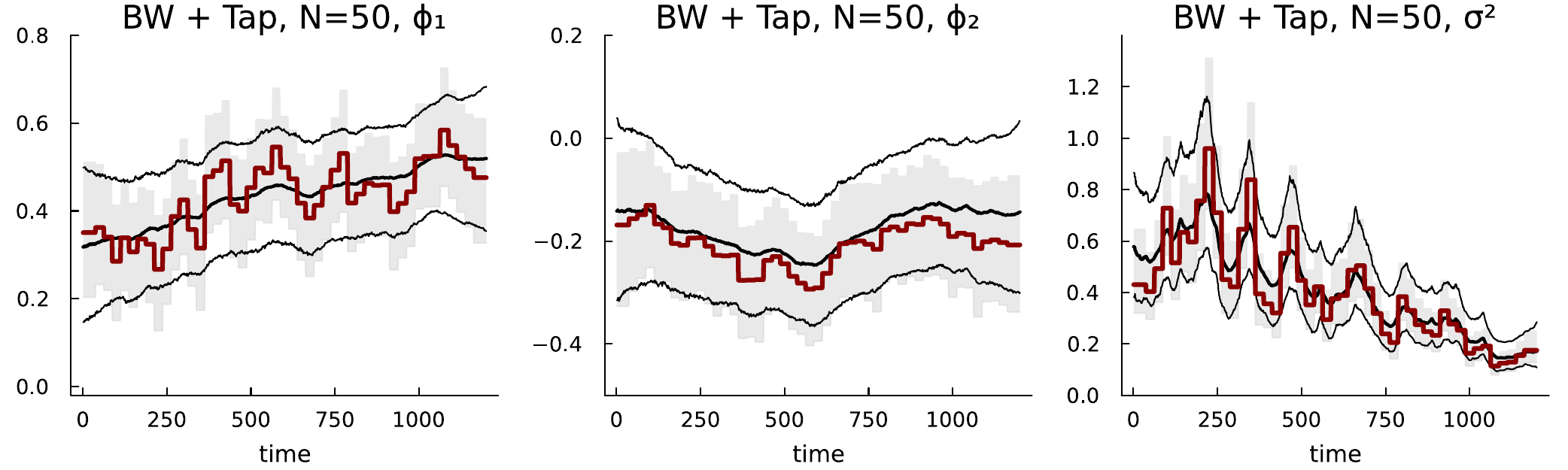}
    \includegraphics[width=0.99\linewidth]{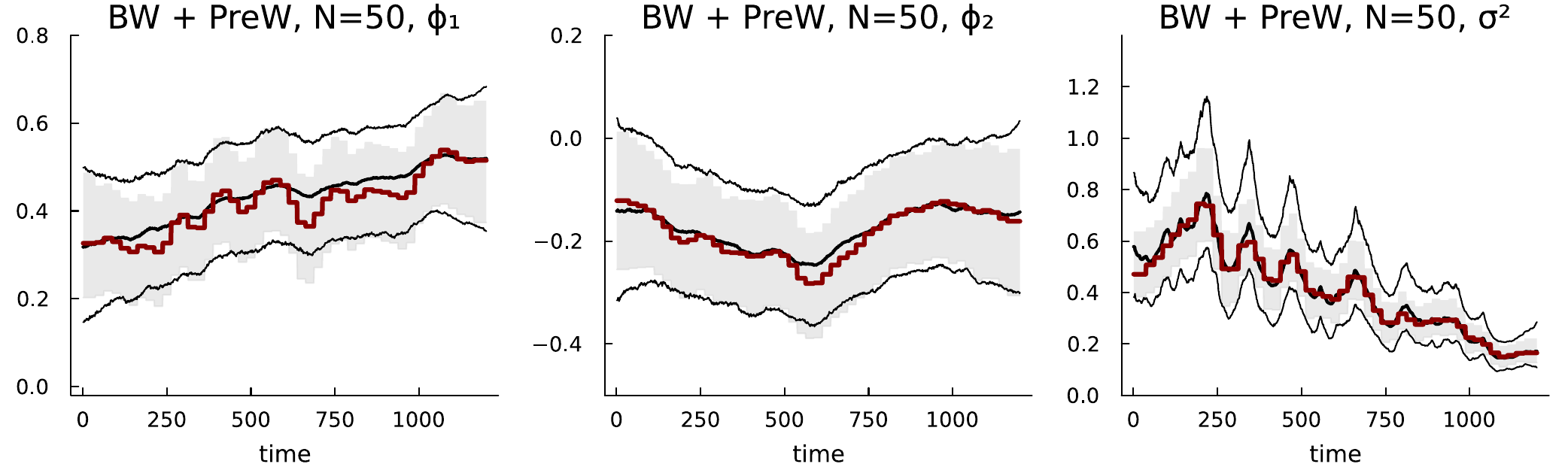}
    \includegraphics[width=0.99\linewidth]{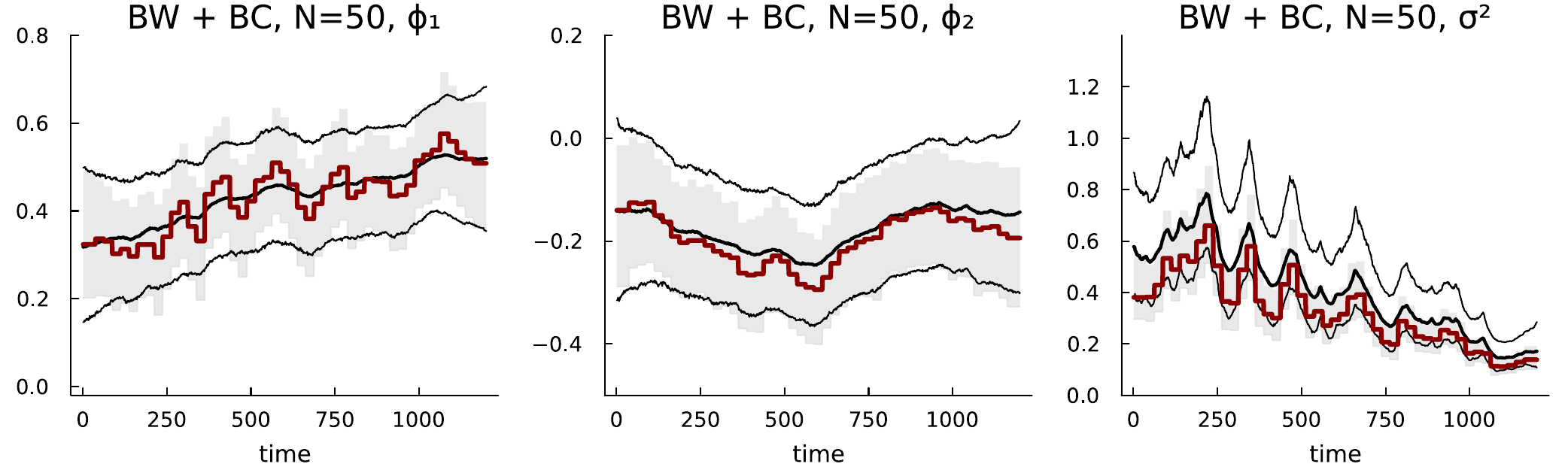}
     \caption{Egg price data. Posterior median (red) and 95\% probability bands (light-shaded grey) for the variance and AR parameter evolutions estimated from the block Whittle likelihood compared to the corresponding measures for the time domain likelihood (black lines).}\label{fig:EggParEvolBW50}
\end{figure}

\begin{figure}
    \includegraphics[width=0.99\linewidth]{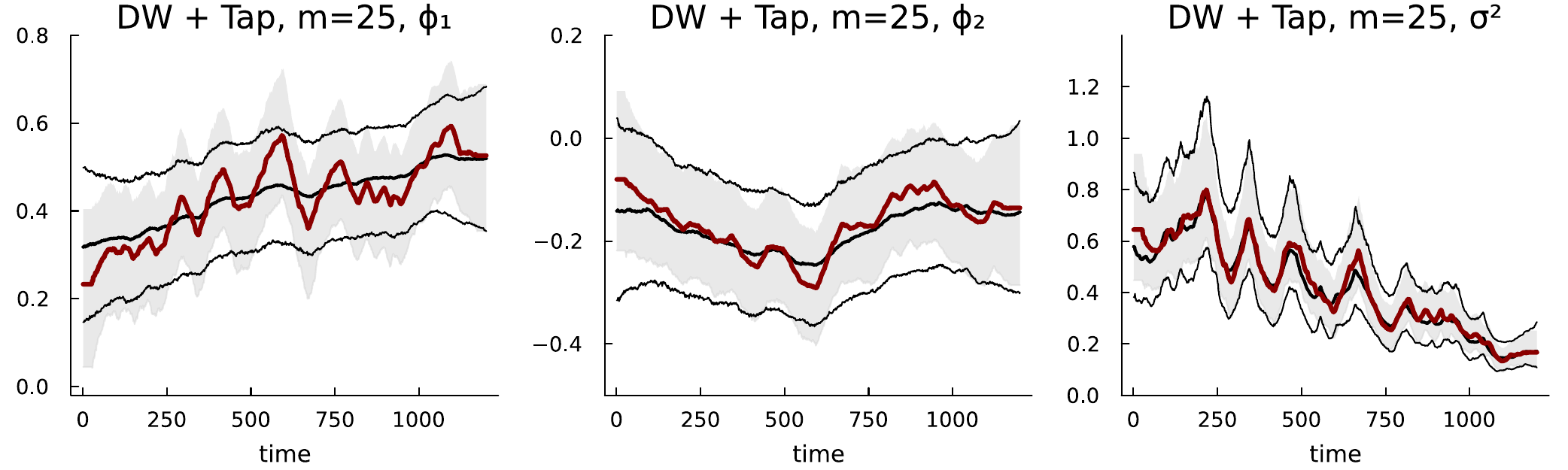}
    \includegraphics[width=0.99\linewidth]{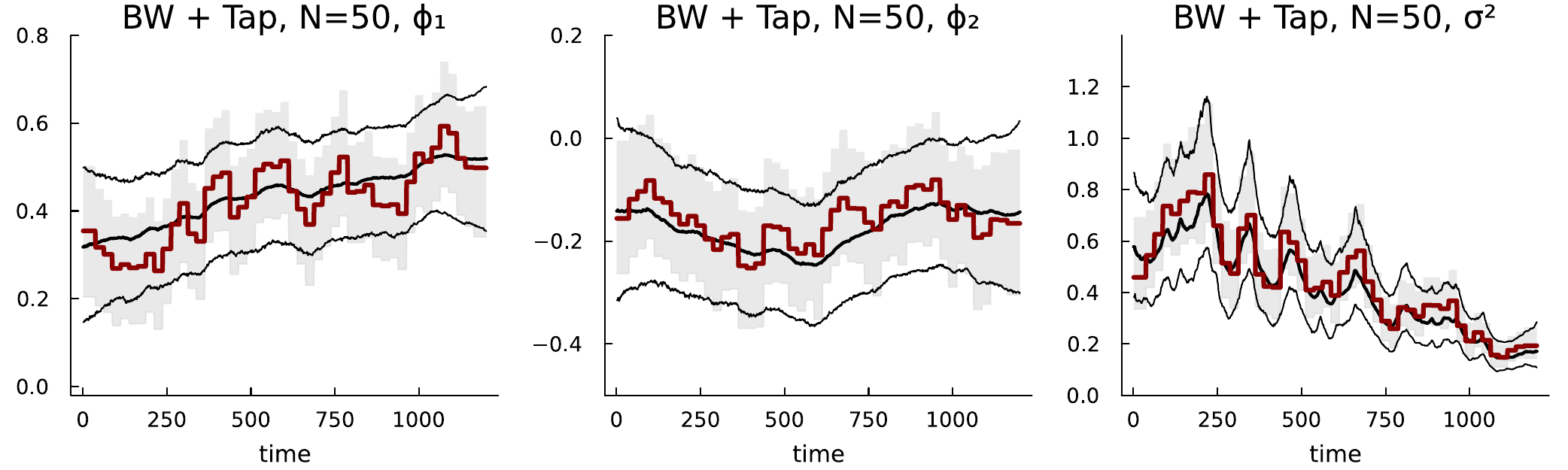}
     \caption{Egg price data. Posterior median (red) and 95\% probability bands (light-shaded grey) for the variance and AR parameter evolutions estimated from the tapered dynamic and block Whittle likelihood, with rescaling on each segment, compared to the corresponding measures for the time domain likelihood (black lines).}\label{fig:EggRescaleParEvolDW25}
\end{figure}

\begin{figure}
    \includegraphics[width=0.99\linewidth]{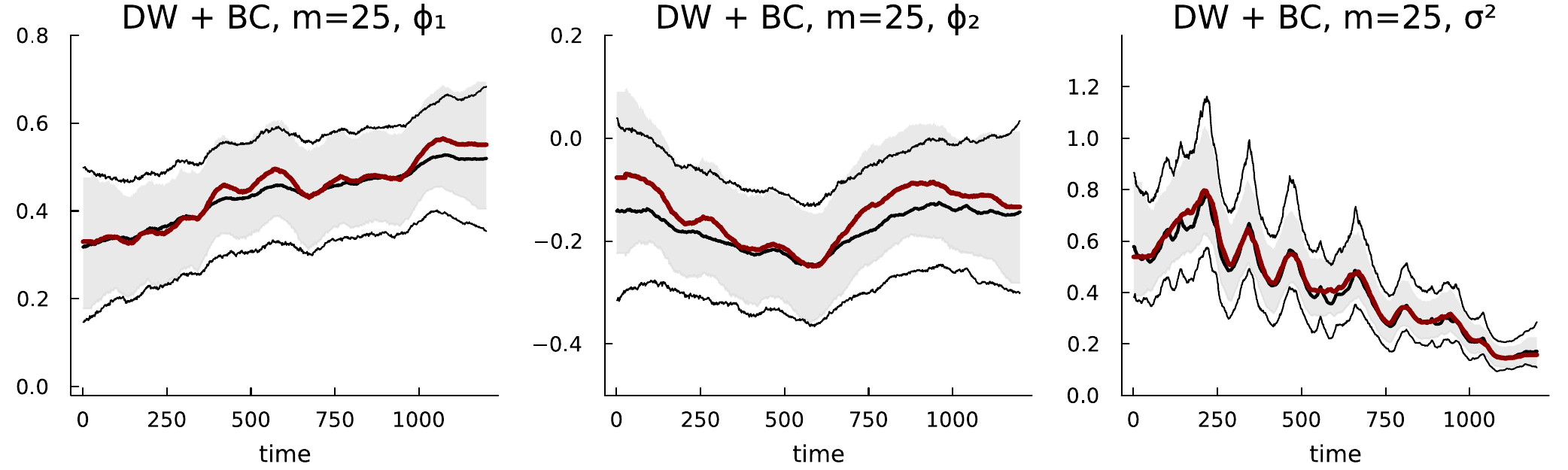}
    \includegraphics[width=0.99\linewidth]{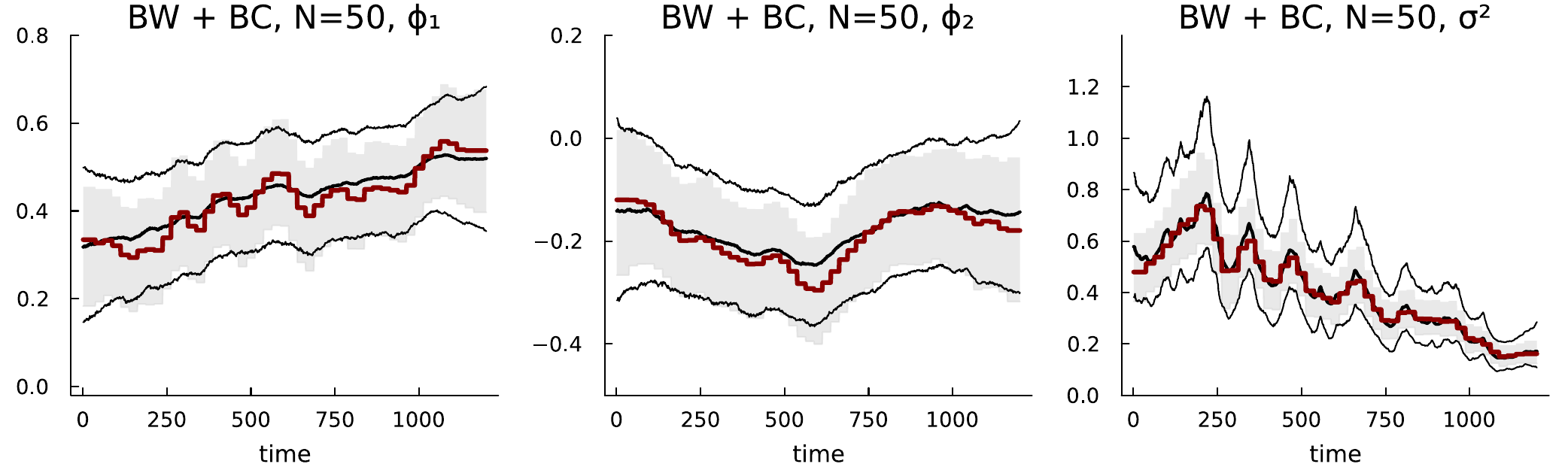}
     \caption{Egg price data. Posterior median (red) and 95\% probability bands (light-shaded grey) for the variance and AR parameter evolutions estimated from the boundary corrected dynamic and block Whittle likelihood, without tapering, compared to the corresponding measures for the time domain likelihood (black lines).}\label{fig:EggBCnoTapParEvolDW25}
\end{figure}

Figures \ref{fig:EggParEvolDW25} - \ref{fig:EggBCnoTapParEvolDW25} displays result for the egg prices data in Section \ref{sec:realdata} using the smaller segment size.

\end{document}